\documentclass[12pt,preprint]{aastex}

\usepackage{lineno}

\slugcomment{}

\shorttitle{{\it Fermi}'s BL Lacs}
\shortauthors{Ajello et al.}

\usepackage{calrsfs,euscript,mathrsfs}

\begin{document}

%% LaTeX will automatically break titles if they run longer than
%% one line. However, you may use \\ to force a line break if
%% you desire.

\title{The Cosmic Evolution of {\it Fermi} BL Lacertae Objects}

%% Use \author, \affil, and the \and command to format
%% author and affiliation information.
%% Note that \email has replaced the old \authoremail command
%% from AASTeX v4.0. You can use \email to mark an email address
%% anywhere in the paper, not just in the front matter.
%% As in the title, use \\ to force line breaks.

\author{
M.~Ajello\altaffilmark{1},
R.~W.~Romani\altaffilmark{2}, 
D.~Gasparrini\altaffilmark{3,4}, 
M.~S.~Shaw\altaffilmark{2}
%%% Alphabetical list now
J.~Bolmer\altaffilmark{5},
G.~Cotter\altaffilmark{6}, 
J.~Finke\altaffilmark{7}, 
J.~Greiner\altaffilmark{8}, 
S.~E.~Healey\altaffilmark{2}, 
O.~King\altaffilmark{9}, 
W.~Max-Moerbeck\altaffilmark{9}, 
P.~F.~Michelson\altaffilmark{2}, 
W.~J.~Potter\altaffilmark{6}, 
A.~Rau\altaffilmark{8}, 
A.~C.~S.~Readhead\altaffilmark{9}, 
J.~L.~Richards\altaffilmark{9}, 
P.~Schady\altaffilmark{8}
}
\altaffiltext{1}{Space Sciences Laboratory, 7 Gauss Way, University of California, Berkeley, CA 94720-7450, USA}
\altaffiltext{2}{W. W. Hansen Experimental Physics Laboratory, Kavli Institute for Particle Astrophysics and Cosmology, Department of Physics and SLAC National Accelerator Laboratory, Stanford University, Stanford, CA 94305, USA}
\altaffiltext{3}{Agenzia Spaziale Italiana (ASI) Science Data Center, I-00044 Frascati (Roma), Italy}
\altaffiltext{4}{Istituto Nazionale di Astrofisica - Osservatorio Astronomico di Roma, I-00040 Monte Porzio Catone (Roma), Italy}
\altaffiltext{5}{Technische Universit\"{a}t M\"{u}nchen, Physik Dept., James-Franck-Str., 85748 Garching, Germany}

\altaffiltext{6}{Department of Astrophysics, University of Oxford, Oxford OX1 3RH, UK}
\altaffiltext{7}{Space Science Division, Naval Research Laboratory, Washington, DC 20375-5352, USA}
\altaffiltext{8}{Max-Planck Institut f\"ur extraterrestrische Physik, 85748 Garching, Germany}
\altaffiltext{9}{Department of Astronomy, California Institute of Technology, Pasadena, CA 91125, USA}

\email{majello@ssl.berkeley.edu,rwr@astro.stanford.edu,gasparrini@asdc.asi.it
msshaw@stanford.edu}

\begin{abstract}
{\it Fermi} has provided the largest sample of $\gamma$-ray selected
blazars to date. In this work we use a uniformly selected set of 211 BL Lacertae
(BL Lac) objects detected by {\it Fermi} during its first year of operation.
We have obtained redshift constraints for 206 out of the 211 BL Lacs
in our sample making it the largest and most complete sample of BL Lacs
available in the literature. We use this sample to determine the luminosity
function of BL Lacs and its evolution with cosmic time. We find that for
most BL Lac classes, the evolution is positive with a space density
peaking at modest redshift (z$\approx$1.2). 
The low-luminosity, high-synchrotron peaked (HSP) 
BL Lacs are an exception, showing strong negative evolution, with number density
increasing for z$\lesssim$0.5. Since this rise corresponds to a drop-off in
the density of flat-spectrum radio quasars (FSRQs), a possible interpretation 
is that these HSPs represent an accretion-starved end-state of an earlier 
merger-driven gas-rich phase. We additionally find that the known BL Lac
correlation between luminosity and photon spectral index persists after
correction for the substantial observational selection effects
with implications for the so called `blazar sequence'.
Finally, 
estimating the beaming corrections to the luminosity function, 
we find that BL Lacs have an average Lorentz factor of $\gamma=6.1^{+1.1}_{-0.8}$, 
and that most are seen within 10$^{\circ}$ of the jet axis.
\end{abstract}


%%%%%%%%%%%%%%%%%%%%%%%%%%%%%%%%%%%%%%%%%%%%%%%%%%%%%%%%%%%%%%%%%%
\section{Introduction}

BL Lacertae (BL Lac) objects are a sub-population of blazars,
an extreme class of active galactic nuclei (AGN),
displaying highly variable emission
likely due to a relativistic jet 
pointing close to our line of sight \citep[e.g.][]{blandford78}.
They are distinguished from their siblings, the flat-spectrum radio
quasars (FSRQs) by an optical spectrum lacking any emission lines with 
equivalent width $>$5\,$\AA$
 \citep[e.g.][]{urry95,marcha96}.
The optical spectra of BL Lac objects are power-law dominated
indicating either especially strong non-thermal continuum (jet aligned 
very close to our line of sight) or unusually weak thermal disk/broad line 
emission \citep[plausibly attributed to low accretion activity;][]{giommi12}.

The synchrotron component{\footnote{BL Lacs and blazars in general can be classified according to the frequency, in the rest frame, of the peak
of the synchrotron component  as
low-synchrotron-peaked (LSP, $\nu_{peak}$$<$$10^{14}$\,Hz),
intermediate-synchrotron-peaked (ISP, $10^{14}$$<$$\nu_{peak}$$<$$10^{15}$\,Hz),
 and high-synchrotron-peaked (HSP, $\nu_{peak}$$>$$10^{15}$\,Hz).}} of BL Lacs shows a range of peak frequencies 
from $\nu\approx$10$^{13}$\,Hz up to $\nu\approx$10$^{17}$\,Hz 
\citep[e.g.][]{2LAC}.  At the high end, these synchrotron peaks 
imply that BL Lacs are able to accelerate electrons beyond 100\,TeV 
\citep[e.g.][]{costamante01,tavecchio11}, making BL Lacs among the most 
powerful accelerators in the Universe.

The lack of strong emission lines hampers traditional
optical spectroscopic measurements of the redshifts of most
BL Lac objects. Indeed, roughly 55\,\% of the 395
BL Lac objects detected in the second {\it Fermi} AGN  catalog
\citep[2LAC, ][]{2LAC} lacked a spectroscopic redshift.
This limitation is also serious at lower frequencies \citep{padovani07} and
the large redshift incompleteness of most BL Lac samples
has hampered so far the determination of a reliable luminosity
function. In turn this handicaps studies of the growth and evolution of BL Lac
objects in the Universe and the relationship between BL Lacs and the FSRQ class.

While it is clear that FSRQs evolve positively at all frequencies 
\citep[i.e. there were more blazars 
in the past,][]{dunlop90} up to a redshift cut-off which depends on luminosity 
\citep[e.g.][]{padovani07,wall08b,ajello09b,fsrq12},
the evolution of BL Lacs remains a matter of debate.
Indeed, various studies have found that BL Lac objects evolved
negatively \citep[e.g.][]{rector00,beckmann03},
positively \citep[e.g.][]{marcha13} or
not at all \citep{caccianiga02,padovani07}. These discrepancies might be 
due to small samples, biases in the set of BL Lacs  and substantial 
redshift incompleteness in these works.

At gamma-ray energies the need for a reliable Luminosity Function (LF) is
particularly acute. Indeed, the present lack of a secure LF makes it 
impossible to estimate the contribution of faint (below detection threshold) 
BL Lacs to the isotropic gamma-ray background \citep[IGRB,][]{lat_edb}.
At GeV-TeV energies BL Lacs are characterized by a harder spectrum
than FSRQs and are found to outnumber (by a factor $>$3) the latter 
particularly above 10\,GeV \citep{pop_pap}. Thus at high energies these sources may well
dominate the cosmic gamma-ray background.

Thanks to the excellent sensitivity,
the Large Area Telescope (LAT) on board {\it Fermi} has
detected $\sim$395 BL Lac objects in the first 2\,years of operations \citep{2LAC}.
To study this sample many different techniques have been employed to obtain
redshift estimates or constraints for these blazars
\cite[see ][]{rau12,shaw13,shaw13b}, yielding the rather surprising detection
of several BL Lacs up to redshift z$\approx$2. These high--z objects
often show a very hard (photon index of $\sim$2) GeV spectrum making
them the most luminous BL Lacs of the high-synchrotron peaked (HSP)
kind ever detected. How these objects fit within the scheme of
the blazar population and blazar sequence is still highly debated
\citep{padovani12,ghisellini12}.

In this work we study the cosmological properties of BL Lacs
focusing on a complete set of 211 BL Lacs detected by {\it Fermi}-LAT during
the first year of operation \citep{agn_cat}. Using the full range of techniques
\cite[see ][]{rau12,shaw13}, we have obtained spectroscopic redshifts or
limits for the great majority ($\sim$98\,\%) of the sources.
This has let us derive the first detailed models for the luminosity function and evolution
of BL Lacs at GeV energies. The large sample size and unusually high redshift completeness
allow new inferences about the nature of the BL Lac population, as a whole.
This paper is organized as follows: $\S$~\ref{sec:sample} and $\S$~\ref{sec:analysis} 
present the properties of the sample, discuss the available redshift
constraints and describe the method used to derive the luminosity function.
The results are presented and discussed in $\S$~\ref{sec:results}, \ref{sec:subclasses}, and  \ref{sec:discussion}.
Throughout this paper, a standard concordance cosmology was assumed
(H$_0$=71\,km s$^{-1}$ Mpc$^{-1}$, $\Omega_M$=1-$\Omega_{\Lambda}$=0.27).

%%%%%%%%%%%%%%%%%%%%%%%%%%%%%%%%%%%%%%%%%%%%%%%%%%%%%%%%%%%%%%%%%%
\section{The Sample}
\label{sec:sample}
The First Fermi LAT Catalog \citep[1FGL,][]{cat1} presented 
more than 1400 sources  detected by {\it Fermi}-LAT during 
its first year of operation. The first LAT AGN catalog \citep[1LAC,][]{agn_cat}
associates $\sim$700 of the high-latitude 1FGL sources ($|b|\geq10^{\circ}$)
with AGN of various types, most of which are blazars.
The sample used for this analysis consists of sources detected by the pipeline developed
by \cite{pop_pap} with a test statistic \footnote{The test statistics (or TS) is defined as:
TS=$-2({\rm ln} L_0 - {\rm ln} L_1)$. Where $L_0$ and $L_1$
are the likelihoods of the background (null hypothesis) and 
the hypothesis being tested (e.g. source plus background).
According to \cite{wilks38}, the TS is expected to be asymptotically
distributed as $\chi_n^2$ in the null hypothesis, where $n$ is the additional
number of free parameters that are optimized for the alternative hypothesis.
Given the 4 degrees of freedom required for source detection
(position and spectral parameters),
a TS of 50 corresponds to $\sim$6.3\,$\sigma$ of a Gaussian distribution.}
(TS)  greater (or equal) than 50 
and with $|b|\geq$15$^{\circ}$.  For these sample cuts we have produced a set
of Monte Carlo simulations that can be used to determine and correct for
the selection effects. This sample contains 486 objects, 211 of which are 
classified as BL Lacs in 1LAC. 
The composition of this sample  is reported in Table~\ref{tab:sample}.
The source classifications reported in Table~\ref{tab:sample} are originally
drawn from the 1LAC and 2LAC catalogs \citep{agn_cat}, and have been 
complemented  with newer observations reported in \cite{shaw12} and \cite{shaw13}.

The 211 BL Lacs detected by {\it Fermi} with TS$\geq50$, $|b|\geq15^{\circ}$,
constitute the sample that will be used in this analysis.
All these objects are reported together with their properties
in the Table reported in the Appendix ($\S$~\ref{sec:table}).
We note that fluxes and photon indices reported
there are those measured  with the pipeline developed
by \cite{pop_pap} and thus, while compatible with the values reported
in the 1FGL catalog \citep{cat1}, they are not exactly the same.
These values are meant to be used with the results of the Monte Carlo
simulations to correctly account for selection effects \citep[see $\S$~4 and $\S$~5 in][]{pop_pap}.

Of the 38 sources remaining unclassified in 
1FGL, three objects now have pulsar identifications, two sources have been dropped
as spurious composites and 10 are flagged as pulsar candidates based
on their variability and spectral properties \citep{unassociated12}.
This leaves 23 objects which might be blazars yet to be identified.
Recent radio observations \citep{petrov13}
find compact source counterparts for 11 of these, so it is likely that these
11 represent missing BL Lacs. Moreover, cross-correlating the list
of 23 objects with the WISE sources whose colors are typical of blazars
\citep{massaro12_gammastrip,dabrusco12}  we found  an additional
8 blazar candidates. Thus a total of 19 sources display properties
of blazars on the basis of their IR colors or radio properties.
Conservatively we assume that all these sources might be BL Lacs, and
that the incompleteness (due to missing identification) in our BL Lac
sample is 19/211=9\,\%. The total incompleteness (due to missing redshifts
and identifications) is thus $\sim$11\,\%.
As it will be shown later this incompleteness does not constitute a problem
for the analysis.

\begin{deluxetable}{lc}
\tablewidth{0pt}
\tablecaption{Composition of the $|b|\geq$15$^{\circ}$, TS$\geq$50, 
sample used in this analysis. 
\label{tab:sample}}
\tablehead{
%%%%%%%% column names
\colhead{Class} & \colhead{\# objects }}
\startdata
Total                         & 486 \\
BL Lacs                       & 211\\
FSRQs                         & 186\\
Pulsars                       & 31 \\
Dropped by 2FGL		      & 2 \\
Other\tablenotemark{a}        & 33 \\ % incl 17 low z AGN, Gal, LINERss. 16 other radio sources no z.
%Radio Associations\tablenotemark{b}          & 17 \\ % incl 17 low z AGN, Gal, LINERss. 16 other radio sources no z.
% ----- The ones below are the UNID
Unassociated  sources         & 23 \\  % no associated radio source.

\enddata
\tablenotetext{a}{Includes starburst galaxies, LINERS,
narrow line Seyfert 1 objects, Seyfert galaxy candidates and
{\it Fermi} sources with a radio counterpart, but no
% Marco, if we count radio counterparts, then we get 33--> 40 (7 more from Petrov et al....)
optical type or redshift measurement.
}
\end{deluxetable}

%%%%%%%%%%%%%%%%%%%%%%%%%%%%%%%%%%%%%%%%%%%%%%%%%%%%%%%%%%%%%%%%%%
\section{Analysis}
\label{sec:analysis}

\subsection{Method}
\label{sec:ml}

In order to derive the LF of BL Lacs we rely on the maximum likelihood (ML) method first introduced by \cite{marshall83} 
and used recently  for the study of blazars detected by {\it Swift} \citep{ajello09b} and FSRQs
detected by {\it Fermi} \citep{fsrq12}.
The aim of this analysis is to determine the space
density of BL Lacs as a function of rest-frame 0.1--100\,GeV luminosity
(L$_{\gamma}$), redshift (z) and photon index ($\Gamma$), by fitting to the functional form:
\begin{equation}
\label{eq:1}
\frac{\partial^3 N}{\partial L_{\gamma} \partial z \partial \Gamma}= \frac{\partial^3N}{\partial L_{\gamma} \partial V \partial \Gamma}%\times 
%\frac{dN}{d\Gamma} 
\times \frac{dV}{dz} 
=\Phi(L_{\gamma},V(z),\Gamma)% \times  \frac{dN}{d\Gamma} 
\times \frac{dV}{dz}
\end{equation}
where $\Phi(L_{\gamma},V(z),\Gamma)$ is the luminosity function,  and $dV/dz$ is the co-moving volume element per unit redshift and unit solid 
angle \citep[see e.g.][]{hogg99}.

 The best-fit LF is found by comparing, through a maximum-likelihood estimator, the 
number of expected objects (for a given model LF) to the observed
number while accounting for selection effects in the 
detection of gamma-ray sources.
In this method, the space of luminosity, redshift, and photon index is divided 
into small intervals of size $dL_{\gamma}dz\,d\Gamma$. In each
element, the expected number of blazars with luminosity $L_{\gamma}$,
redshift $z$ and photon index $\Gamma$ is:
\begin{equation}
\lambda(L_{\gamma},z, \Gamma)dL_{\gamma}dz d\Gamma   = 
\Phi(L_{\gamma},V(z),\Gamma)\cdot \Omega(L_{\gamma},z,\Gamma)\ \frac{dV}{dz}\  dL_{\gamma} dz d\Gamma
\label{eq:lambda}
\end{equation}
where $\Omega(L_{\gamma},z,\Gamma)$ is the sky coverage and represents the probability of 
detecting in this survey a blazar with luminosity $L_{\gamma}$,
redshift $z$ and photon index $\Gamma$. This probability was derived for 
the sample used here by \cite{pop_pap} and the reader is referred to that
paper for more details. With sufficiently fine sampling of
the $L_{\gamma}-z-\Gamma$ space the infinitesimal element will either contain 0 or 1 BL Lac.
In this regime one has a likelihood function based on joint Poisson
probabilities:
\begin{equation}
L = \prod_i \lambda(L_{\gamma,i},z_i,\Gamma_i) dL_{\gamma} dz  d\Gamma
e^{-\lambda(L_{\gamma,i},z_i,\Gamma_i) dL_{\gamma} dz d\Gamma}
\times \prod_j e^{-\lambda(L_{\gamma,j},z_j,\Gamma_j) dL_{\Gamma} dz d\Gamma}
\end{equation}
This is the combined probability of detecting one blazar in each bin of 
$(L_{\gamma,i},z_i,\Gamma_i)$ populated by one observed {\it Fermi} BL Lac and 
zero BL Lacs for all other $(L_{\gamma,j},z_j,\Gamma_j)$.
Transforming to the standard expression $S=-2\ln\ L$ and dropping
terms which are not model dependent, we obtain:
\begin{equation}
S = -2\sum_i  {\rm \ln} \frac{\partial^3 N}{\partial L_{\gamma} \partial z \partial \Gamma}+ 2 
\int^{\Gamma_{max}}_{\Gamma_{min}} \int^{L_{\gamma,max}}_{L_{\gamma,min}} 
\int^{z_{max}}_{z_{min}} \lambda(L_{\gamma},\Gamma,z) dL_{\gamma} dz d\Gamma
\label{eq:s}
\end{equation}
The limits of integration of Eq.~\ref{eq:s} and subsequent equations, unless otherwise stated, are:
$L_{\gamma,min}=7\times 10^{43}$\,erg s$^{-1}$, $L_{\gamma,max}$=10$^{52}$\,erg s$^{-1}$, 
$z_{min}=$0.03, $z_{max}$=6, $\Gamma_{min}=$1.45 and $\Gamma_{max}=$2.80.
The results of this analysis are independent of the choice of the  maximum
redshift and luminosity. All other limits correspond to those spanned
by the set of sources analyzed here.
The best-fit parameters are determined by minimizing\footnotemark{}
\footnotetext{The MINUIT minimization package, embedded in ROOT (root.cern.ch),
has been used for this purpose.}
$S$ and the associated 1\,$\sigma$ errors are computed via bootstrap analysis
(see later).
While computationally intensive, Eq.~\ref{eq:s} has the advantage that each 
source has its appropriate individual detection efficiency and k-correction\footnote{The k-correction is the ratio of source rest-frame luminosity to observed luminosity and allows to transform an observed luminosity into a rest-frame one.}
treated independently.

	To test whether the best-fit LF provides a good description
of the data we compare the {\it observed} redshift, luminosity, index
and source count distributions against the prediction of the LF.
The first three distributions  can be obtained from the LF as:
\begin{eqnarray}
\frac{dN}{dz} & = & \int^{\Gamma_{max}}_{\Gamma_{min}} \int^{L_{\gamma,max}}_{L_{\gamma,min}} \lambda(L_{\gamma},\Gamma,z) dL_{\gamma}  d\Gamma \\
\frac{dN}{dL_{\gamma}} & = & \int^{\Gamma_{max}}_{\Gamma_{min}} 
\int^{z_{max}}_{z_{min}} \lambda(L_{\gamma},\Gamma,z) dz d\Gamma \\
\frac{dN}{d\Gamma}  & = &  \int^{L_{\gamma,max}}_{L_{\gamma,min}} 
\int^{z_{max}}_{z_{min}} \lambda(L_{\gamma},\Gamma,z) dL_{\gamma} dz 
\end{eqnarray}
where the limits of integration are the same as in Eq.~\ref{eq:s}.
The source count distribution can be derived as:
\begin{equation}
N(>F) =  \int^{\Gamma_{max}}_{\Gamma_{min}} 
\int^{z_{max}}_{z_{min}}  \int^{L_{\gamma,max}}_{L_{\gamma}(z,F)}
 \Phi(L_{\gamma},V(z),\Gamma) \frac{dV}{dz} d\Gamma dz dL_{\gamma}
\label{eq:logn}
\end{equation}
where $L_{\gamma}(z,F)$ is the luminosity of a source at redshift $z$
having a flux of $F$.

To display the LF we rely on the ``N$^{obs}$/N$^{mdl}$'' method
devised by \cite{lafranca97} and \cite{miyaji01} and 
employed in several recent works \citep[e.g.][]{lafranca05,hasinger05}.
Once a best-fit function for the LF has been found, it is possible
to determine the value of the observed LF in a given bin of luminosity
and redshift:
\begin{equation}
\label{eq:nmdl}
\Phi(L_{\gamma,i},V(z_i),\Gamma_i) = \Phi^{mdl}(L_{\gamma,i},V(z_i),\Gamma_i) \frac {N^{obs}_i}{N^{mdl}_i}
\end{equation}
where $L_{\gamma,i}$, $z_i$ and $\Gamma_i$ are the luminosity, redshift 
and photon index of the i$^{th}$
bin, $\Phi^{mdl}(L_{\gamma,i},V(z_i),\Gamma_i)$ is the best-fit LF model and $N^{obs}_i$ 
and $N^{mdl}_i$ are the observed and the predicted numbers of BL Lacs in that bin. 
These two techniques (the \cite{marshall83} ML method and the ``N$^{obs}$/N$^{mdl}$'' 
estimator) provide a minimally biased estimate of the luminosity function 
\citep[cf.][]{miyaji01}.

%%%%%%%%%%%%%%%%%%%%%%%%%%%%%%%%%%%%%%%%%%%%%%%%%%%%%%%%%%%%%%%%%%%%%%%%%%
%%%%%%%%%%%%%%%%%%%%%%%%%%%%%%%%%%%%%%%%%%%%%%%%%%%%%%%%%%%%%%%%%%%%%%%%%%
\subsection{Parametrization of the Luminosity Function}
\label{sec:lf}

We model the intrinsic distribution of photon indices with a Gaussian, which
implies that for a given redshift $z$ and luminosity $L_{\gamma}$ the LF is:
\begin{equation}
\label{eq:index}
\Phi(L_{\gamma},z,\Gamma) \propto  e^{-\frac{ (\Gamma-\mu(L_{\gamma}))^2}{2\sigma^2}}
\end{equation}
where $\mu$ and $\sigma$ are, respectively, the Gaussian mean and dispersion.
To test possible correlation
of the photon index with luminosity, as previously noted in the literature
\citep[see e.g.][]{ghisellini09b,meyer12b},
we allow the mean\footnote{We also
tested a scenario for which $\sigma$ depends on the source luminosity
or the redshift, but we did not find any evidence for such
trends.} $\mu$ to be a function
of the source luminosity:
\begin{equation}
\label{eq:blazseq}
\mu(L_{\gamma}) = \mu^* + \beta \times ( Log_{10}(L_{\gamma}) - 46).
\end{equation}

The LF at redshift z=0 is modeled as a smoothly-joined  double power law
multiplied by the photon index distribution of Eq.~\ref{eq:index}:

\begin{equation}
\Phi(L_{\gamma},z=0, \Gamma) =\frac{A}{\ln(10)L_{\gamma}}
\left[\left(\frac{L_{\gamma}}{L_{*}}\right)^{\gamma_1}+
\left(\frac{L_{\gamma}}{L_{*}}\right)^{\gamma2} 
\right]^{-1}\cdot  e^{-\frac{ (\Gamma-\mu(L_{\gamma}))^2}{2\sigma^2}}
\label{eq:lf0}
\end{equation}
To parametrize the evolution of the LF we employ three commonly assumed
evolutionary trends: a pure-density evolution (PDE), a pure luminosity evolution
(PLE) and a luminosity-dependent density evolution (LDDE).

For both the PDE and PLE case we rely on an evolutionary factor defined as:
\begin{equation}
e(z) = (1+z)^{k_d} e^{z/\xi}.
\label{eq:ez}
\end{equation}
where 
\begin{equation}
k_d=k^* + \tau \times (Log_{10}(L_{\gamma})-46).
\label{eq:tau}
\end{equation}
For the PDE the evolution is defined as:
\begin{equation}
\Phi(L_{\gamma},z,\Gamma) = \Phi(L_{\gamma},z=0,\Gamma) \times e(z)
\label{eq:pde}
\end{equation}
while for the PLE case it is:
\begin{equation}
\Phi(L_{\gamma},z,\Gamma) = \Phi(L_{\gamma}/e(z),\Gamma).
\label{eq:ple}
\end{equation}

The PLE and PDE models have  10 free parameters ($A$, $\gamma_1$, $L_*$,
$\gamma_2$, $k^*$, $\tau$, $\xi$, $\mu^*$, $\beta$, and $\sigma$).

For the LDDE we adopt the same parametrization reported in \cite{fsrq12}:
\begin{equation}
\Phi(L_{\gamma},z,\Gamma) = \Phi(L_{\gamma},z=0,\Gamma) \times e(z,L_{\gamma})
\end{equation}
where
\begin{equation}
e(z,L_{\gamma})= \left[ 
\left( \frac{1+z}{1+z_c(L_{\gamma})}\right)^{p1(L_{\gamma})} + 
\left( \frac{1+z}{1+z_c(L_{\gamma})}\right)^{p2} 	
 \right]^{-1}
\label{eq:evol}
\end{equation}
\begin{equation}
z_c(L_{\gamma})= z_c^*\cdot (L_{\gamma}/10^{48})^{\alpha}  .
\label{eq:zpeak}
\end{equation}
\begin{equation}
p1(L_{\gamma}) = p1^* + \tau \times (Log_{10}(L_{\gamma})-46)
\label{eq:p1}
\end{equation}
Here $\Phi(L_{\gamma},z=0,\Gamma)$ is the same double power law used in Eq.~\ref{eq:lf0} and
$z_c(L_{\gamma})$ corresponds to the 
(luminosity-dependent) redshift where the evolution changes sign (positive to 
negative), with $z_c^*$ being the redshift peak for a BL Lac with a luminosity 
of 10$^{48}$\,erg s$^{-1}$. The LDDE model has a total of 12 free parameters
($A$, $\gamma_1$, $L_*$, $\gamma_2$, $z^*_c$, $p1^*$, $\tau$, $p2$, $\alpha$, 
$\mu^*$, $\beta$, and $\sigma$). Note that the evolutionary term
$e(z,L_{\gamma})$ in Eq.~\ref{eq:evol}
is not equal to one at redshift zero (see also $\S$~\ref{sec:ldde}).

%%%%%%%%%%%%%%%%%%%%%%%%%%%%%%%%%%%%%%%%%%%%%%%%%%%%%%%%%%%%%%%%%%%%%%%%%%
%%%%%%%%%%%%%%%%%%%%%%%%%%%%%%%%%%%%%%%%%%%%%%%%%%%%%%%%%%%%%%%%%%%%%%%%%%
\subsection{Dealing with Redshift Constraints}
\label{sec:z}

Only 103 of the 211 BL Lacs in our sample have a spectroscopic
redshift measurement \citep{2LAC}. However, for another 104 BL Lac objects
we were able to provide quantitative constraints on the redshift.
The constraints are:

\begin{itemize}
\item {\bf Photometric Redshift Estimates:}
The neutral hydrogen along the line of sight to the source efficiently absorbs photons 
with a rest-frame wavelength blue-wards of the Lyman-limit. This results in a 
flux depression that can be used to estimate the absorber's redshift 
via spectral energy distribution (SED) template fitting.  The absence of any drop-out provides
an upper limit to the source redshift limited by the bluest available pass band 
\citep[e.g., $z\le1.3$ based on {\it Swift/UVOT} in the study of][]{rau12}.
In our sample, three sources have a photometric redshift estimate while
34 have a photometric-redshift upper limit.
%%%%%%%%
%%%%%%%%
\item {\bf Redshift lower limits via intervening absorption systems:}
Metal line absorption systems (i.e. MgII, FeII, CIV etc.) in the optical spectra 
caused by intervening systems provide a firm lower limit to the source redshift 
\citep{shaw13}. In our sample 39 sources have a spectroscopic redshift
lower limit.
%%%%%%%%
%%%%%%%%
\item {\bf Spectroscopic Redshift Upper Limits:} \cite{shaw13} used
the absence of individual Lyman-$\alpha$ absorptions to provide statistically-based
upper limits for all the BL Lacs without redshifts. As reported there
the exclusion $z_{\rm max}$ falls in the 1.65$<$z$<$3.0 range. Although not as constraining
as the UV-based SED bounds from \citet{rau12}, we can extract these limits for
all objects with spectra.  All but 5 of our BL Lacs were in the \cite{shaw13} sample and thus
have a $z_{\rm max}$ estimate.
%%%%%%%%%
%%%%%%%%%
\item {\bf Host Galaxy Spectral Fitting:}
According to, e.g., \cite{urry00} and \cite{sbarufatti05} BL Lacs are hosted
by giant ellipticals with bright absolute magnitude of $M_{R}=-22.9\pm0.5$.
If one assumes that these objects are standard candles then the 
host {\it non-detection} places a lower limit on the source redshift.
\cite{shaw13} have improved this technique by fitting spectral templates
of elliptical galaxies to their BL Lac optical spectra and re-calibrating the host magnitudes
against the spectroscopically measured set.
For each trial redshift $z_i$ they are able to test the hypotheses
of whether the optical spectrum is compatible (aside from
the featureless BL Lac emission) with the red-shifted emission
of the host galaxies. Thus, for every object they are able 
to provide {\it exclusion probabilities} for the source redshift
as a function of redshift.  Again all but 5 of our BL Lacs lacking 
spectroscopic redshifts have exclusion probabilities from \cite{shaw13}.
\end{itemize}

The five sources not included in \cite{shaw13} and thus without redshift
constraints are: 1FGL J0006.9+4652,
1FGL J0322.1+2336,        
1FGL J0354.6+8009,        
1FGL J1838.6+4756, and        
1FGL J2325.8-4043.   
All available constraints (with the exception of the exclusion functions)
are listed in the Table  in the Appendix.
For each source, the available redshift constraints are 
combined. %Some examples are reported in Fig.~\ref{fig:pdf}.
The most constraining cases are those where
there is either a spectroscopic redshift lower limit (always coupled
to a $z_{max}$ limit) or a photometric upper limit (typically z$\lesssim$1.3).
Lower and upper limits on the redshift are treated as step functions
and we tested that the results reported in the next sections
are robust against the exclusion of a fraction ($\sim$10\,\%) of these
limits.

\begin{figure*}[ht!]
  \begin{center}
  \begin{tabular}{cc}
\hspace{-1cm}
    \includegraphics[scale=0.45]{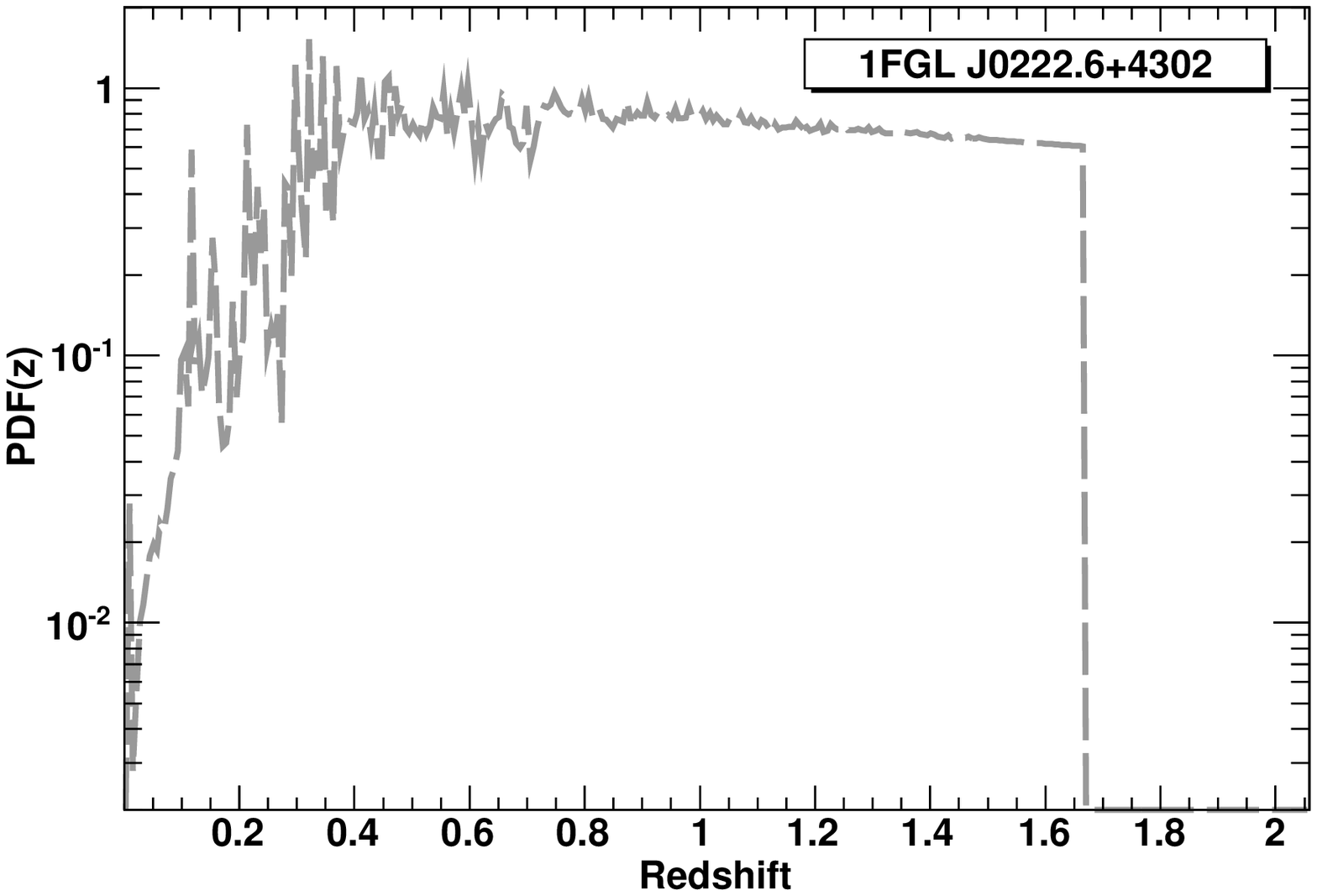} &
\hspace{-1cm}
  	 \includegraphics[scale=0.45]{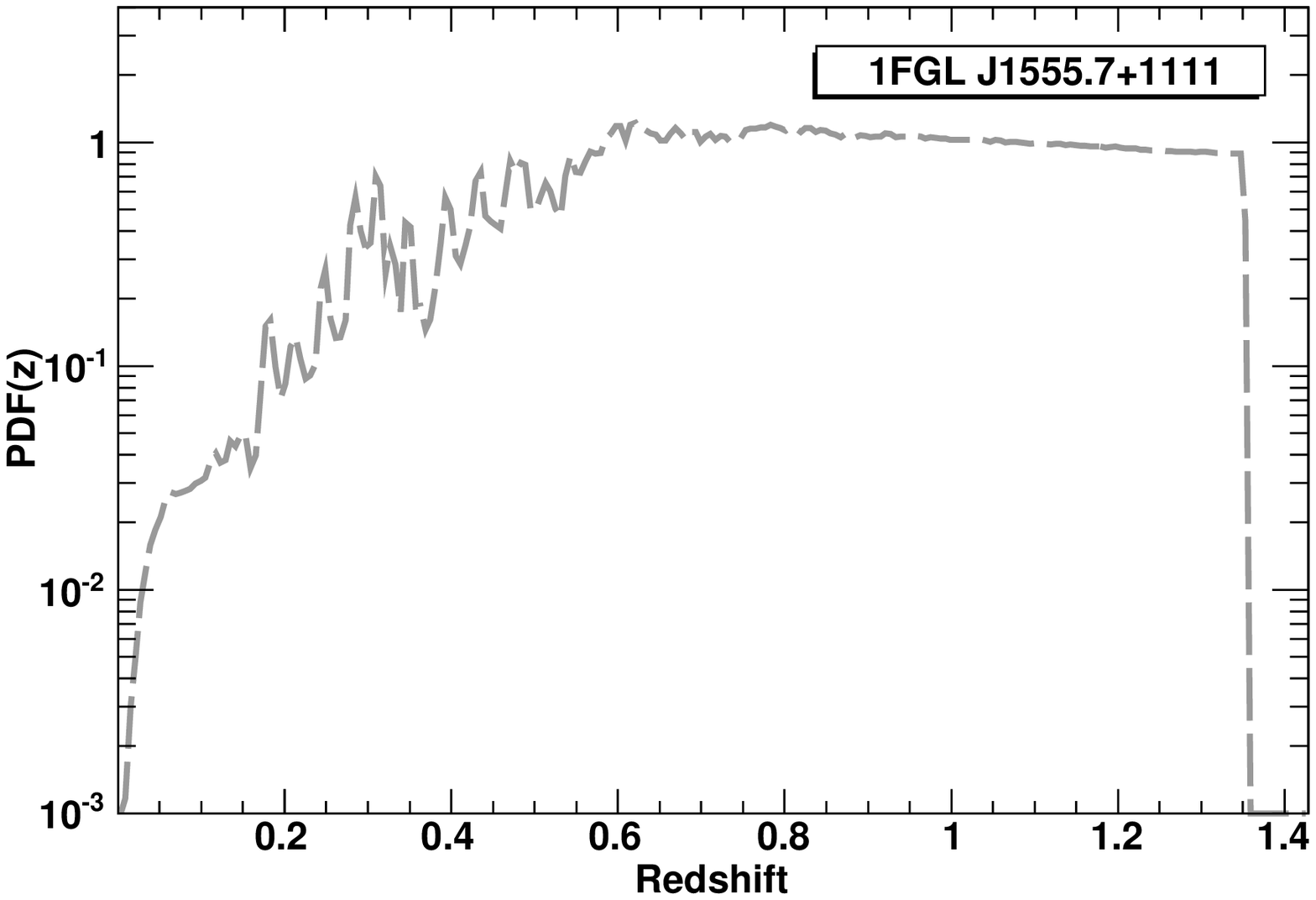} \\
\hspace{-1cm}
 	 \includegraphics[scale=0.45]{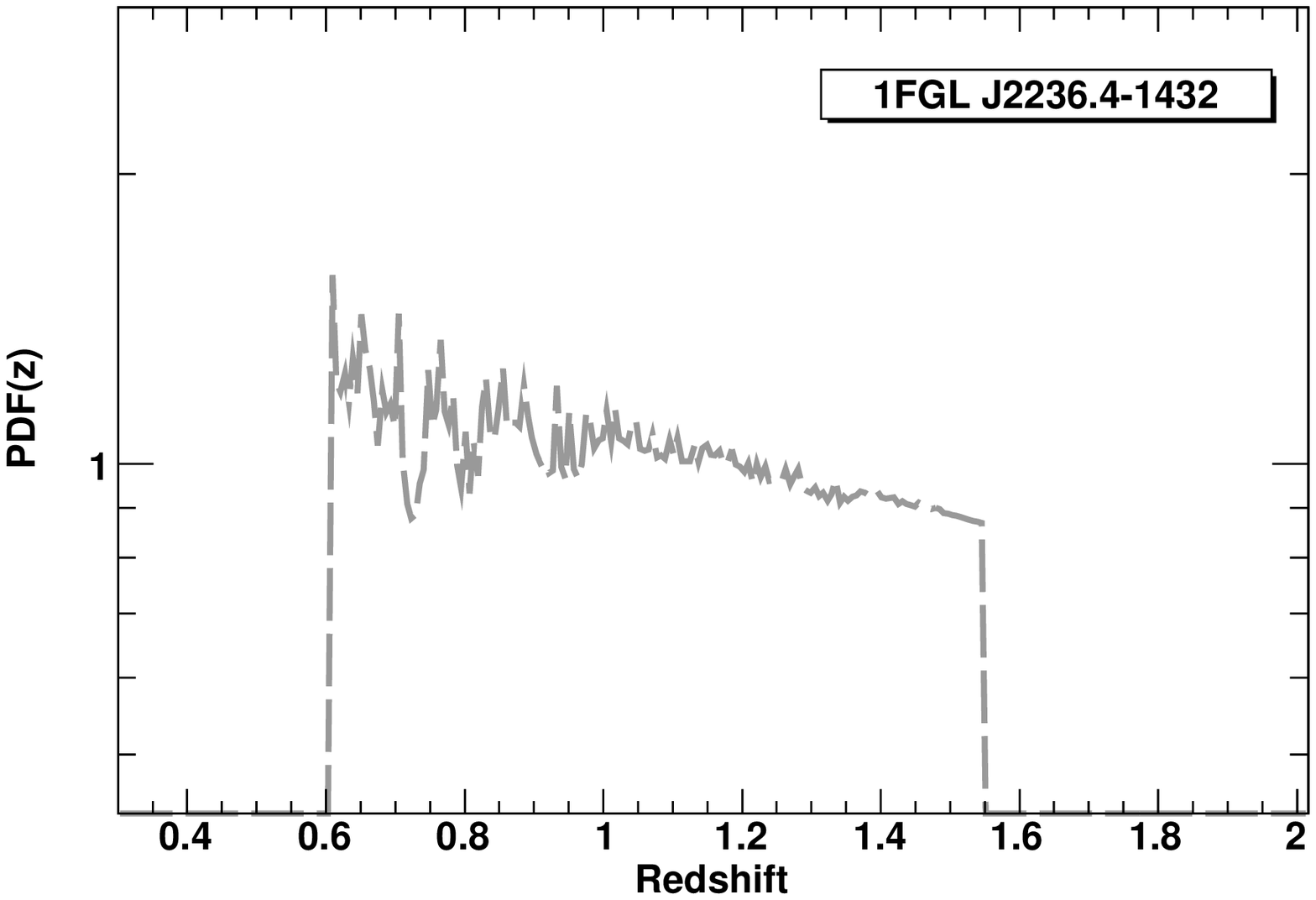} &
\hspace{-1cm}
	 \includegraphics[scale=0.45]{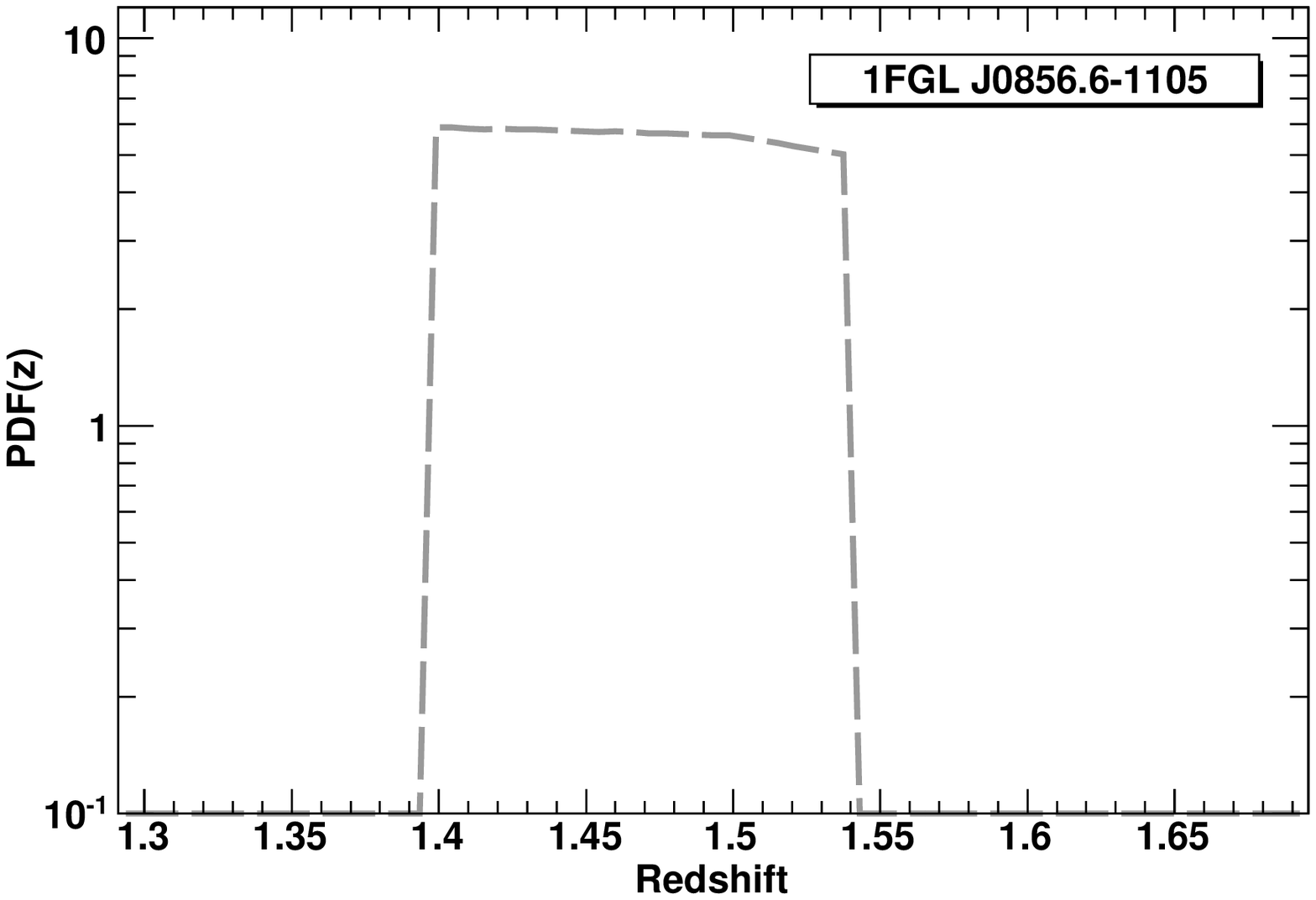}
\end{tabular}
  \end{center}
\caption{Examples of probability density functions (PDFs) for the redshifts
of 4 sources. The upper plots show the case of sources with upper limits 
(spectroscopic and photometric respectively in the left and right plots)
coupled to exclusion probabilities and a prior function as discussed in
$\S$~\ref{sec:z}. The bottom panels show the case of sources with
both spectroscopic lower limits and photometric upper limits.
Both PDFs were combined (as above) with the
exclusion probabilities and the prior function.
\label{fig:pdf}}
\end{figure*}

We combine these constraints to produce, for each object, the observationally
allowed probability density function (PDF) for the source redshift. However, for the LF
analysis we need the redshift PDF, subject to these observational constraints, for
the source as a representative member of the {\it Fermi}-detected BL Lacs. Accordingly,
we assume a {\it prior} function that represents the $dN/dz$ distribution if one could 
measure the spectroscopic redshift for {\it all} the BL Lacs in our {\it Fermi} sample.
This is multiplied by the observational PDF to derive the final PDF for each 
{\it Fermi}-detected BL Lac. If, for example, only $z_{\rm min}$ and $z_{\rm max}$
constraints were available for a given source, its final PDF would follow the prior 
$dN/dz$ between these limits. As noted below, the {\it prior} has only a mild
effect on the luminosity function. 
For each source, then, the PDF is obtained as:
\begin{equation}
{\rm PDF}(z) = \frac{dN}{dz} \cdot \prod_{i}^{n} C_i(z)
\end{equation}
where the $C_i(z)$ are the redshift constraints available for that source.
Sample PDFs are shown in Fig.~\ref{fig:pdf}.

	Drawing possible redshifts from these final PDFs for each source, we 
compute the sample LF as described above and then use this to predict the
{\it observed} $dN/dz$ using Eq.~6, which represents the redshift distribution
expected if all sources could have spectroscopic redshift measurements. In general, this
will differ from the initial assumed prior. We replace the prior with this predicted
$dN/dz$ and iterate to convergence. Since the $dN/dz$ distribution is rather flat in 
the range $0.02<$z$<$2, we find that the initially assumed prior has very little effect.
In practice we find robust convergence to the same final LF for an initial prior 
$dN/dz \propto z^{-t}$ with $-0.3<t<0.6$. In all cases the derived distribution
shows a clear drop in the number of observed BL Lacs at $z>2$ (see below).
However, as this may be an important evolutionary
effect that we wish to measure without bias, we conservatively 
assume a $dN/dz \propto z^{-t}$ prior extending to all $z$ allowed by the constraints.
%this may be an 
%important BL Lac evolutionary effect (see below). 
We adopt a computation with an
initial $t=0.2$ prior which is shown in the upper left panel of Fig.~\ref{fig:ple}.

%%%%%%%%%%%%%%%%%%%%%%%%%%%%%%%%%%%%%%%%%%%%%%%%%%%%%%%%%%%%%%%%%%%%%%%%%%
%%%%%%%%%%%%%%%%%%%%%%%%%%%%%%%%%%%%%%%%%%%%%%%%%%%%%%%%%%%%%%%%%%%%%%%%%%
\subsection{Summary of the Analysis Chain}
\label{sec:chain}
We use a Monte Carlo approach in order to derive the LF and its uncertainty.
The steps of the analysis are as follows:
\begin{enumerate}
\item An initial {\it prior} function (see $\S$~\ref{sec:z})
 is chosen to approximate the 
$dN/dz$ distribution of the {\it Fermi} BL Lacs.
%%%%%%%%%
\item We then create 1000 samples of 206 BL Lacs whose redshifts
are extracted {at random} from the PDF of each source.  The 206 BL Lacs\footnote{Including
or excluding the 5 BL Lacs without  redshift information does not change
the result of our analysis. When those objects are included their
redshifts are randomly extracted from the {\it prior} function.} are drawn
with replacement from the objects { reported in the Appendix.}
%%%%%
\item We use the ML method
described in $\S$~\ref{sec:ml} with one of the parametrizations
in $\S$~\ref{sec:lf} to derive the best-fit LF. This is done independently
for each Monte Carlo. The final LF is built as the average of the Monte Carlo
LFs and its uncertainty takes into account the spread of all the Monte Carlo
LFs. This allows us to quantify naturally the uncertainty in the LF
due to the sample size and the spread in the redshift measurements. The LF
is used to predict the observed $dN/dz$ through Eq.~5.
%%%%%%
\item The $dN/dz$ is compared to the {\it prior} function used at step 1):
if the two functions are different\footnote{A chi-square 
fit in the 0.02$<$z$<$2 redshift interval is used to assess the compatibility 
between the {\it prior} function and the $dN/dz$.}, then a new {\it prior}
function based on the latest $dN/dz$ (step 3) is created and substituted
to the one of step 1).
%%%%%%%
\item Steps 1-4 are repeated until the {\it prior} and the predicted $dN/dz$ are compatible with each other. 
\end{enumerate} 

We note that a change in the {\it prior} function
causes a change in the redshift PDFs of all sources, 
and thus new PDFs have to be created and the entire analysis (steps 2-4) has
to be repeated.

%%%%%%%%%%%%%%%%%%%%%%%%%%%%%%%%%%%%%%%%%%%%%%%%%%%%%%%%%%%%%%%%%%%%%%%%%%
%%%%%%%%%%%%%%%%%%%%%%%%%%%%%%%%%%%%%%%%%%%%%%%%%%%%%%%%%%%%%%%%%%%%%%%%%%
\section{Results}
\label{sec:results}

In this section we present results on the best-fitting LF models.
Particular attention is given to whether adding the $\beta$ and $\tau$
parameters (representing respectively the luminosity-dependent photon index
and  a luminosity-dependent speed of evolution, see Eq.~\ref{eq:blazseq} and 
\ref{eq:tau}) significantly improves the quality of the fit.

%%%%%%%%%%%%%%%%%%%%%%%%%%%%%%%%%%%%%%%%%%%%%%%%%%%%%%%%%%%%%%%%%%%%%%%%%%
%%%%%%%%%%%%%%%%%%%%%%%%%%%%%%%%%%%%%%%%%%%%%%%%%%%%%%%%%%%%%%%%%%%%%%%%%%
\subsection{Density and Luminosity Evolution}
\label{sec:resultpde}

\begin{figure*}[ht!]
  \begin{center}
  \begin{tabular}{cc}
\hspace{-1cm}
    \includegraphics[scale=0.45]{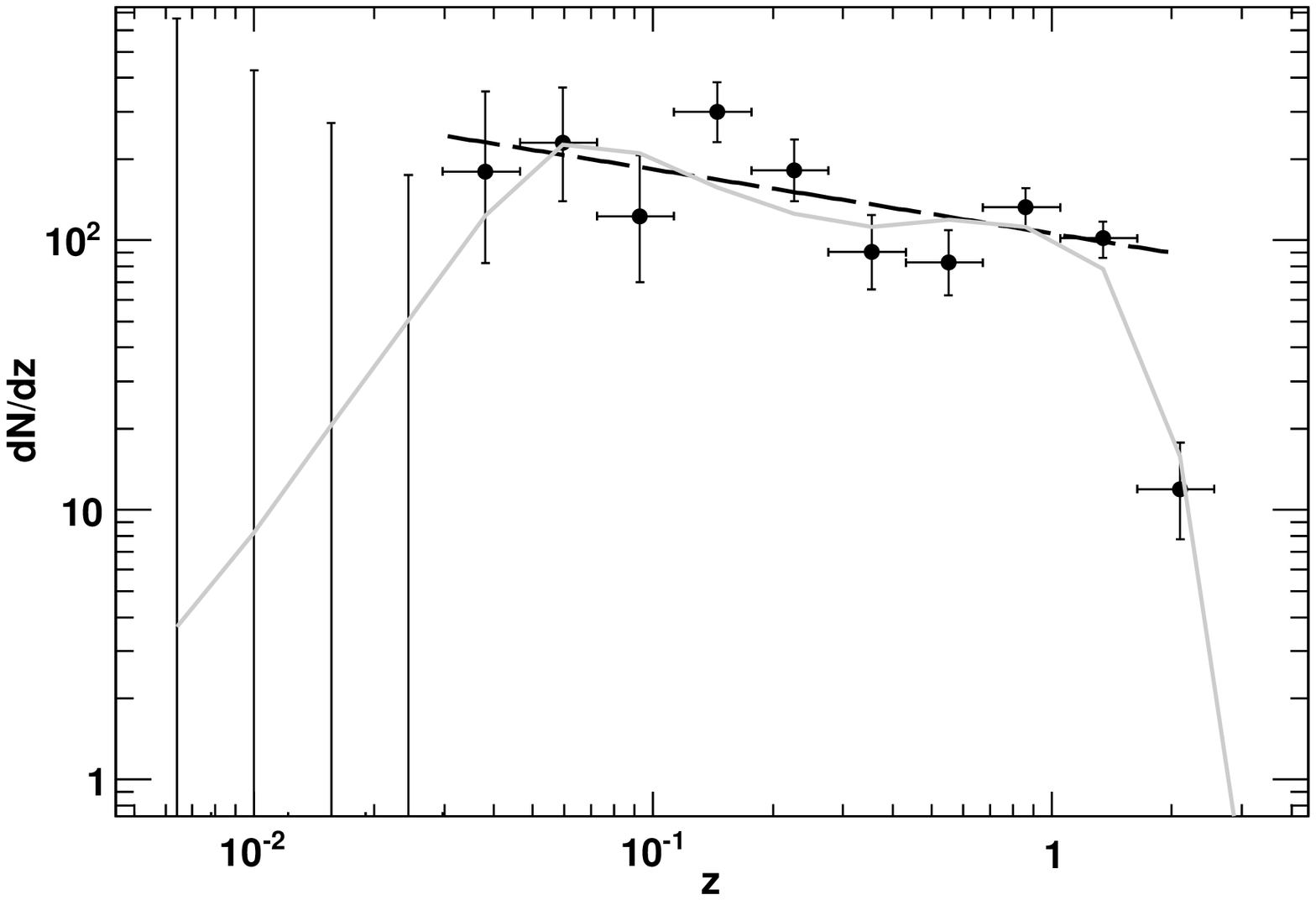} &
\hspace{-1cm}
  	 \includegraphics[scale=0.45]{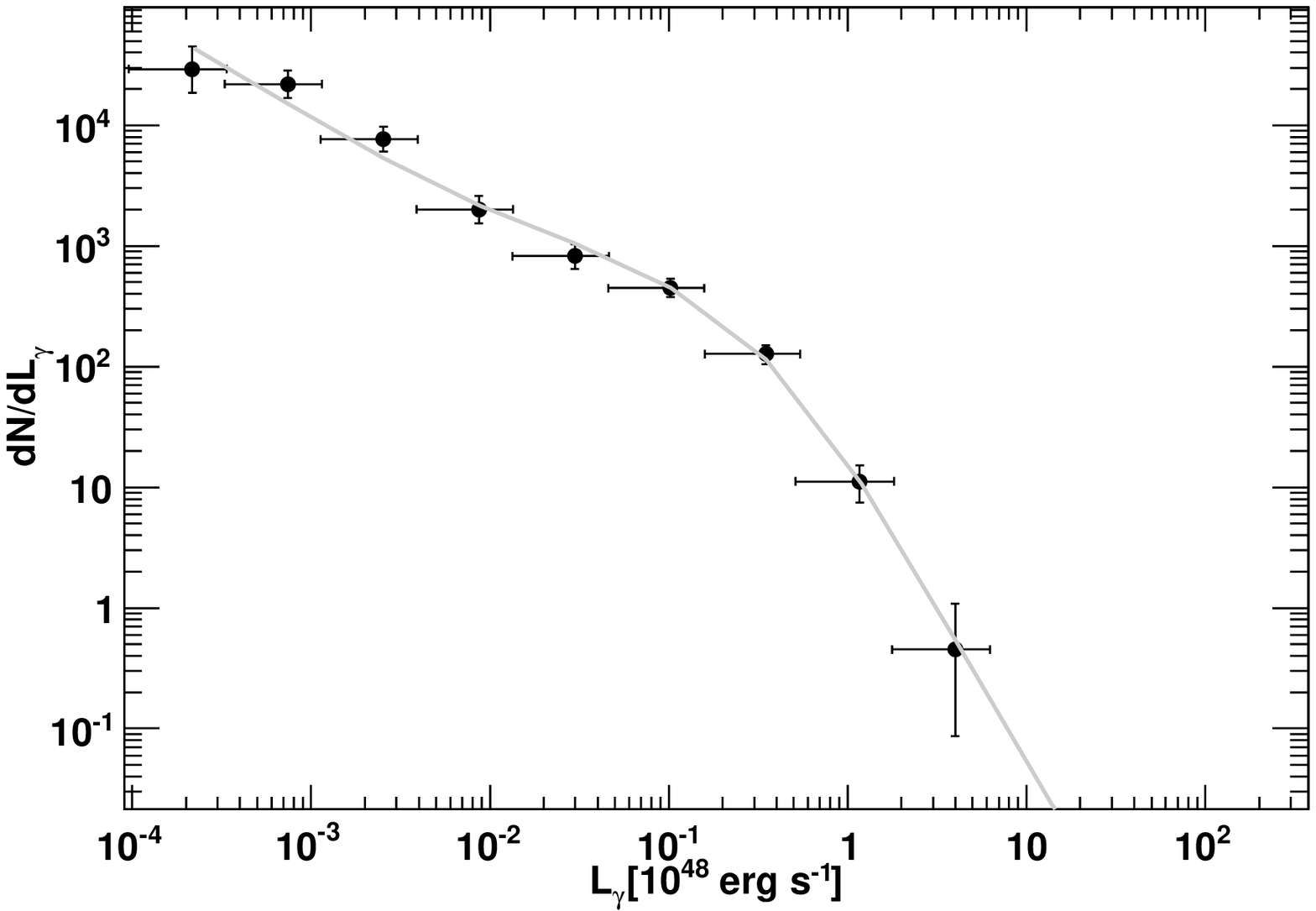} \\
\hspace{-1cm}
 	 \includegraphics[scale=0.45]{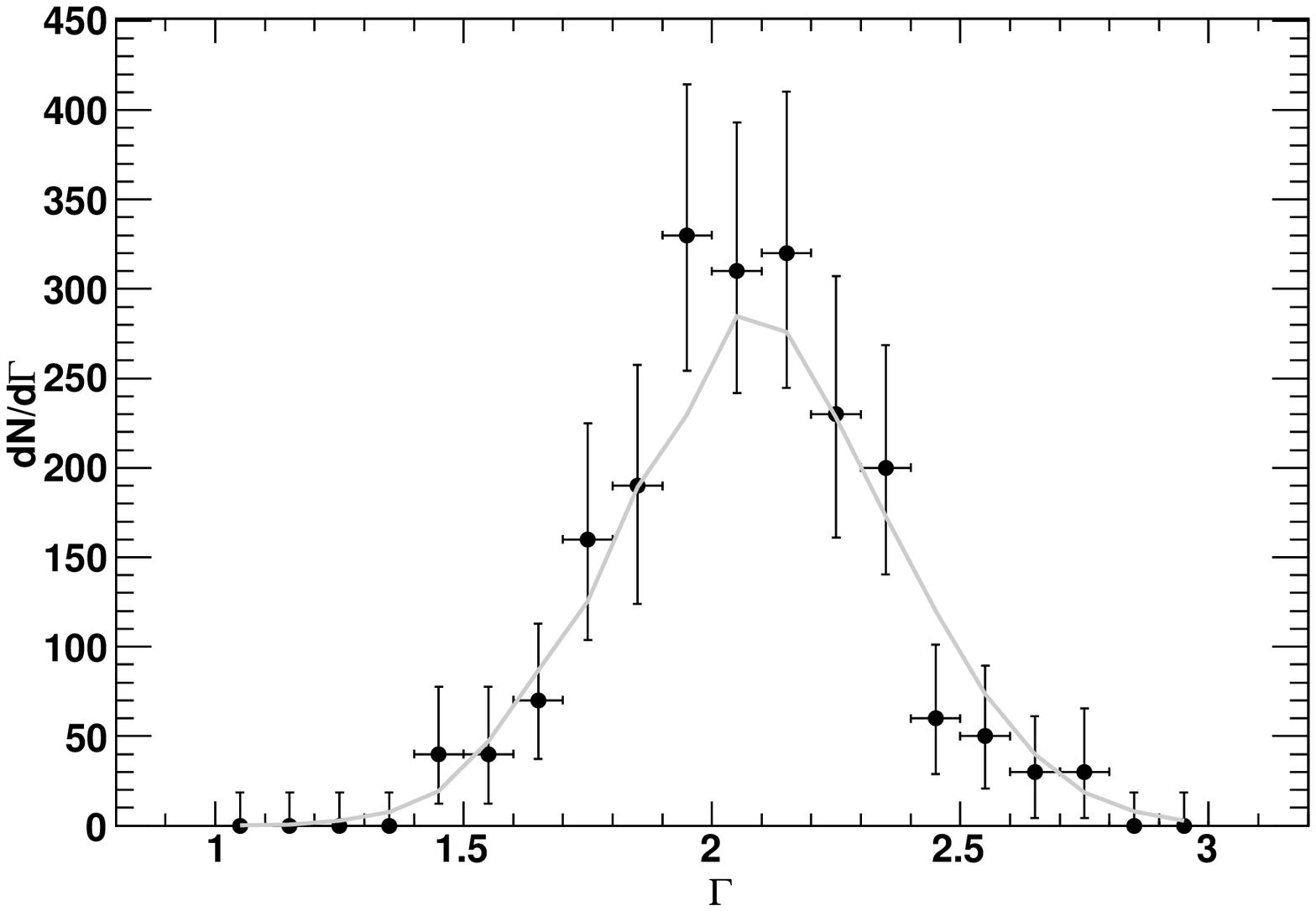} &
\hspace{-1cm}
	 \includegraphics[scale=0.45]{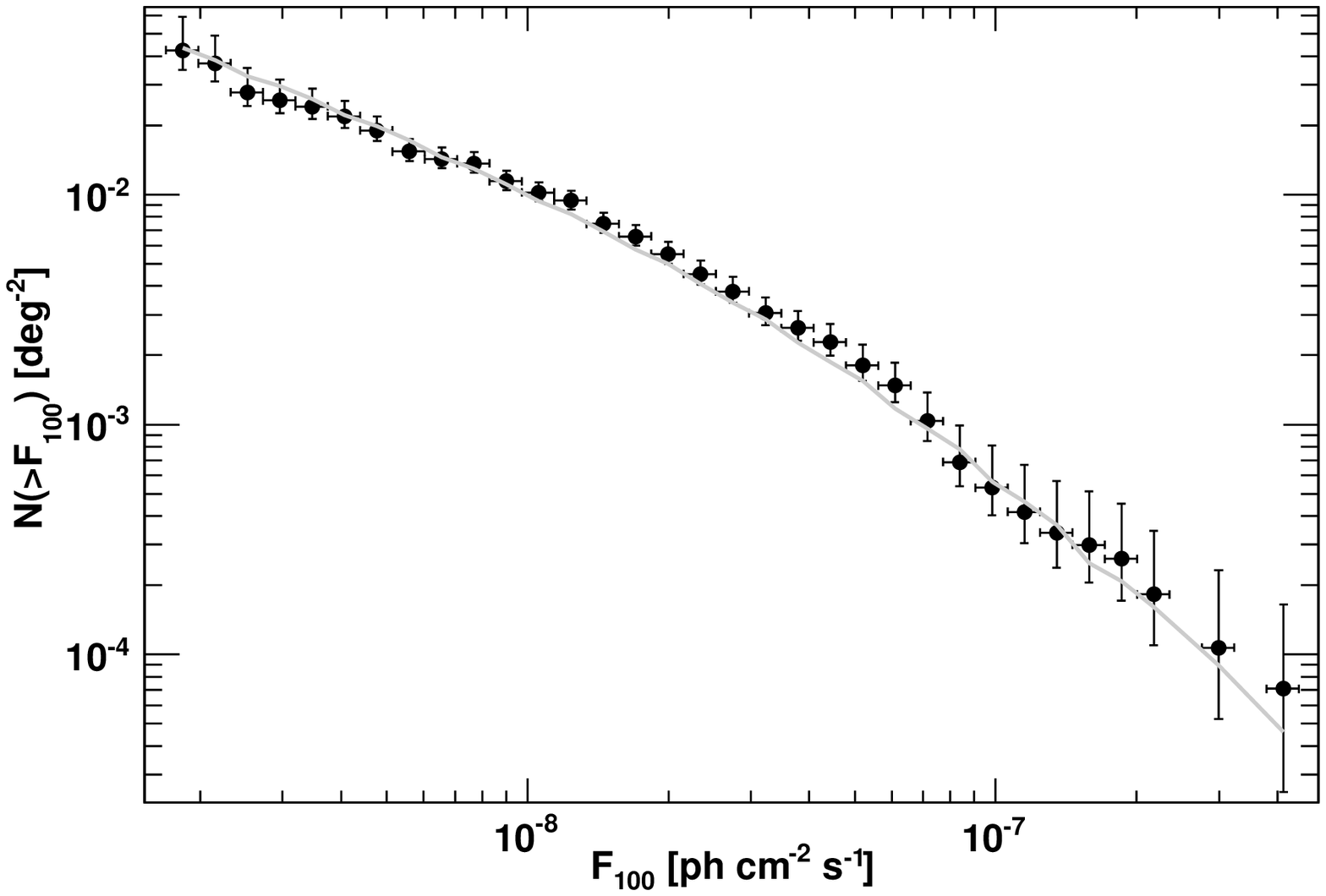}
\end{tabular}
  \end{center}
\caption{Observed redshift (upper left), luminosity (upper right), 
photon index (lower left),  and source count (lower right)
distributions of LAT BL Lacs. 
The continuous solid line is  the best-fit PLE model convolved
with the selection effects of {\it Fermi}. The error
bars reflect the statistical uncertainty including (for the upper plots) 
the uncertainty in the sources' redshifts. 
Error bars compatible with zero are 1\,$\sigma$ upper limits for the case
of observing zero events in a given bin \cite[see][]{gehrels86}.
The dashed
line in the redshift distribution shows one of the {\it prior} functions
used in $\S$~\ref{sec:z}.
\label{fig:ple}}
\end{figure*}

Tab.~\ref{tab:ple} reports the results of the best fits using 
a PDE or a PLE parametrization, including cases for which $\beta$ and
$\tau$ are allowed to vary.
Both the PLE and PDE LFs provide adequate representations of the 
{\it Fermi} data when
$\beta$ and $\tau$ are allowed to vary (see PLE$_3$ and PDE$_3$ models
in Tab.~\ref{tab:ple}). In all cases the PLE model provides
a better representation of the {\it Fermi} data than the PDE model
as indicated by the value of the log-likelihood ($S$ in Eq:~\ref{eq:s}).  
As shown in Fig.~\ref{fig:ple}, the best-fit PLE model (model PLE$_3$ in
Tab.~\ref{tab:ple}) reproduces accurately the distribution in luminosity,
redshift, photon index and source counts of the {\it Fermi} blazars.
The model PLE$_3$ provides the best representation of the LF of BL Lacs.

The improvement in the log-likelihood when $\beta$ and $\tau$
are allowed to vary can be used to quantify the improvement of the fit
with the standard formula TS= - 2(ln L$_0$ - ln L$_1$), where $L_1$ is the 
hypothesis tested again the null one (L$_0$) and TS is the likelihood test
statistic.
We find that allowing the parameter $\beta$ to vary produces an improvement
in the fit of { TS$>$10 (see Tab.~\ref{tab:ple}) which corresponds to 
$>3$\,$\sigma$ for the case of one additional degree of freedom}. The $\tau$ parameter which governs
the speed of the evolution as a function of luminosity produces 
an improvement in the fit of { TS=52} ($\sim$7.2\,$\sigma$)
 for the PLE and { TS=12} ($\sim$3.4\,$\sigma$) for the PDE model.

If we take the luminosities of 10$^{45}$, 10$^{46}$, and 10$^{47}$\,erg s$^{-1}$
as { reference luminosities for the {\it Fermi} populations} of  HSPs, ISPs and LSPs   then 
we find that the redshift peaks of the luminosity evolution
are respectively  z$_c$=0.5, 0.8, and 1.2 for these three luminosities.
The maximum-likelihood value of the 
speed of the evolution (parameter $k_d$) also changes 
from 4.7 to 5.8 and 7.0, respectively.  It thus seems clear that the 
evolution depends on the luminosity class.

\begin{deluxetable}{lccccccccccc}
\tablewidth{0pt}
\tabletypesize{\scriptsize}
\rotate
\tablecaption{Best-fit parameters of the Pure Luminosity and Pure Density
Evolution LFs. Parameters without an error estimate were kept fixed during the fit. Parameter values were computed as the median of all the best-fit parameters
to the Monte Carlo sample, while the uncertainties represent the 68\,\%
containment regions around the median value.
\label{tab:ple}}
\tablehead{\colhead{Model}   & 
\colhead{A\tablenotemark{a}} & \colhead{$\gamma_1$} & 
\colhead{L$_*$\tablenotemark{b}}              & \colhead{$\gamma_2$} &
\colhead{k}                  & \colhead{$\tau$}     &
\colhead{$\xi$}              &
\colhead{$\mu^*$}              & \colhead{$\beta$}    & 
\colhead{$\sigma$}           & -2$\ln$L\tablenotemark{c}
}
\startdata 

PLE$_1$   & $7.29^{+31.80}_{-7.13}\times10^{3}$ & $1.26^{+0.08}_{-0.20}$ & $1.42^{+89.33}_{-0.94}\times10^{-2}$ & $1.31^{+1.78}_{-0.09}$ & $4.87^{+0.78}_{-5.39}$ & $0$ & $-0.48^{+3.48}_{-0.08}$ & $2.15^{+0.03}_{-0.03}$ & $0$ & $0.27^{+0.02}_{-0.02}$ & -$690.1$\\

PLE$_2$   & $2.89^{+30.91}_{-2.70}\times10^{3}$ & $1.22^{+0.09}_{-0.42}$ & $2.16^{+73.16}_{-1.67}\times10^{-2}$ & $1.37^{+2.10}_{-0.14}$ & $4.61^{+0.75}_{-5.13}$ & $0$ & $-0.48^{+3.48}_{-0.10}$ & $2.12^{+0.03}_{-0.03}$ & $6.48^{+2.28}_{-2.09}\times10^{-2}$ & $0.26^{+0.02}_{-0.02}$ & -$699.9$\\

PLE$_3$   & $9.68^{+6.88}_{-4.75}\times10^{2}$ & $1.47^{+0.14}_{-0.12}$ & $4.48^{+2.32}_{-1.20}\times10^{-2}$ & $4.45^{+1.08}_{-0.93}$ & $5.89^{+0.99}_{-0.95}$ & $1.18^{+0.16}_{-0.22}$ & $-0.31^{+0.05}_{-0.06}$ & $2.11^{+0.03}_{-0.03}$ & $6.47^{+2.23}_{-2.40}\times10^{-2}$ & $0.26^{+0.03}_{-0.02}$ & -$752.1$\\

\hline
PLE$_{no-z}$ &  9.12$^{+0.90}_{-0.60}\times10^5$ & 2.07$\pm0.52$ & 0.12$\pm0.22$ & 0.77$\pm0.67$ & 8.60$\pm1.07$ & 1.41$\pm0.33$ & -0.17$\pm0.04$ & 2.19$\pm0.04$ & 0.16$\pm0.04$ & 0.30$\pm0.04$ & \nodata \\
\hline

PDE$_1$   & $78.53^{+906.10}_{-73.82}$ & $1.32^{+18.68}_{-0.10}$ & $0.58^{+3.01}_{-0.47}$ & $1.25^{+0.09}_{-0.08}$ & $11.47^{+1.44}_{-1.94}$ & $0$ & $-0.21^{+0.02}_{-0.04}$ & $2.15^{+0.03}_{-0.03}$ & $0$ & $0.27^{+0.02}_{-0.02}$ & -$695.8$\\

PDE$_2$   & $62.22^{+989.87}_{-55.53}$ & $1.32^{+18.68}_{-0.10}$ & $1.10^{+2.34}_{-1.01}$ & $1.24^{+0.07}_{-0.07}$ & $10.72^{+1.50}_{-2.23}$ & $0$ & $-0.24^{+0.03}_{-0.06}$ & $2.12^{+0.03}_{-0.03}$ & $6.33^{+2.31}_{-2.00}\times10^{-2}$ & $0.26^{+0.03}_{-0.02}$ & -$711.9$\\

PDE$_3$   & $18.78^{+65.86}_{-14.69}$ & $3.43^{+0.78}_{-0.42}$ & $0.38^{+0.46}_{-0.17}$ & $1.56^{+0.16}_{-0.12}$ & $16.69^{+3.52}_{-2.77}$ & $3.23^{+0.85}_{-0.79}$ & $-0.11^{+0.02}_{-0.02}$ & $2.10^{+0.03}_{-0.03}$ & $6.45^{+2.31}_{-2.31}\times10^{-2}$ & $0.26^{+0.02}_{-0.03}$ & -$724.8$\\

\enddata
\tablenotetext{a}{In units of $10^{-13}$\,Mpc$^{-3}$  erg$^{-1}$ s.}
\tablenotetext{b}{In units of $10^{48}$\,erg s$^{-1}$.}
\tablenotetext{c}{Value of the -2$\times$log-likelihood when the function
is minimized.}
\end{deluxetable}

%%%%%%%%%%%%%%%%%%%%%%%%%%%%%%%%%%%%%%%%%%%%%%%%%%%%%%%%%%%%%%%%%%
%%%%%%%%%%%%%%%%%%%%%%%%%%%%%%%%%%%%%%%%%%%%%%%%%%%%%%%%%%%%%%%%%%
\subsection{Luminosity-Dependent Density Evolution}
\label{sec:ldde}

\begin{figure*}[ht!]
  \begin{center}
  \begin{tabular}{cc}
\hspace{-1cm}
    \includegraphics[scale=0.45]{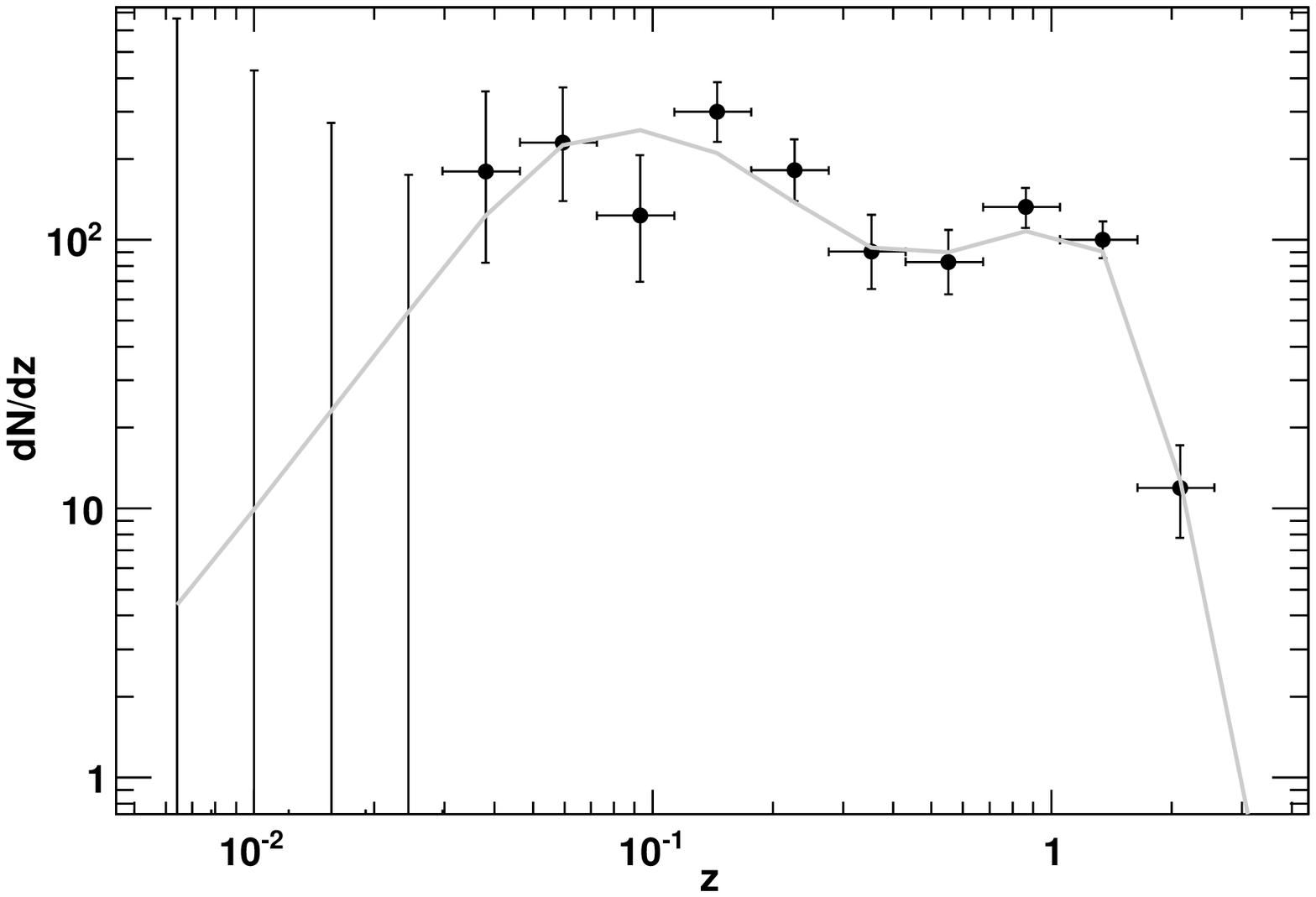} &
\hspace{-1cm}
  	 \includegraphics[scale=0.45]{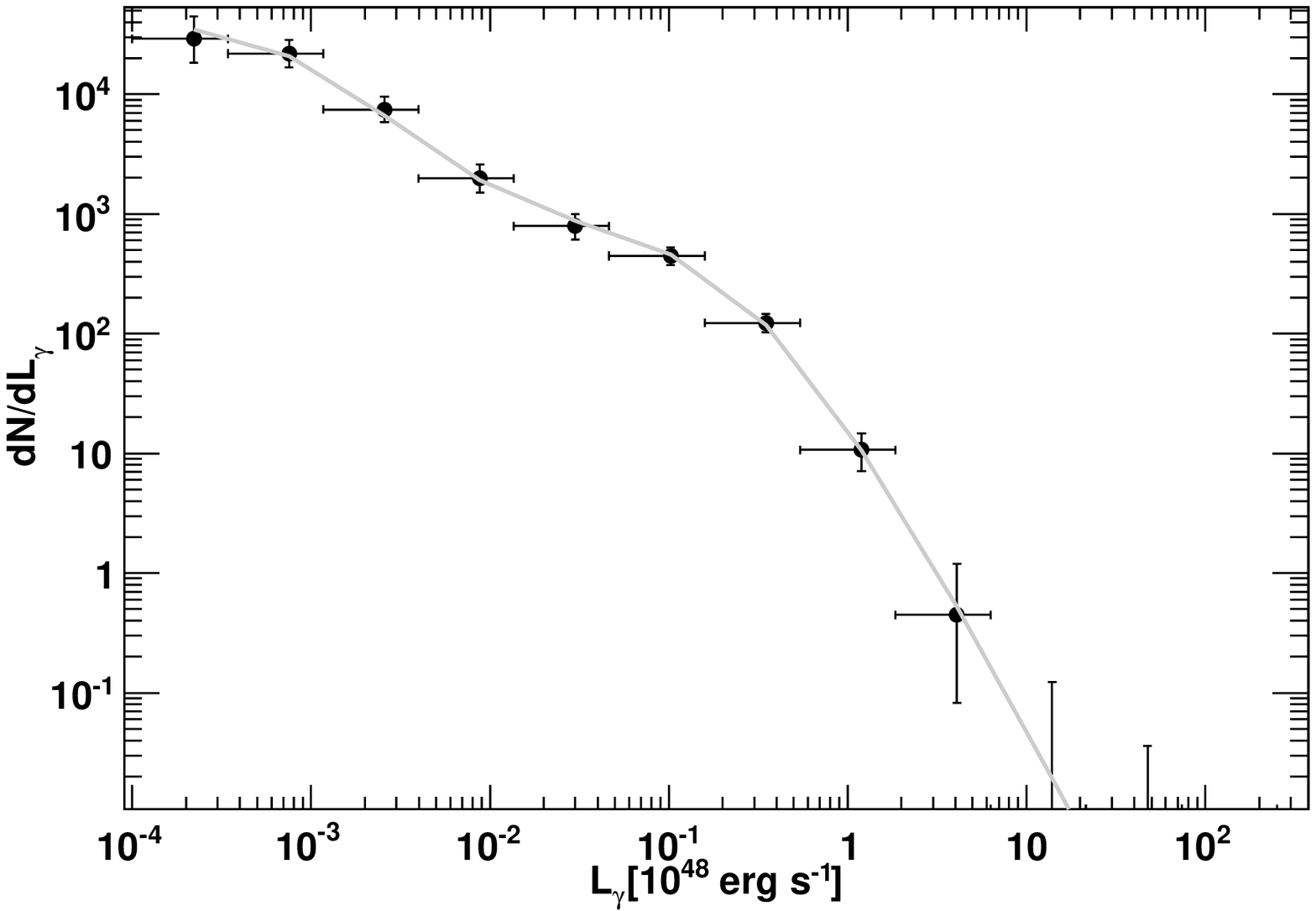} \\
\hspace{-1cm}
 	 \includegraphics[scale=0.45]{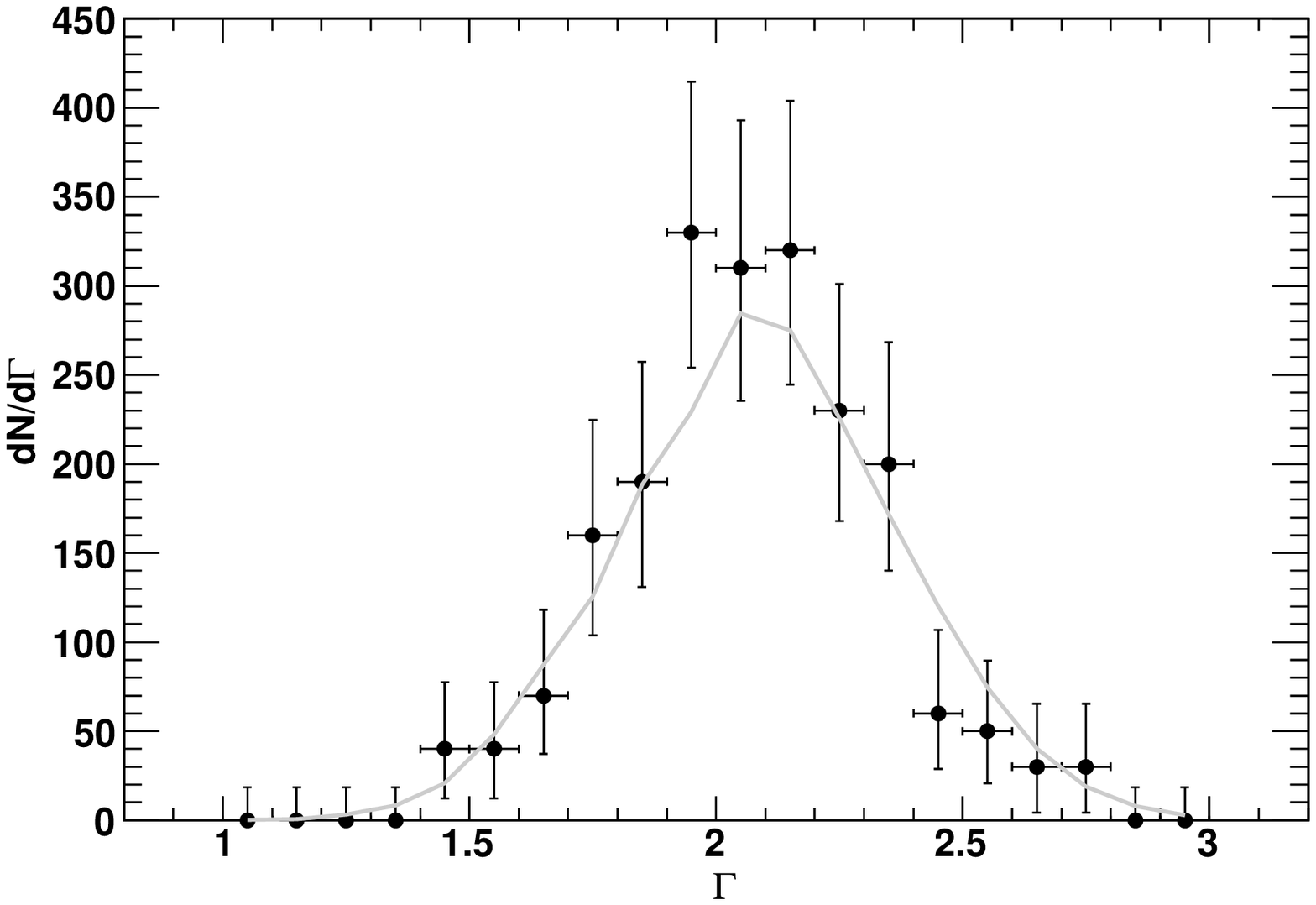} &
\hspace{-1cm}
	 \includegraphics[scale=0.45]{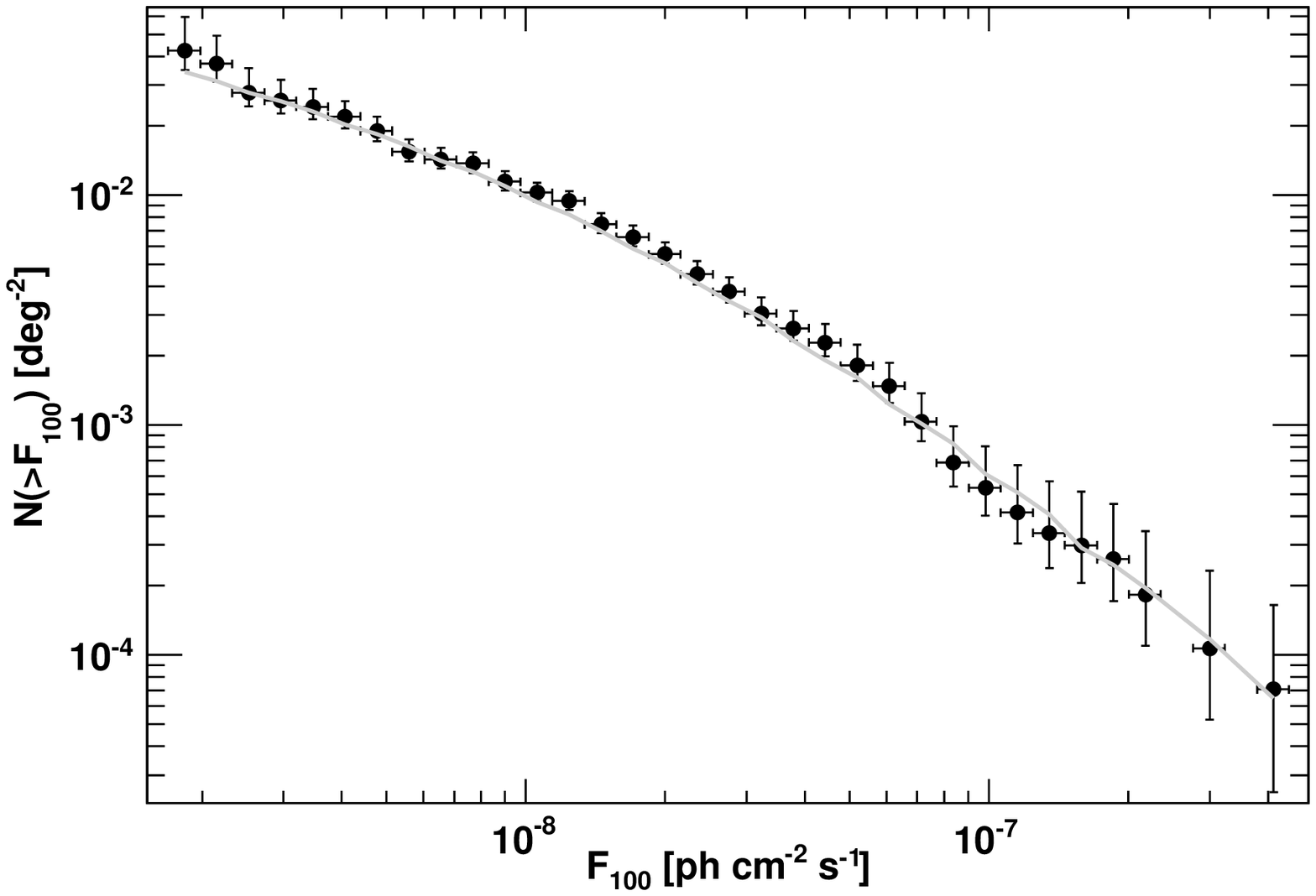}
\end{tabular}
  \end{center}
\caption{Observed redshift (upper left), luminosity (upper right), 
photon index (lower left),  and source count (lower right)
distributions of LAT BL Lacs. 
The continuous solid line is  the best-fit LDDE model convolved
with the selection effects of {\it Fermi}. 
 The error
bars reflect the statistical uncertainty including (for the upper plots) 
the uncertainty in the sources' redshifts. Error bars consistent with zero
represent 1\,$\sigma$ upper limits for the case of observing zero events
in a given bin \citep[see][]{gehrels86}.
\label{fig:ldde}}
\end{figure*}

Given the clear luminosity dependence of the evolution
found in the previous section we try to fit the LDDE model
of $\S$~\ref{sec:lf}. This model has two additional parameters
with respect to the PLE and PDE models. The fit with $\tau$=0 (all
luminosity classes evolve in the same way) already provides 
a representation of the data which is as good as the best-fit PLE
model (see Tab.~\ref{tab:ldde}). If we allow $\tau$ to vary the fit
improves further with respect to the baseline LDDE$_1$ model (TS=30, { i.e. $\sim$5.5\,$\sigma$}). Figure~\ref{fig:ldde} shows how the LDDE$_3$ model
reproduces the observed distributions.

{ The improvement of the LDDE$_2$ model with respect to the PLE$_3$ model
can be quantified using the Akaike information criterion \citep[AIC,][]{akaike74,wall12}. For each model, one can define the quantity $AIC_i=2n_{par} - 2ln L$ where $n_{par}$
is the number of free parameters and $-2ln L$ is twice the log-likelihood value
as reported in Tab.~\ref{tab:ple} and \ref{tab:ldde}.
The relative likelihood of a model with respect to another one can be evaluated
as $p=e^{0.5(AIC_{min}-AIC_{i})}$ where $AIC_{min}$ comes from the model
providing the minimal $AIC$ value.
According to this test the PLE$_3$ model has a relative likelihood
with respect to the LDDE$_2$ model of $\sim$0.0024.
Thus, the model LDDE$_2$ whose parameters
are reported in Tab.~\ref{tab:ldde}  fits the 
{\it Fermi} data  better ($\sim$3\,$\sigma$) than the PLE$_3$ model.
}

\begin{figure}[h!]
\begin{centering}
	\includegraphics[scale=0.9]{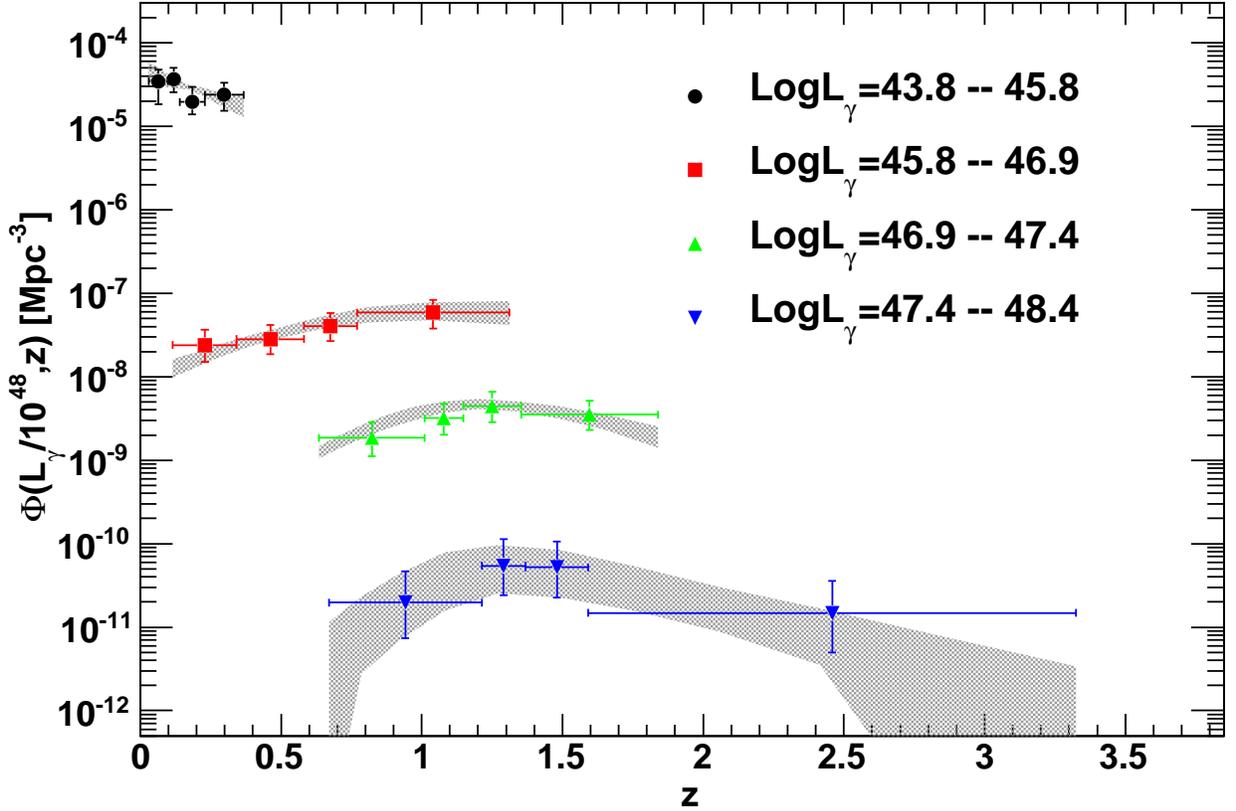} 
	\caption{Growth and evolution of BL Lacs, separated by luminosity class. 
The gray bands represent 68\,\% confidence regions
around the best fitting LDDE LF model (for each Monte Carlo sample).
Both data points and band errors include uncertainties for the source redshifts
as well as statistical uncertainty. All but the least luminous class
have a redshift peak near z$\approx$1.5; the lowest luminosity BL Lacs increase
toward z=0.
	\label{fig:spacedensity}}
\end{centering}
\end{figure}

In this representation low-luminosity (L$_{\gamma}$=10$^{44}$\,erg s$^{-1}$)
sources are found  to evolve negatively ($p1$=-7.6).
On the other hand high-luminosity (L$_{\gamma}$=10$^{47}$\,erg s$^{-1}$)
sources are found to evolve positively ($p1$=7.1).
{ Both evolutionary trends are correctly represented also
in the best-fit PLE model (PLE$_3$ in Tab.~\ref{tab:ple}), but the LDDE
model provides a slightly better representation of the data.}
The different evolution of low-luminosity and high-luminosity
sources can be readily appreciated in Fig.~\ref{fig:spacedensity}
which shows the space density of different luminosity classes
of BL Lacs as a function of redshift. This figure was created
taking into account the dispersion in both redshift and luminosity
introduced by the uncertainty in the redshift of many of our BL Lacs.
A noteworthy fact is that the least-luminous BL Lacs are 10$^3$ times
more numerous than the least luminous FSRQs detected by {\it Fermi}
\citep[see Fig.~4 in][]{fsrq12}. The data points were deconvolved
with the method described in $\S$~\ref{sec:ml} (see Eq.~\ref{eq:nmdl})
while the LF is displayed as the region enclosing 68\,\% of all
the best-fit LDDE models to the 1000 Monte Carlo samples.

\begin{figure}[h!]
\begin{centering}
	\includegraphics[scale=0.9]{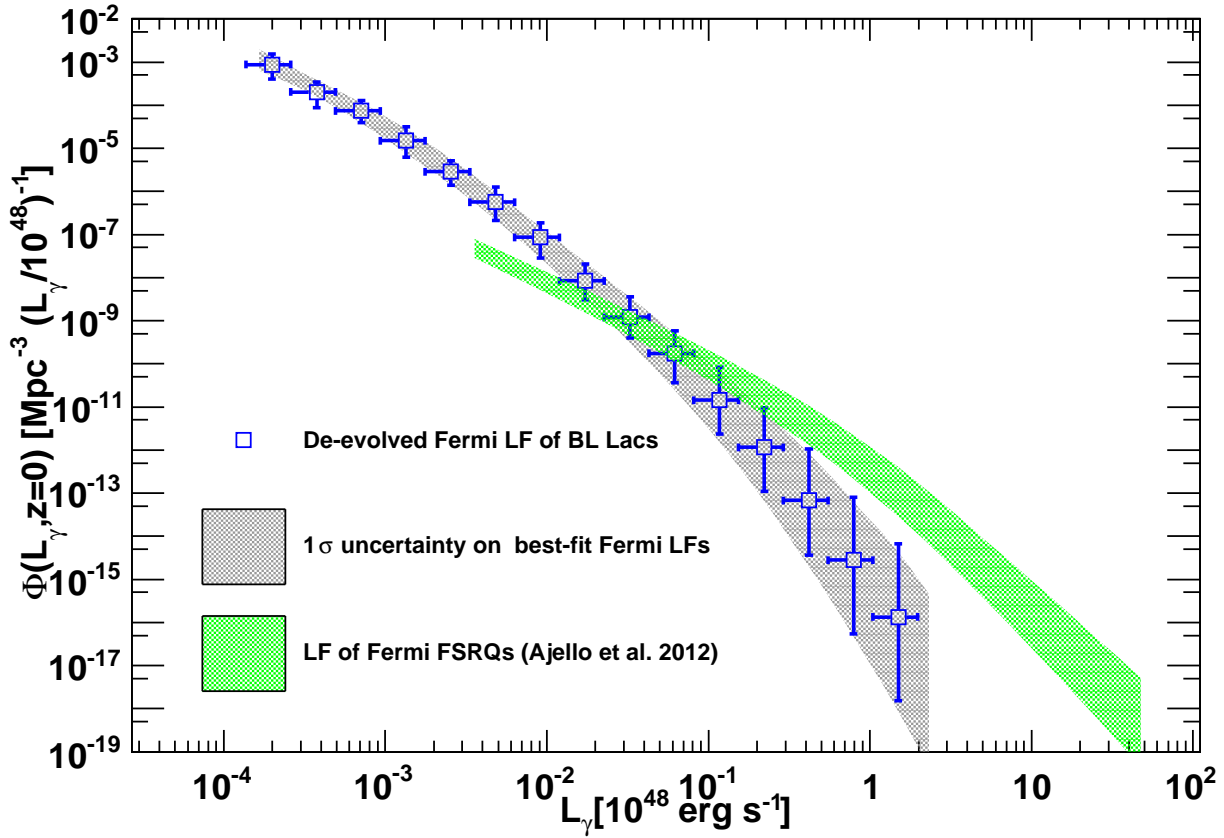} 
	\caption{Local (z=0) LF as derived from the best-fit LDDE model ($\S$~\ref{sec:ldde}). 
BL Lacs dominate the local luminosity function for 
$L_{\gamma} < 10^{46}$\, erg s$^{-1}$.
The gray band represents the confidence region
enclosing 68\,\% of the realizations of the best-fit LF to the Monte Carlo
samples.
	\label{fig:localglf}}
\end{centering}
\end{figure}

The local LF is the luminosity function at redshift zero.
For an evolving population, the local LF is obtained by 
de-evolving the luminosities (or the densities) according to
the best-fit model. We follow two approaches to derive the local LF.
First, we de-evolve the luminosities using the 1/V$_{\rm MAX}$ method of \cite{schmidt68} but weighting the maximum volume (V$_{\rm MAX}$) by
the density evolution implied (for a given source luminosity) by our
best-fit LDDE model. Following \cite{dellaceca08b} and \cite{fsrq12},
the maximum allowed volume for a given source is defined as:
\begin{equation}
V_{\rm MAX} = \int^{z_{max}}_{z_{min}} \Omega(L_i,z,\Gamma) \frac{e(z,L_i)}{e(z_{min},L_i)}
  \frac{dV}{dz}dz
\end{equation}
where $L_i$ is the source luminosity, $\Omega(L_i,z,\Gamma)$ 
is the sky coverage,
$z_{max}$ is the redshift above which the source drops out of the survey,
and $e(z,L_i)$ is the evolution term of Eq.~\ref{eq:ez} normalized
(through $e(z_{min},L_i)$) at the redshift $z_{min}$ to which
the LF is to be de-evolved. The LF de-evolved at $z_{min}$ ($z_{min}$=0
in this case) is built using the standard 1/V$_{\rm MAX}$ method \citep{schmidt68}. 
This is reported (data points) in Fig.~\ref{fig:localglf}.
To estimate the uncertainties that  different methods might introduce
in the local LF we also extrapolated to $z=0$ from the best-fit LDDE models 
to all the Monte Carlo samples to measure the 68\,\% range for the local LF. This is 
shown in Fig.~\ref{fig:localglf} as a gray band. It is apparent that the two methods give
consistent results. 

\begin{figure}[h!]
\begin{centering}
	\includegraphics[scale=0.7]{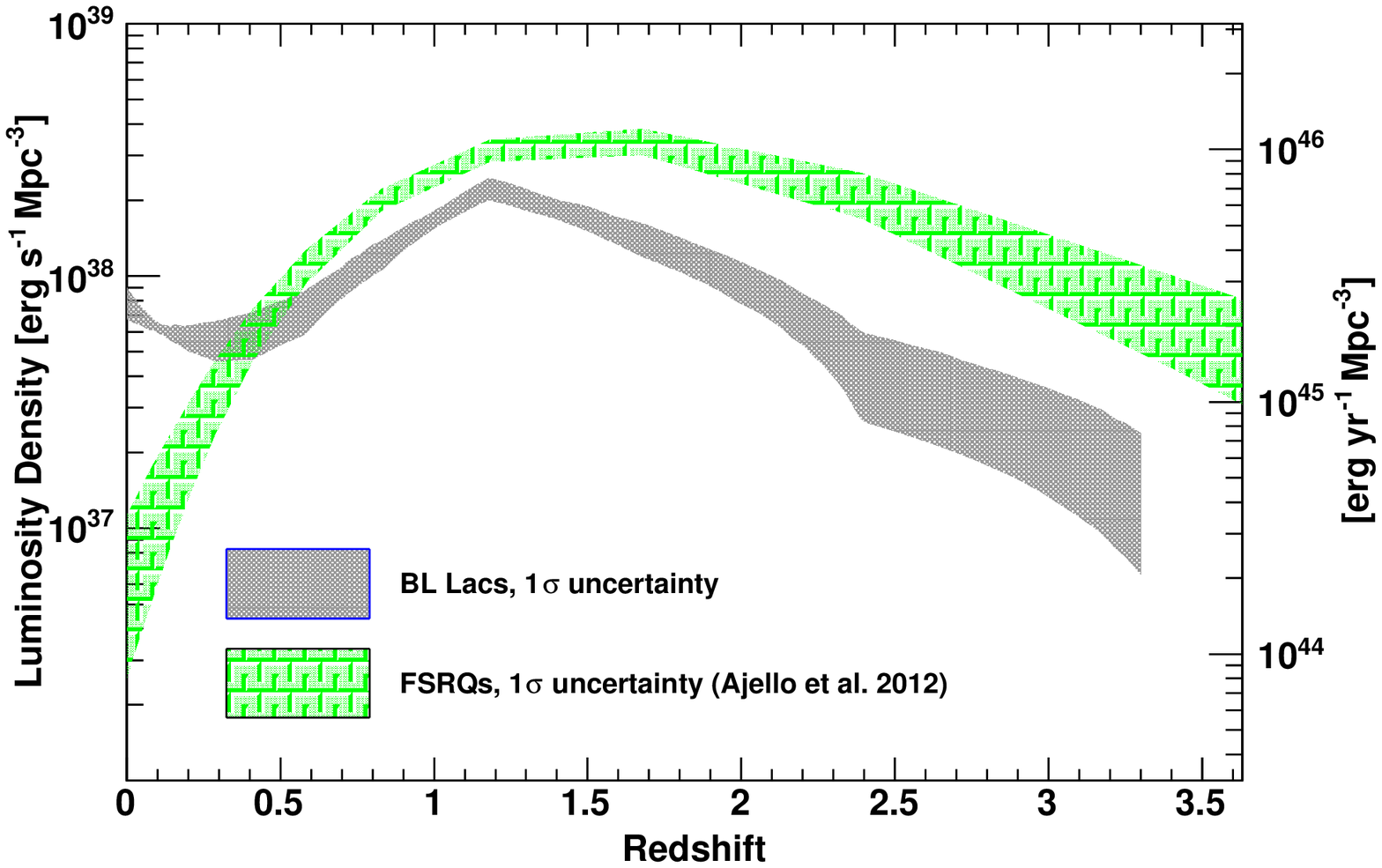} 
	\caption{Luminosity density as a function of redshift produced by
the {\it Fermi} BL Lacs.
 The gray band represents the confidence region
enclosing 68\,\% of the realizations of the best-fit LF to the Monte Carlo
samples. 
	\label{fig:lumindensity}
}
\end{centering}
\end{figure}

The local LF is found to have a rather steep power law 
($dN/dL\propto L^{-3.5}$) down to luminosities of 10$^{46}$\,erg s$^{-1}$,
flattening ($dN/dL\propto L^{-2.0}$)  below this value. 
{ Because of their steeper local LF and their lower luminosity,
BL Lacs reach higher densities than FSRQs (whose local LF is shown for comparison in Fig.~\ref{fig:localglf}). }
Fig.~\ref{fig:lumindensity} shows the evolution of the luminosity density of BL Lacs
compared to that of FSRQs.  With their larger luminosity,
FSRQs dominate at all redshifts $z>0.3$. Yet the extreme growth in BL Lac
numbers at low $z$ allows them to produce $>$10$^{45}$\,erg yr$^{-1}$ Mpc$^{-3}$,
or $\sim 90$\% of the local luminosity density.

\begin{deluxetable}{lccccccccccccc}
\tablewidth{0pt}
\tabletypesize{\tiny}
\rotate
\tablecaption{Best-fit parameters of the LDDE LFs. Parameters without an error
estimate were kept fixed during the fit. Parameter values were computed as the median of all the best-fit parameters
to the Monte Carlo sample, while the uncertainty represent the 68\,\%
containment region around the median value.
\label{tab:ldde}}
\tablehead{\colhead{Model}   & 
\colhead{A\tablenotemark{a}}          & \colhead{$\gamma_1$} & 
\colhead{L$_*$\tablenotemark{b}}      & \colhead{$\gamma_2$} &
\colhead{z$_c^*$}                     & 
\colhead{p1$^*$}                  & \colhead{$\tau$}     &
\colhead{p2}              & \colhead{$\alpha$} &
\colhead{$\mu^*$}              & \colhead{$\beta$}    & 
\colhead{$\sigma$}           & -2$\ln$L\tablenotemark{c}
}
\startdata

LDDE$_1$ & $9.20^{+20.60}_{-8.77}\times10^{2}$ & $1.12^{+0.13}_{-0.16}$ & $2.43^{+2.25}_{-1.30}$ & $3.71^{+16.29}_{-2.39}$ & $1.67^{+0.14}_{-0.10}$ & $4.50^{+0.75}_{-0.61}$ & $0.0$ & $-12.88^{+3.66}_{-2.12}$ & $4.46^{+6.47}_{-5.24}\times10^{-2}$ & $2.12^{+0.03}_{-0.03}$ & $6.04^{+2.15}_{-2.02}\times10^{-2}$ & $0.26^{+0.02}_{-0.02}$ & $-734.1$\\

LDDE$_2$ & $3.39^{+7.44}_{-2.13}\times10^{4}$ & $0.27^{+0.26}_{-0.46}$ &$0.28^{+0.43}_{-0.21}$ & $1.86^{+0.86}_{-0.48}$ & $1.34^{+0.22}_{-0.27}$ & $2.24^{+1.25}_{-1.07}$ & $4.92^{+1.45}_{-2.12}$ & $-7.37^{+2.95}_{-5.43}$ & $4.53^{+4.98}_{-6.52}\times10^{-2}$ & $2.10^{+0.03}_{-0.03}$ & $6.46^{+2.34}_{-2.07}\times10^{-2}$ & $0.26^{+0.02}_{-0.02}$ & $-764.6$\\

\hline
LDDE$_{noProb}$ & 1.04$^{+14.90}_{-0.74} \times10^{4}$& 0.58$^{+0.18}_{-0.75}$ & 0.50$^{+0.75}_{-0.47}$ & 1.99$^{+1.70}_{-0.70}$& 1.18$^{+0.38}_{-0.27}$ & 2.30$^{+2.11}_{-1.17}$ & 4.62$^{+5.38}_{-1.73}$ & -4.30$^{+2.07}_{-4.50}$ &  8.62$^{+5.55}_{-13.30}\times10^{-2}$ & 2.11$^{+0.03}_{-0.03}$ & 6.64$^{+1.84}_{-2.05}\times 10^{-2}$ & 0.26$^{+0.02}_{-0.02}$ & $-985$ \\

\enddata
\tablenotetext{a}{In unit of $10^{-13}$\,Mpc$^{-3}$  erg$^{-1}$ s.}
\tablenotetext{b}{In unit of $10^{48}$\,erg s$^{-1}$.}
\tablenotetext{c}{Value of the -2$\times$log-likelihood when the function
is minimized.}
\end{deluxetable}

%%%%%%%%%%%%%%%%%%%%%%%%%%%%%%%%%%%%%%%%%%%%%%%%%%%%%%%%%%%%%%%%%%
%%%%%%%%%%%%%%%%%%%%%%%%%%%%%%%%%%%%%%%%%%%%%%%%%%%%%%%%%%%%%%%%%%
\subsection{The Effect of Neglecting Redshift Constraints}
\label{sec:noz}

Neglecting redshift constraints and relying only on spectroscopic redshifts 
reduces the completeness of our sample to only $\sim$48\,\%.
As we show in the following this has dramatic effects on the reliability
of the luminosity function.

\begin{figure}[h!]
\begin{centering}
	\includegraphics[scale=0.7]{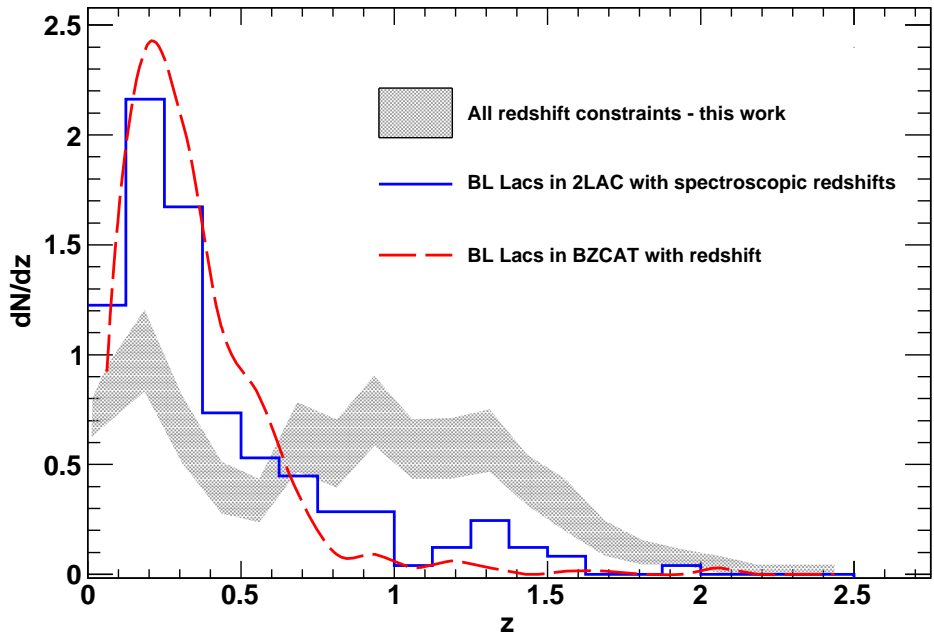} 
	\caption{Redshift distribution of {\it Fermi}'s BL Lac derived using
all constraints of $\S$~\ref{sec:z} compared to spectroscopic redshift
distributions of BL Lacs in the 2LAC catalog \citep{2LAC} 
and the Roma blazar catalog \citep[BZCAT,][]{massaro09}. The gray band
encloses the 68\,\% of all realizations of the redshift distribution
 of the Monte Carlo samples.
\label{fig:zdist}}
\end{centering}
\end{figure}

{ The main reason is that the distribution of spectroscopic redshifts approximates poorly the redshift distribution of BL Lacs inferred using all the redshift constraints presented in $\S$~\ref{sec:z}. This can clearly be seen in Fig.~\ref{fig:zdist}
which compares the BL Lac redshift distribution taking all constraints into account
compared to known BL Lac redshift distributions based solely on spectroscopic redshifts. These latter are biased to find low redshift BL Lacs, while it is clear from recent works \citep{rau12,shaw13b,shaw13,furniss13} that there is a relevant population of BL Lacs
at intermediate (z$\approx$0.5--1.5) redshift. This is not a spurious effect caused
by any of the techniques presented in $\S$~\ref{sec:z}, but an evidence
that comes from all of them. In order to test this, we removed the
exclusion probabilities from the used constraints and re-derived the LF.
The exclusion probability is available for all but 5 BL Lacs without
redshift and on average constrains a given object to be at z$\gtrsim$0.3-0.5.
If wrong, it might artificially push the average redshift of BL Lacs
to higher values. We find this is not the case. Indeed, even removing
the exclusion probabilities the redshift distribution of BL Lacs still
shows an increase at z$>0.5$ which is this time mostly due to the redshift lower limits. Moreover as reported in Tab.~\ref{tab:ldde}, the LF derived from discarding only the exclusion probabilities (see model LDDE$_{noProb}$)
is still in agreement with the best-fitting model (LDDE$_2$) that relies
on all constraints.

As expected from the above discussion
if we neglect all redshift constraints and rely only on the 103 BL Lacs
with spectroscopic redshifts, the best-fit LF (reported as model PLE$_{no-z}$
in Tab.~\ref{tab:ple}) changes fairly dramatically with respect to the
best fit LDDE$_2$ model. Indeed, instead of showing a change in the
evolution with source luminosity, it displays a very mild positive evolution
for all luminosity classes.
This would lead to a biased estimate of the evolution of BL Lacs.
We thus believe that results based on BL Lac samples with 
scarce redshift coverage are unreliable.
}

%%%%%%%%%%%%%%%%%%%%%%%%%%%%%%%%%%%%%%%%%%%%%%%%%%%%%%%%%%%%%%%%%%
%%%%%%%%%%%%%%%%%%%%%%%%%%%%%%%%%%%%%%%%%%%%%%%%%%%%%%%%%%%%%%%%%%
\subsection{The Intrinsic Luminosity Function of BL Lac Objects}
\label{sec:beaming}

Beaming is known to alter the shape of the intrinsic luminosity function 
\citep[e.g.,][]{urry84,urry91}.
In this Section we correct for this effect, recovering the intrinsic luminosity function
of the {\it Fermi} BL Lacs and their Lorentz  and Doppler
factor distributions. Here we adopt the formalism and symbols
already used in \cite{fsrq12}.

The observed 0.1--100\,GeV
luminosities $L$ defined in the present work are apparent isotropic luminosities (expressed in erg s$^{-1}$).
Since the jet material is moving at relativistic speed, the observed,
Doppler boosted, luminosities are related to the intrinsic values by:
\begin{equation}
L = \delta^p \mathscr{L}
\label{eq:int}
\end{equation}
where $\mathscr{L}$ is the intrinsic (unbeamed) luminosity and 
$\delta$ is the kinematic Doppler factor 
\begin{equation}
\delta = \left( \gamma -\sqrt{\gamma^2 -1}\cos\ \theta  \right)^{-1}
\end{equation}
where $\gamma=(1-\beta^2)^{-1/2}$ is the Lorentz factor, $\beta=v/c$ is the 
velocity of the emitting plasma and $\theta$ is the angle between the line of sight and the jet axis.  We will assume that our sources have Lorentz factors 
$\gamma$ in the range $\gamma_a\leq\gamma\leq\gamma_b$. Then the minimum Doppler 
factor is $\delta_{min}=\gamma_b^{-1}$ (when $\theta$=90$^{\circ}$) and the 
maximum is $\delta_{max}=(\gamma_a+\sqrt{\gamma_a^2 -1})^{-1}$ (when $\theta=0^{\circ}$).
We  adopt a value of $p$=4 which is appropriate if the observed emission is dominated 
by the SSC component of ejected plasma blobs
and discuss also the case of $p=3$ which applies to the case of
continuous jet emission.

We define the intrinsic luminosity function as:
\begin{equation}
\Phi(\mathscr{L}) = k_{1} \mathscr{L}^{-B}
\end{equation}
valid in the $\mathscr{L}_1\leq \mathscr{L}\leq \mathscr{L}_2$ range.
The joint probability of observing a  beamed luminosity $L$
and Doppler factor $\delta$ is \citep[see also][]{lister03}:
\begin{equation}
P(L,\delta) =  P_{\delta}(\delta) \cdot \Phi(\mathscr{L}) \frac{d\mathscr{L}}{dL}
\end{equation}
where $P_{\delta}(\delta)$ is the probability density for
the Doppler $\delta$ and $d\mathscr{L}/dL$=$\delta^{-p}$. Assuming a random distribution for the jet angles 
(i.e. $P_{\theta}=\sin\ \theta$), % the probability density function $P_{\delta}(\delta)$ 
this results in 
\begin{equation}
P_{\delta}(\delta) = \int P_{\gamma}(\gamma) P_{\theta}(\theta) \left|\frac{d\theta}{d\delta}\right| d\gamma = \int P_{\gamma}(\gamma) \frac{1}{\gamma \delta^2 \beta}d\gamma,
\end{equation}
since 
\begin{equation}
\left|\frac{d\theta}{d\delta}\right| = \frac{1}{sin(\theta)\delta^2\sqrt{\gamma^2-1}}=
\frac{1}{sin(\theta)\delta^2 \gamma \beta}
\end{equation}

From here it follows that
\begin{equation}
P_{\delta}(\delta) = \delta^{-2} \int^{\gamma_b}_{f(\delta)}
\frac{P_{\gamma}(\gamma)}{\sqrt{\gamma^2-1}}\ d\gamma
\end{equation}
where $P_{\gamma}(\gamma)$ is the probability density for $\gamma$ and
the lower limit of integration $f(\delta)$ depends on the Doppler
factor value and is reported in Eq.~A6 in \cite{lister03}.
Integrating over $\delta$ yields the observed luminosity function
of the Doppler beamed BL Lacs:
\begin{equation}
\Phi(L) = k_1 L^{-B} \int^{\delta_2(L)}_{\delta_1(L)} P_{\delta}(\delta) \delta^{p(B-1)}d\delta
\label{eq:unbeamed}
\end{equation}
where, as in \cite{cara08}, the limits of integration are
\begin{eqnarray}
\delta_1(L) = {\rm min} \{  \delta_{max},{\rm max}\left(\delta_{min},(L/\mathscr{L}_2)^{1/p}\right) \} \\
\delta_2(L) = {\rm max} \{   \delta_{min},{\rm min}\left(\delta_{max},(L/\mathscr{L}_1)^{1/p}\right) \} 
\end{eqnarray}
In this way, by fitting Eq.~\ref{eq:unbeamed} to the {\it Fermi} Doppler
boosted LF, it is possible to determine the parameters of the
intrinsic luminosity function and of the Lorentz-factor distribution. 

We assume that the probability density distribution for $\gamma$ is a power law
of the form
\begin{equation}
P_{\gamma}(\gamma)=C \gamma^k
\label{eq:gammadist}
\end{equation}
where C is a normalization constant and the function is valid for $\gamma_a\leq \gamma \leq \gamma_b$. We set the largest intrinsic luminosity $\mathcal{L}_2=10^4\mathcal{L}_1$, but this choice has hardly any impact
on the results.
Fits with parameters similar to those of FSRQs 
($p=4$, $\gamma_a=5$, $\gamma_b=40$ and $\mathcal{L}_1=10^{40}$\,erg s$^{-1}$)
are ruled out ($\chi^2$/dof $>$2.5). In order to obtain acceptable fits
we find that  $\mathcal{L}_1$ has to be set to 
$\leq10^{40}$\,erg s$^{-1}$ or $\leq10^{38}$\,erg s$^{-1}$
for the $p=3$ and $p=4$ case respectively.
% Marco, I thought p=4 was preferred. Should we not list it first?
Moreover, in agreement with the observation of BL Lacs in radio
\cite[e.g.][]{lahteenm03,lister09,savolainen10}, we set $\gamma_a=2$
which is lower than the minimum value used (and found) for FSRQs 
\cite[see e.g.][]{fsrq12}. In order to allow for a population
of highly beamed BL Lacs we set $\gamma_b=90$.

\begin{figure}[h!]
\begin{centering}
	\includegraphics[scale=0.7]{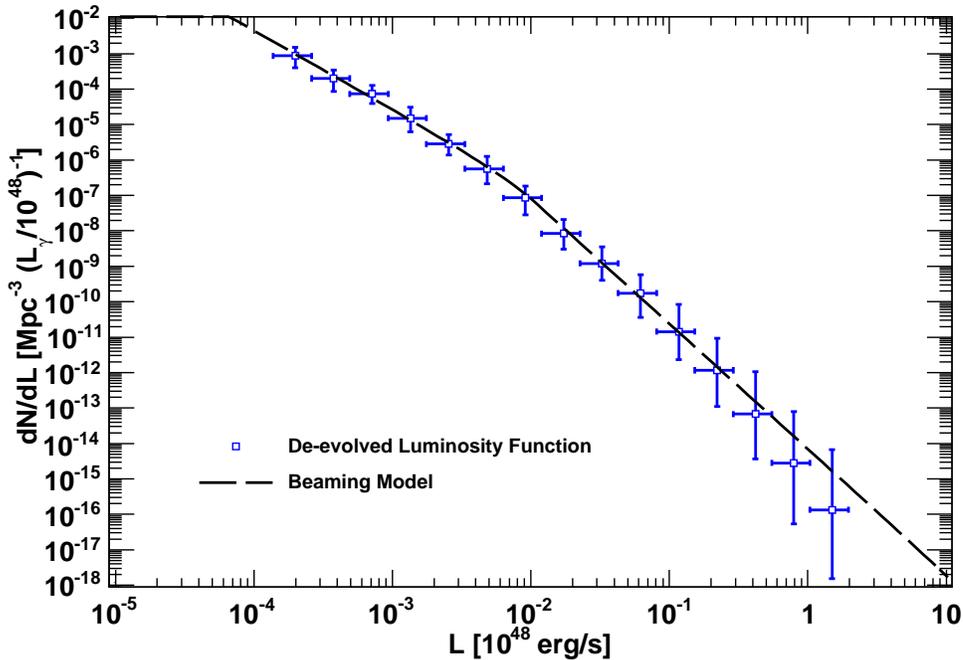} 
	\caption{{\it Fermi}'s LF de-evolved at redshift zero and best-fit 
beaming model (for $p$=4, dashed line, see text) described in $\S$~\ref{sec:beaming}. 
\label{fig:beaming}
}
\end{centering}
\end{figure}

\begin{figure}[h!]
\begin{centering}
	\includegraphics[scale=0.7]{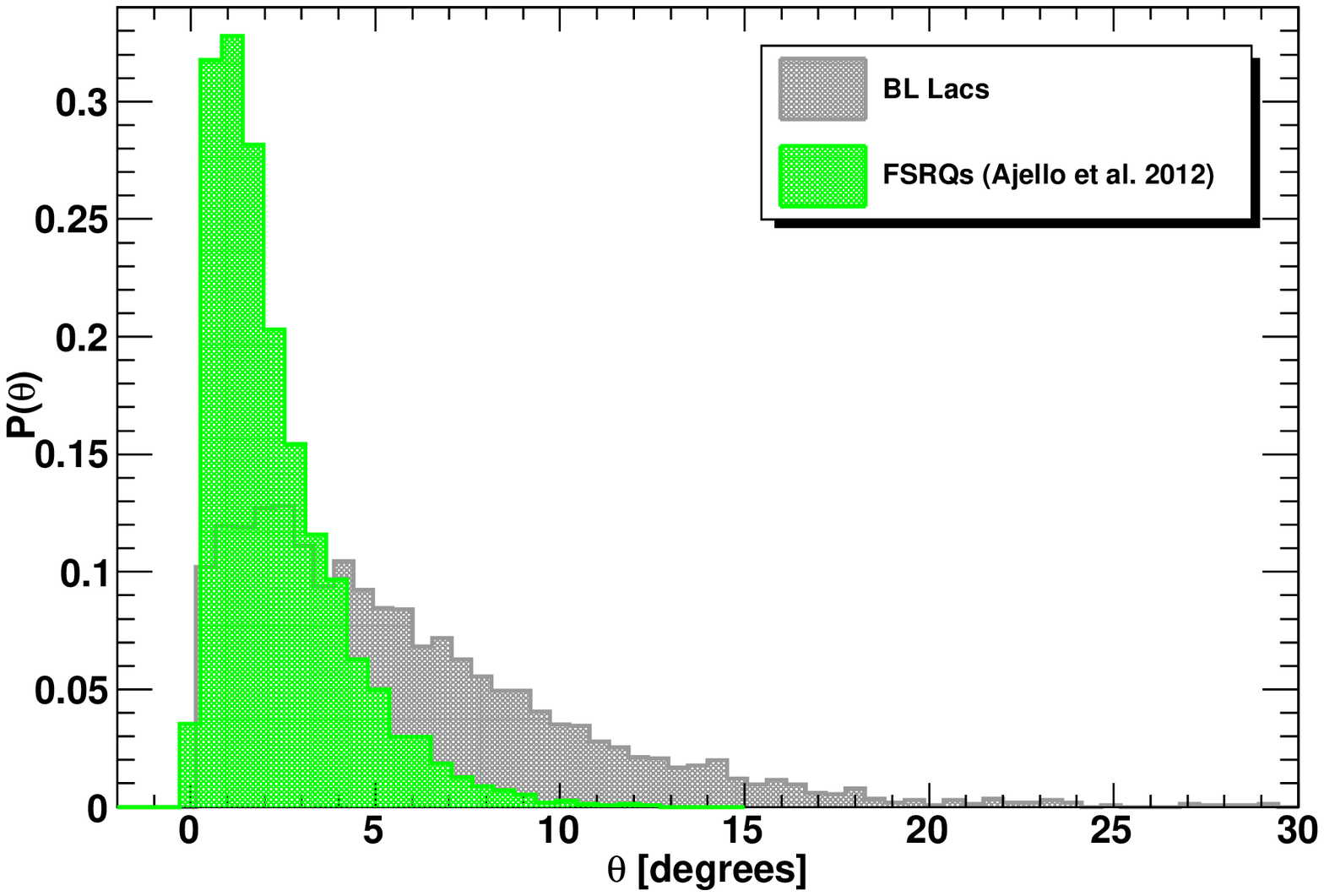} 
	\caption{Normalized distributions of viewing angles with respect to
the jet axis for  {\it Fermi} BL Lacs and FSRQs.
\label{fig:los}
}
\end{centering}
\end{figure}

The free parameters of the problem are the normalization ($k_1$) and, the slope
($B$) of the intrinsic LF and the slope $k$ of the Lorentz factor distribution.
We have fitted Eq.~\ref{eq:unbeamed} to the {\it Fermi} LF de-evolved at 
redshift zero derived in $\S$\ref{sec:ldde}.
Fig.~\ref{fig:beaming} shows how the best-fit beaming model reproduces the 
local LF of BL Lacs measured by {\it Fermi}. For the $p=4$ case we can use 
the fit values to derive an intrinsic LF slope of $B=3.30\pm0.30$ and a
Lorentz-factor distribution index of $k=-2.32\pm0.51$.  The parameters for 
the $p=3$ case are similar.  Our distribution of Lorentz factors
is somewhat steeper than (but compatible with, within the uncertainties) 
that found by \cite{lister97} who report a slope of $-1.75<k<-1.5$.
The fit values are summarized in Table~\ref{tab:beaming}. 
The Lorentz-factor distributions (for the $p=3$ and $p=4$ cases)
imply an average Lorentz factor $\gamma\approx6$ for the {\it detected}
{\it Fermi} blazars. 
This is in agreement with past inferences for radio and X-ray selected
BL Lacs \citep[see discussion in ][]{urry95,morganti95}.
The average Lorentz-factor depends on the value adopted for 
$\gamma_a$ (and to lesser extent on $\gamma_b$). Within the errors, the slope $k$
is the same for BL Lacs and FSRQs ($-2.32\pm0.51$ versus $-2.03\pm0.70$ 
respectively).
The fact that it is not possible to produce a good fit
to the data adopting the same $\gamma_a$ for both populations implies
that a population of BL Lacs exists
 with jets slower than those of FSRQs.
This yields a smaller value for the average Lorentz factor 
($\gamma^{BL\ Lac}\approx6$ versus $\gamma^{FSRQ}\approx12$)
and that BL Lacs are seen under larger angles 
($\sim$5\,$^{\circ}$ versus $\sim$2\,$^{\circ}$ for FSRQs, see Fig.~\ref{fig:los}).

{ Finally, we also tested different parametrizations of the distribution
of Lorentz factors (Eq.~\ref{eq:gammadist}). We used a linear, an exponential,
and a Gaussian distribution. None of these models provides an acceptable fit
to the data ($\chi^2/dof>3$). We thus conclude that parametrizing the Lorentz factor 
distribution 
with a power-law model
\citep[as done also in the literature, e.g.][]{urry84,cara08}  is a reasonable assumption.} 

\begin{deluxetable}{lc|c}
\tablewidth{0pt}
\tablecaption{Parameters of the beaming models described in the text. 
Parameters without
an error estimate were kept fixed during the fitting stage.
\label{tab:beaming}}
\tablehead{
%%%%%%%% column names
\colhead{Parameter} & \colhead{Value} & \colhead{Value}}
\startdata
$k$              &  -2.26$\pm0.20$  & -2.32$\pm0.51$ \\
$k_1$            &   4.3$\pm0.5$\tablenotemark{a} & 2.7$\pm0.5$\tablenotemark{a}\\
$B$              &   3.96$\pm0.08 $ & 3.30$\pm0.30$\\ 
$\gamma_a$       &   2  & 2 \\
$\gamma_b$       &   90 & 90 \\
$\mathscr{L}_1$  & 10$^{40}$ & 10$^{38}$\\
$\mathscr{L}_2$  & 10$^{44}$ & 10$^{42}$\\
$p$              &   3       & 4\\
$\chi^2/dof$     & 0.3& 0.21\\
Average $\gamma$ & 6.1$^{+1.1}_{-0.8}$ & 5.8$^{+3.6}_{-1.6}$

\enddata

\tablenotetext{a}{In units of 10$^{-27}$.}
%\tablenotetext{b}{In units of 10$^{-26}$.}
\end{deluxetable}

%%%%%%%%%%%%%%%%%%%%%%%%%%%%%%%%%%%%%%%%%%%%%%%%%%%%%%%%%%%%%%%%%%
%
%         Subclasses
%
%%%%%%%%%%%%%%%%%%%%%%%%%%%%%%%%%%%%%%%%%%%%%%%%%%%%%%%%%%%%%%%%%%
\section{Sub-classes of BL Lac Objects}
\label{sec:subclasses}

Our sample can be subdivided into 96 HSPs, 64 ISPs and 45 LSPs
on the basis of the frequency of the synchrotron peak \citep[see][]{2LAC}.
For only 6 objects there is not enough multiwavelength coverage
to define accurately the position of the synchrotron peak.
It is thus possible to test whether the different sub-classes of BL Lacs
have different evolution. In particular we are interested in testing
the following two scenarios: 1) whether HSPs have a different evolution
with respect to ISPs and LSPs, and 2) whether LSPs have a different
evolution with respect to HSPs and ISPs.
{ For completeness the best-fit parameters of all the models
described in $\S$~\ref{sec:hsp} and \ref{sec:lsp} are reported in
the Appendix ($\S$~\ref{sec:app}).}

%%%%%%%%%%%%%%%%%%%%%%%%%%%%%%%%%%%%%%%%%%%%%%%%%%%%%%%%%%%%%%%%%%
%
%         Subclasses : HSP
%
%%%%%%%%%%%%%%%%%%%%%%%%%%%%%%%%%%%%%%%%%%%%%%%%%%%%%%%%%%%%%%%%%%
\subsection{The Evolution of HSP Objects}
\label{sec:hsp}

Using the same best-fit models (namely the PLE and LDDE models of 
$\S$~\ref{sec:lf}) we next examine separately the HSP objects.
The LDDE model is slightly preferred to the PLE model (TS$\approx$12).
Both models indicate that the evolution of the HSP is negative: i.e.
the density is growing with decreasing redshift.

For the PLE model, the relevant parameters are: 
$k=3.82^{+1.29}_{-1.17}$, $\tau=1.35^{+0.17}_{-0.32}$, 
and $\gamma=-0.40^{+0.07}_{-0.14}$.
For all the HSPs with L$_{\gamma}\leq$10$^{46}$\,erg s$^{-1}$
the evolution is negative $z_c\leq0$ and $k_d\leq$0.
The same trend is confirmed by the LDDE model whose relevant
parameters are $p1^*=0.48^{+1.63}_{-0.48}$, and
$\tau=6.76^{+2.33}_{-1.82}$ (see Eq.~\ref{eq:p1}).

For the class of ISP and LSP objects the LDDE model produces
a very small improvement over the PLE model (TS$\approx$3).
Both models indicate positive evolution for the ISPs and LSPs considered together.
For the PLE, the parameters that govern the evolution
are: $k=7.86^{+1.41}_{-1.86}$, $\tau=0.98^{+0.28}_{-0.31}$, 
and $\xi=-0.25^{+0.05}_{-0.08}$. In this scenario low-luminosity
sources are characterized by a slow positive evolution consistent with
no evolution.

\begin{figure}[h!]
\begin{centering}
	\includegraphics[scale=0.7]{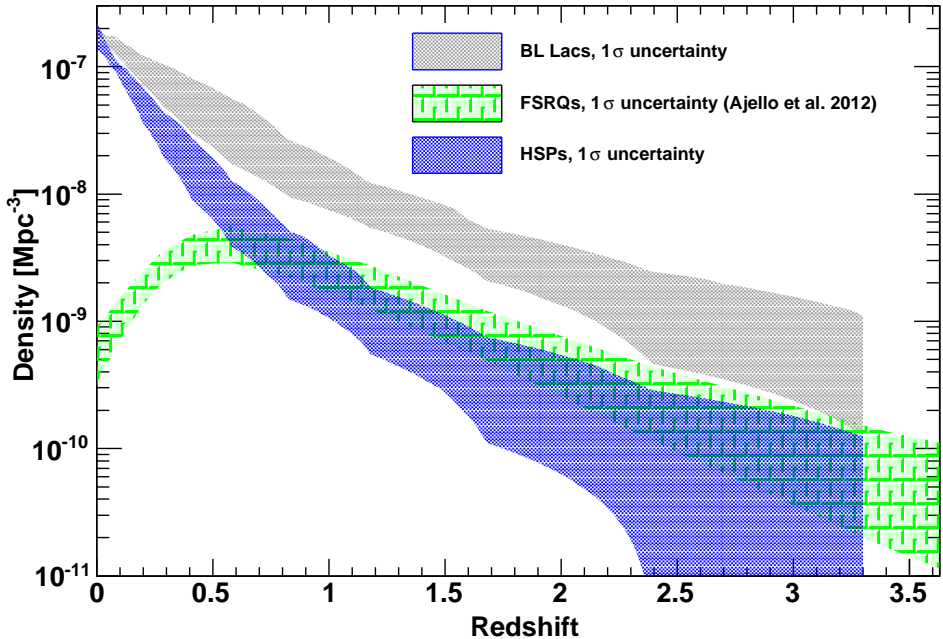} 
	\caption{Number density (per unit co-moving volume) of BL Lacs,
FSRQs and HSPs.
\label{fig:hsp}
}
\end{centering}
\end{figure}

\begin{figure*}[ht!]
  \begin{center}
  \begin{tabular}{c}
%\hspace{-1cm}
    \includegraphics[scale=0.88]{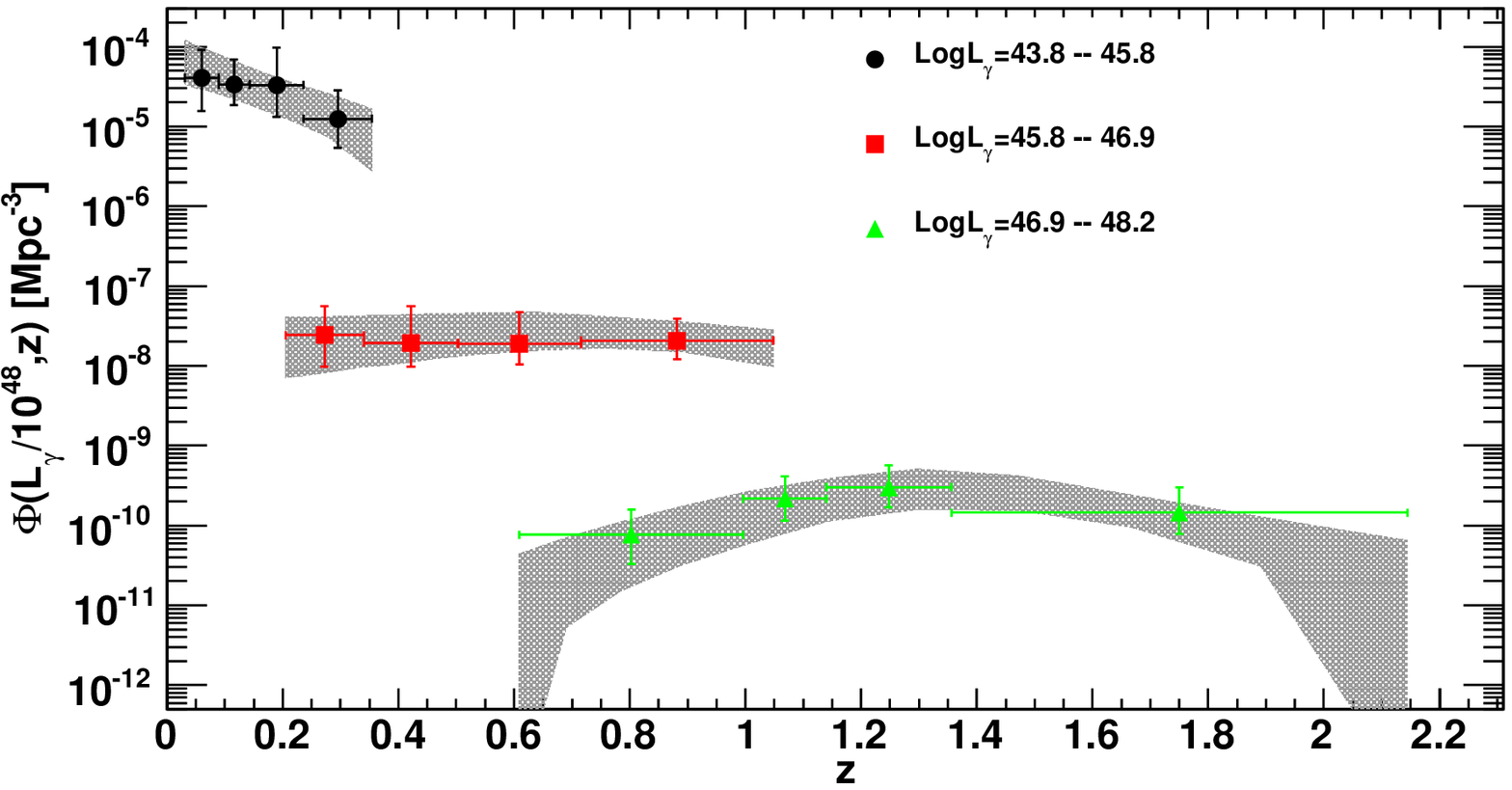} \\
%\hspace{-1cm}
  	 \includegraphics[scale=0.88]{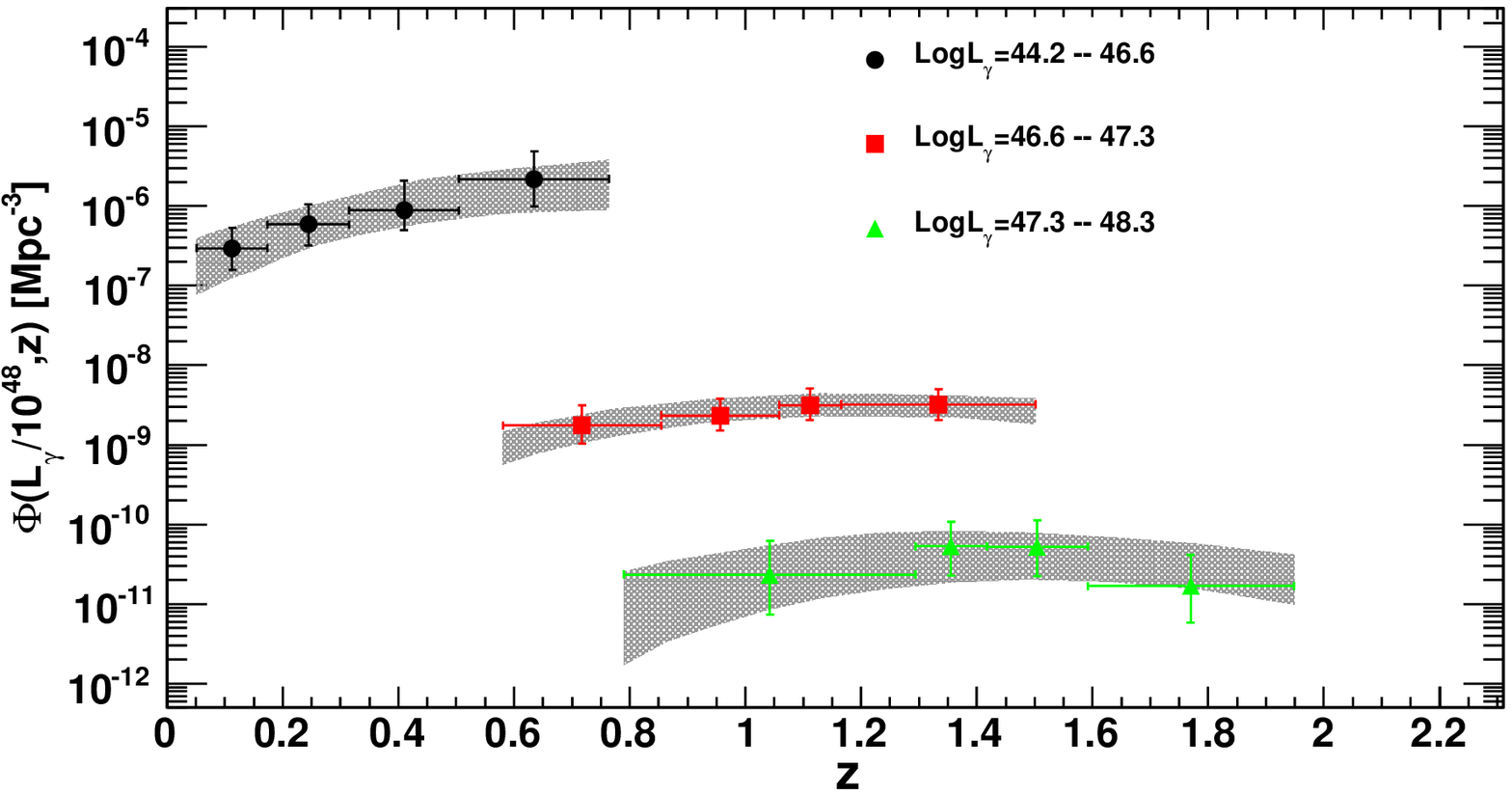} \\
\end{tabular}
  \end{center}
\caption{Evolution of different luminosity classes of HSPs (top)
and ISPs+LSPs (bottom). Note the different evolutionary behavior
(negative for HSPs versus positive for ISPs+LSPs evolution) of low-luminosity sources.
\label{fig:classes}}
\end{figure*}

The different evolutionary behavior of HSPs with respect to
all other blazar classes can be appreciated in Fig.~\ref{fig:hsp}
which shows that the dramatic rise in the number density 
of BL Lacs at z$\leq$1 is driven almost entirely by the HSP population.
The fact that low-luminosity HSP objects are the only ones experiencing
negative evolution can also be seen directly in Fig.~\ref{fig:classes}.

%%%%%%%%%%%%%%%%%%%%%%%%%%%%%%%%%%%%%%%%%%%%%%%%%%%%%%%%%%%%%%%%%%
\subsection{The Evolution of LSP Objects}
\label{sec:lsp}

LSP objects are the class of BL Lac objects that most closely resemble the FSRQ class.
Their synchrotron component peaks at frequencies $<10^{14}$\,Hz
\citep{2LAC}, they can show rather large
values of the Compton dominance\footnote{The Compton dominance
is the ratio between the  Compton peak luminosity to the synchrotron 
peak luminosity.} \citep{finke13}, and their average redshift
is larger than that of the rest of BL Lacs.
A number of LSPs might be FSRQs whose jet is aligned along our line-of-sight
and whose non-thermal radiation reduces the equivalent width of optical
lines. Indeed, \citet{shaw13} find that many BL Lac sources 
(especially LSPs) are
spectrally classified as FSRQ when seen in low states.
Since the FSRQ class is known to evolve positively \citep{fsrq12}, the close connection 
between FSRQs and high luminosity BL Lacs might be responsible for the positive
evolution detected for the high-luminosity objects in $\S$~\ref{sec:ldde}.

{ The model that best describes the LF of LSP objects is the
PLE model. The best-fit evolutionary parameters of the PLE model
($k=7.59^{+1.78}_{-2.09}$, $\tau=1.30^{+0.26}_{-0.39}$, 
and $\xi=-0.23^{+0.05}_{-0.08}$) imply a strong positive evolution for 
LSP objects of all luminosities as it is the case for FSRQs \citep[see e.g.][]{fsrq12}.}

The LDDE model\footnote{The PLE model produces a worse
fit (TS=-18) than the LDDE.} applied to HSPs and ISPs
yields: p$_1^*=1.98^{+1.46}_{-1.20}$, and
$\tau=6.38^{+1.58}_{-1.66}$. These parameters are in agreement
with those of the full sample (reported in Tab.~\ref{tab:ldde}
and imply negative evolution for low-luminosity objects 
(L$_{\gamma}\leq$10$^{46}$\,erg s$^{-1}$) and  positive
evolution for high-luminosity objects (L$_{\gamma}>$10$^{46}$\,erg s$^{-1}$).

For high-luminosity BL Lacs (L$_{\gamma}\sim$10$^{47}$\,erg s$^{-1}$)
both models described above find a positive evolution with $k_d\approx8.9$ (for the PLE model) and
p$_1\approx 8-9$ (for the LDDE).
As such it is apparent that LSPs are not driving the positive
evolution of the whole BL Lac sample, but that this is a characteristic
of all high-luminosity BL Lacs.

%%%%%%%%%%%%%%%%%%%%%%%%%%%%%%%%%%%%%%%%%%%%%%%%%%%%%%%%%%%%%%%%%%
%
%         Discussion and Conclusions
%
%%%%%%%%%%%%%%%%%%%%%%%%%%%%%%%%%%%%%%%%%%%%%%%%%%%%%%%%%%%%%%%%%%
\section{Discussion and Conclusions}
\label{sec:discussion}

In this work we determined the first luminosity function 
of GeV-detected BL Lacs. This was made possible by the relatively complete redshift
information gleaned from a variety of methods \cite[see e.g.][]{rau12,shaw13},
leaving only 5 of our 211 BL Lacs without redshift constraints.
Previous BL Lac samples selected at other frequencies contained few objects 
(often $<$50) and typically lacked redshift information for $\geq$30\,\% of 
the objects  \citep[see e.g.][]{stickel91,padovani07,marcha13}.
Poor redshift completeness renders the luminosity function unreliable (see 
$\S$~\ref{sec:noz}).
Also, our sample contains a substantial number of BL Lacs from each of the
three spectral peak subclasses and covers a large redshift range.
As such, this sample stands as the largest and most complete 
(redshift wise) set of BL Lacs ever used at any frequency, and has
allowed a greatly improved characterization of the BL Lac population,
beaming and evolution.
The main results of our analysis are discussed below.

%%%%%%%%%%%%%%%%%%%%%%%%%%%%%%%%%%%%%%%%%%%%%%%%%%%%%%%%%%%%%%%%%%
%
%         Evolution of the Luminosity Function
%
%%%%%%%%%%%%%%%%%%%%%%%%%%%%%%%%%%%%%%%%%%%%%%%%%%%%%%%%%%%%%%%%%%
\subsection{The Evolution of the BL Lac Luminosity Function}

In the past, BL Lacs have been found to show a wide range of evolutionary patterns.
\cite{rector00}, \cite{giommi99}, and \cite{beckmann03} 
(whose samples contained large fractions of HSP objects)
found the BL Lacs to evolve negatively, 
\cite{caccianiga02} and \cite{padovani07} found that BL Lacs
do not evolve, and recently \cite{marcha13} reported on a sample with
positive evolution. These different results were likely due
to limited statistics and inadequate redshift coverage
mixed with selection of different classes of BL Lacs.

As clear from this work, the evolution of the BL Lac class is complex.
We found that the evolution of BL Lac objects selected
by {\it Fermi} can be described with a  LDDE model similar to the one
used for FSRQs \citep{fsrq12}. 
Indeed, 
luminous BL Lacs ($\sim$10$^{47}$\,erg s$^{-1}$) evolve
as strongly ($p_1\sim7$) as FSRQs (see  $\S$~\ref{sec:ldde}). However, the evolution of BL Lacs
slows down with luminosity, becoming negative for  objects
with L$_{\gamma}\leq$10$^{45.5}$\,erg s$^{-1}$.

Subdividing the sample in HSP, ISP and LSP objects we find that
the negative evolution is in fact isolated to the HSP population,
while the ISP and LSP evolve positively from the lowest luminosities.
Our analysis thus confirms results based on samples dominated by
HSP objects \citep[e.g.][]{giommi99,beckmann03}.
{ We tested if different slopes of the luminosity function (Eq.~\ref{eq:lf0}) respectively for HSPs and ISPs+LSPs could be compatible with a common (e.g. positive) shape of the evolution. We find that, while it seems HSPs have a slightly flatter luminosity function  (at redshift $\approx$0) than ISPs+LSPs, imposing a common shape of the evolution to the whole population substantially worsen the fit (by $\sim$10\,$\sigma$). On the other hand, allowing HSPs and ISPs+LSPs to have different evolutions reproduces the negative-positive dichotomy\footnote{In this test we defined the luminosity function as the sum of two different functions representing the HSP and ISP+LSP populations. For the HSP population we adopted a single power law in luminosity and a PLE model with $e(z)=(1+z)^k$ while we adopted the PLE model with $\beta=0$ described in Sec.~\ref{sec:lf}. For the HSP and ISP+LSP populations we found a slope (in luminosity) of the luminosity function of 2.04$\pm0.08$ and 2.35$\pm0.10$ respectively, while for the evolutionary factor we found k=-0.9$\pm0.3$ (for HSPs) and k=12.4$\pm0.7$ and $\gamma$=-0.19$\pm$0.01 (for the ISPs+LSPs). These results imply that the two populations have a similar slope in luminosity, but a different form of the evolution which is confirmed to be negative for HSPs and positive (with a redshift peak at $\approx$1.3) for ISPs+LSPs.}.}
We can also exclude that the negative evolution scenario is caused by
inadequate redshift coverage (incompleteness), 
or by the fact that HSPs are not detected
to sufficiently large redshifts (sensitivity limit). 
Indeed, from our set of Monte Carlo
simulations we find that $~\sim$30\,\%  and $~\sim$7\,\%
of all the HSPs detected by {\it Fermi} lie respectively at z$>1$ and z$>1.5$.
Moreover, the effect of the extragalactic background light (EBL, see $\S$~\ref{sec:softening}) is not
severe and does not bias either the measured fluxes or the photon
indices in the 0.1--100\,GeV band.
In order to exclude that the negative evolution of low-luminosity BL Lacs
(and HSPs) is caused by the incompleteness of the sample used here (see $\S$~\ref{sec:sample}), we explore a worst case scenario assuming that all
 $\sim$20 unassociated sources  are
BL Lacs lying in the 0.2--0.7 redshift range. A large population
of BL Lacs at intermediate redshifts (z$\sim$0.5, see left panel of
 Fig.~\ref{fig:classes}) would be needed to invert the negative evolution.
Using actual fluxes and photon indices drawn from the 23 unclassified possible
AGN, assuming that all are HSP and drawing random redshifts in the
critical 0.2-0.7 range, we find that only a relatively small fraction 
($\sim$12\,\%)  could be HSPs with Log L$_{\gamma}<$45.5.
Accordingly, even in this worst-case scenario we find that 
these missing identifications cannot significantly alter our measurement
of negative evolution for this sub-class.

The slowing down of the evolution with decreasing source luminosity
has been observed in many kinds of AGN, including the population of radio galaxies
\citep{longair66,schmidt72,willott01}, but an inversion of the evolution at very low luminosity
as observed here is difficult to interpret.
While the close connection between the FSRQ and LSP classes is quite apparent,
it is less obvious that this trend can be extended to the HSP BL Lacs.
However, one may interpret this spectral sequence as a progression
caused by the gradual depletion of an AGN's gas reservoir 
via accretion \citep[e.g.][]{cavaliere02,bottcher02}. In this context
a LSP object would transition from disk-powered jet production (at high
accretion rates) through the ISP class to  an HSP BL Lac object with low 
accretion rates and a radiatively inefficient accretion flow. In 
LSPs, strong cooling due to the circumnuclear radiation fields would limit the maximum
energy reached by the accelerated electrons.
For the HSPs, due to the decreased cooling efficiency, particles
would be accelerated to much larger energies which would translate
into a peak frequency of the synchrotron component that moves
from 10$^{13}$\,Hz up to 10$^{17}$\,Hz. This reproduces the
paradigm of the blazar sequence \citep{ghisellini98,fossati99}.

The activity of FSRQs, if triggered by galaxy merging events as
is common for high-luminosity quasars, would be short lived
($\tau\sim$0.1\,Gyr), and be followed by the
low-accretion regime of HSP-type BL Lacs which can be sustained for much 
longer times \citep[$\tau\sim$5--7\,Gyr][]{cavaliere02}. 
In the high-redshift Universe, where gas was abundant, galaxy merging
favors the activity of FSRQs. As the Universe expands, galaxy merging
becomes infrequent and most of the FSRQs/LSPs finish consuming their fuel reserve,
 transitioning to a long-lasting low accretion regime.
If the HSPs are indeed starved LSP objects then one should
observe an increase in the space density of BL Lacs with only a slight lag
(since $\tau\sim$0.1\,Gyr for FSRQs) from the decrease in the space density
of FSRQs. Fig.~\ref{fig:hsp} can been seen as supporting this picture.
Indeed at z$\geq$1.5 the number density of HSPs decreases in a similar
way to that of FSRQs and LSPs and ISPs objects. At z$<$0.5 when
the FSRQs turn off, the space density of HSPs, and in particular
the low luminosity HSP, quickly increases.

	This scenario is attractive but still speculative.
At present we lack a quantitative comparison between the
space densities of the FSRQ+LSP objects
and the (possibly remnant) population of HSP.
Certainly different beaming characteristics ({ and their potential
evolution with redshift}) can affect the estimated populations
and complicate this comparison. There may also be differences between the low
and high-peaked sources in the typical  black hole mass or host galaxy environment.
Nevertheless, the correlation of opposing evolutionary trends found
here points to a  possible connection between these AGN populations.

%%%%%%%%%%%%%%%%%%%%%%%%%%%%%%%%%%%%%%%%%%%%%%%%%%%%%%%%%%%%%%%%%%
%
%         Softening of Blazar spectra
%
%%%%%%%%%%%%%%%%%%%%%%%%%%%%%%%%%%%%%%%%%%%%%%%%%%%%%%%%%%%%%%%%%%
\subsection{Softening of Blazar Spectra with Redshift}
\label{sec:softening}

In $\S$~\ref{sec:resultpde} and \ref{sec:ldde} we found that
{\it Fermi} blazar spectra soften with increasing luminosity.
In particular, all the best-fit models have BL Lac spectra softening 
at high luminosity.
The average photon index changes from $\sim$2.0 to $\sim$2.2 when
the luminosity changes from 10$^{44}$\,erg  s$^{-1}$
to 10$^{48}$\,erg s$^{-1}$.

\begin{figure*}[ht!]
  \begin{center}
  \begin{tabular}{c}
%\hspace{-1cm}
    \includegraphics[scale=0.7]{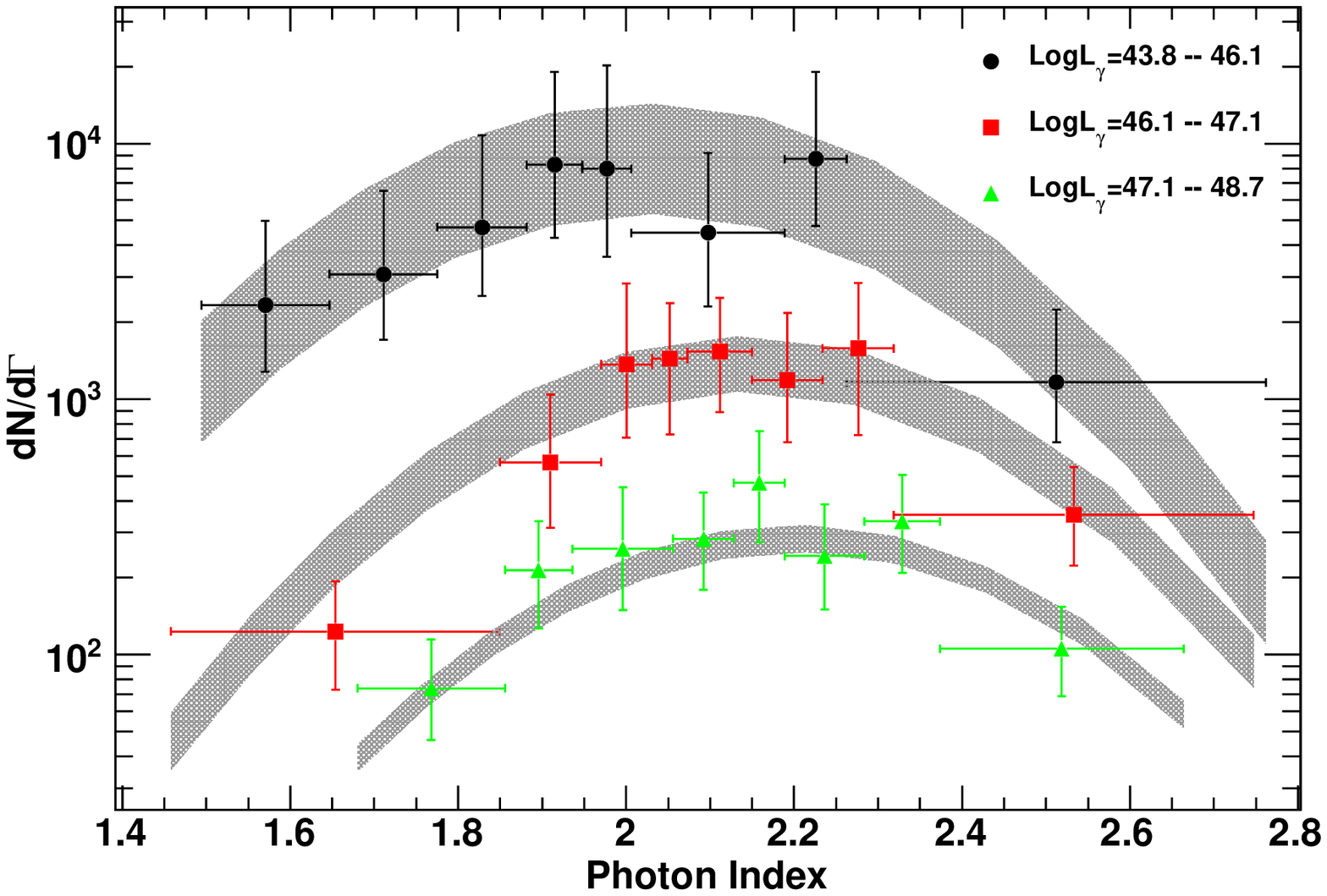} \\
%\hspace{-1cm}
  	 \includegraphics[scale=0.7]{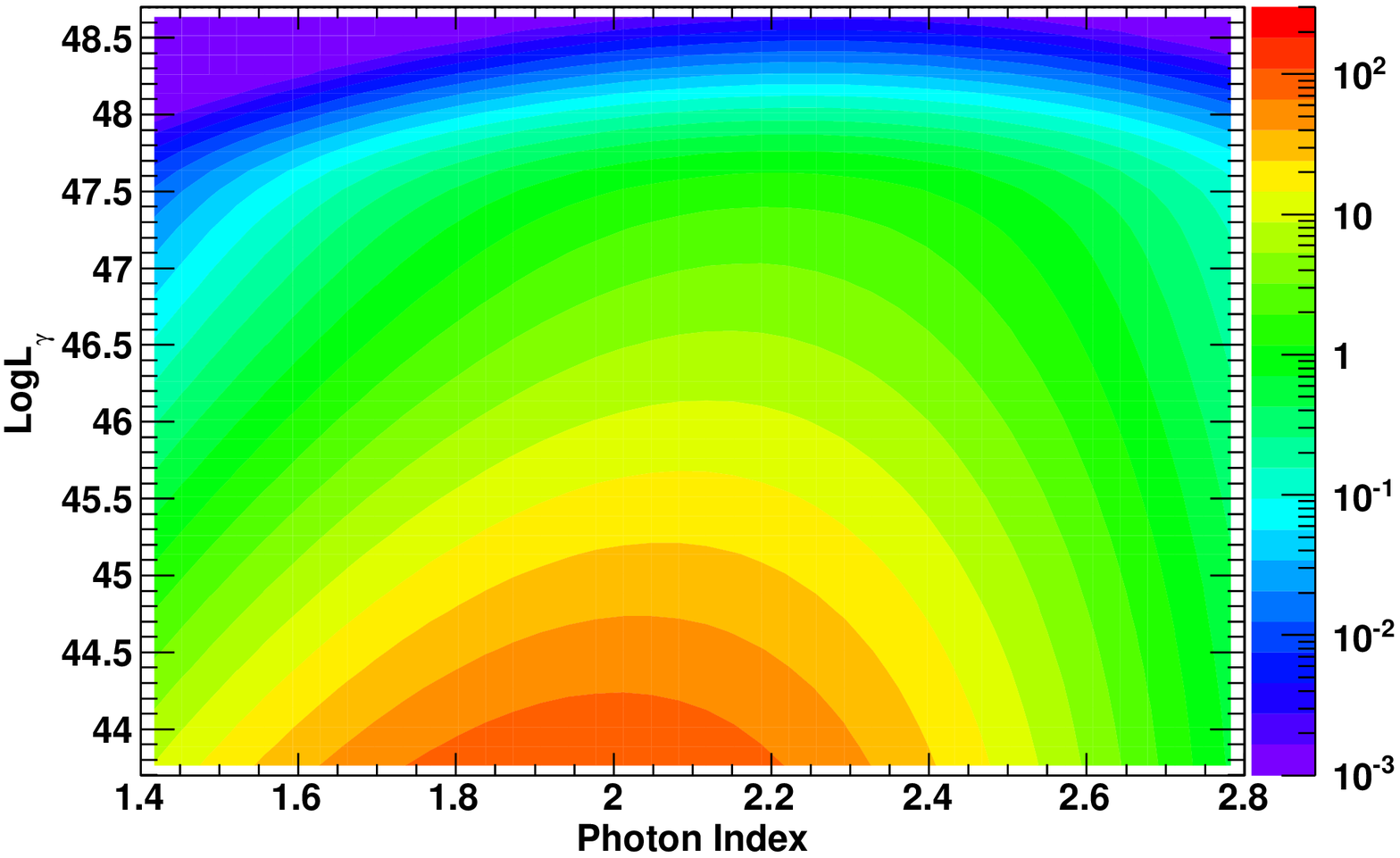} 
\end{tabular}
  \end{center}
\caption{Top Panel: Deconvolved index distribution
for three luminosity classes of {\it Fermi} BL Lacs.
Note the shift in the average of the distribution with luminosity.
Bottom Panel: Deconvolved photon index-luminosity plane for the 
{\it Fermi} BL Lacs (in units of number of BL Lacs per bin
of luminosity and index).
\label{fig:index}}
\end{figure*}

The left panel of Fig.~\ref{fig:index} shows the deconvolved {\it intrinsic}
photon index distributions for three different luminosity classes.
The deconvolution was performed with the method outlined in $\S$~\ref{sec:ml}
(see Eq.~\ref{eq:nmdl}). The y-axis reports the integral
over redshift and luminosity of Eq.~\ref{eq:1} (essentially $dN/d\Gamma$).
The trend of the average softening of the BL Lac spectra with
increasing luminosity is apparent even though the dynamic range
is small: i.e. the index changes by only $\Delta\Gamma\approx0.2$ in 4
orders of magnitudes in luminosity.
The right panel of Fig.~\ref{fig:index} shows the photon index-luminosity
plane as predicted by the best-fit LF\footnote{Each point of the index-luminosity
plane reports the number of BL Lacs that would be visible in the whole sky
from an ideal telescope which suffered no selection effects.}.
The correlation of the photon index with luminosity is very clear.
Both of these figures include corrections for all known selection
effects, so we infer that this trend is directly apparent in the sources
(although it is strongly amplified in the observed sample through selection
effects).
 
If selection effects  are not properly taken into account, a spurious
 index-luminosity correlation can be artificially 
introduced because of the energy dependence
of the {\it Fermi}-LAT point-spread function \citep{atwood09} which
favors the detection at low fluxes of sources with a hard spectrum
\citep[see Fig.~1 in][]{pop_pap}. However, the analysis of the
source count distribution as a function of photon index did not reveal
any significant correlation between flux and photon index 
\citep{pop_pap,singal12}. 

\begin{figure}[h!]
\begin{centering}
	\includegraphics[scale=0.7]{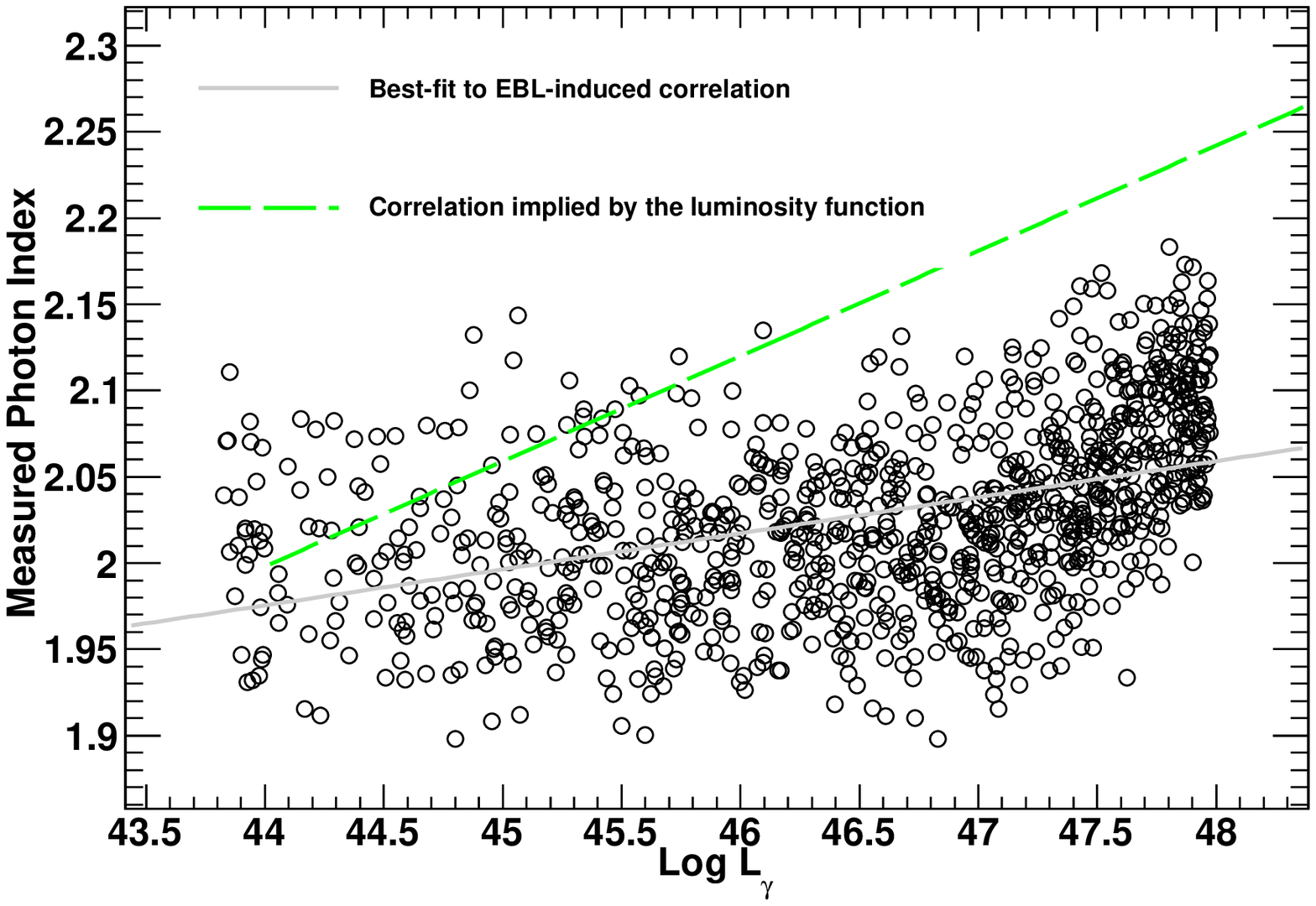} 
	\caption{Measured photon index versus measured luminosity
for a sample of 1000 power-law spectra (with $\Gamma$=2)  with
fluxes and redshifts randomly drawn from the sample of BL Lacs
used here. An EBL attenuation using the model of \cite{franceschini08}
was applied to all the spectra.
The measured quantities
(index, flux and luminosity) were derived fitting source spectra 
with a power law.
The solid line shows the best linear fit to the observed data points.
The dashed line shows the correlation between photon index
and luminosity found by the best-fit LDDE model in $\S$~\ref{sec:ldde}.
\label{fig:ebl}
}
\end{centering}
\end{figure}

Finally, a spurious luminosity-index correlation might
be produced by absorption of high-energy photons by the EBL. 
The EBL attenuation would make measured spectra
steeper than intrinsic, preferentially affecting high-redshift (and thus 
high-luminosity) sources. We checked to see if this produced the observed trend,
by simulating $\sim$1000
spectra in the 0.1--100\,GeV band using a power-law model with a photon
index of 2.0. Fluxes and redshifts were drawn from the observed sample of BL Lacs
and EBL absorption was applied using models 
\citep{franceschini08,finke10,dominguez11} in agreement with {\it Fermi} observations
of the EBL attenuation \citep{ebl12}. The result of this analysis, 
reported in Fig.~\ref{fig:ebl}, shows that the EBL effect on measured
photon indices (in the 0.1--100\,GeV band) is minor.
While the photon index of BL Lacs with Log L$_{\gamma}>$47.5\,erg s$^{-1}$
is modified by the EBL attenuation by $\Delta\Gamma\sim$0.055, the index
of all sources below that luminosity is basically unaffected.
Thus, we conclude  that the observed index-luminosity
correlation is not an artifact of selection effects or cosmic EBL absorption, but
intrinsic to the sources.

\cite{ghisellini09b} was the first to note (although without
accounting for selection effects) that a correlation between
index and $\gamma$-ray luminosity seemed to exist for BL Lacs
and FSRQs detected by {\it Fermi}. They proposed that the 0.1--100\,GeV
luminosity of 10$^{47}$\,erg s$^{-1}$  which separates
{\it hard} BL Lacs from {\it soft} FSRQs could be associated with
a transition in the accretion flow from  radiatively inefficient \citep[e.g.][]{narayan97}
to optically-thick radiatively efficient  \citep{shakura73}.
 
The picture seems to be slightly more complex. FSRQs stand as a monolithic
population for which there is no correlation between photon index and luminosity
\citep{fsrq12}. This is likely due to the fact that at GeV energies
their spectrum is dominated by the external Compton emission \citep{dermer93}.
On the other hand, the index-luminosity correlation for BL Lacs 
as argued above has a significant intrinsic component.
This points towards the fact that particles in
luminous BL Lacs cool more efficiently than in low-luminosity objects, 
in agreement with the results of \cite{finke13}
who finds an anti-correlation between the Compton dominance and the 
frequency at the peak of the synchrotron component.

The correlation between index and luminosity reported by \cite{ghisellini09b}
is much stronger than that which we find here
($\Delta\Gamma$/$\Delta {\rm Log} L_{\gamma}\sim$0.25 versus $\sim$0.06); this
was thus likely dominated by the uncorrected selection effects.
It does not seem to be the case that luminous and hard BL Lacs
exist in such numbers to destroy the correlation as suggested by \cite{giommi13},
although a few such objects are indeed seen in our sample.
Hard luminous BL Lacs exist (see Fig.~\ref{fig:index}), 
as predicted by the {\it Fermi} best-fit
luminosity function, but they are rare, representing the tail of
the $dN/dL_{\gamma}d\Gamma$ distribution.

%%%%%%%%%%%%%%%%%%%%%%%%%%%%%%%%%%%%%%%%%%%%%%%%%%%%%%%%%%%%%%%%%%
\subsection{The Contribution to the Isotropic Gamma-Ray Background}
\label{sec:egb}

This analysis has important consequences for the understanding of the 
isotropic gamma-ray background  \cite[IGRB,][]{fichtel95,sreekumar98,lat_edb} 
whose origin is still unclear \citep{pop_pap,fsrq12}.

A simple integration of the luminosity function yields  the
diffuse emission arising from 
the {\em unresolved} BL Lac class (in the 0.1--100\,GeV band)
as 8.0$^{+2.0}_{-1.3}\times10^{-7}$\,ph cm$^{-2}$ s$^{-1}$ sr$^{-1}$,
which represents 7.7$^{+2.0}_{-1.3}$\,\% of the intensity measured by {\it Fermi}. 
The slightly disfavored PLE model predicts that BL Lacs produce
1.07$^{+0.21}_{-0.17}\times10^{-6}$\,ph cm$^{-2}$ s$^{-1}$ sr$^{-1}$
and thus 10.3$\pm2$\,\% of the IGRB. It thus seems clear that BL Lacs
do not account for more than $\sim$10--15\,\% of the IGRB.

While this might seem to represent a small number, the large density
of hard sources present in the nearby Universe, as predicted by the luminosity
function makes the spectrum of the diffuse emission arising from the BL Lac
class harder than that of the IGRB. Since this depends on the assumed
spectral models for different BL Lac classes and on the EBL model,
the actual contribution from the common extreme HSP sources may be larger.
The exact energy-dependent derivation is left to a future publication.

%%%%%%%%%%%%%%%%%%%%%%%%%%%%%%%%%%%%%%%%%%%%%%%%%%%%%%%%%%%%%%%%%%
\section{Summary}
\label{sec:summary}
{ 
This work relies on a complete sample of 211 BL Lacs, detected by {\it Fermi}
during its first year of operations, to deepen our knowledge of this elusive,
yet very important, blazar population. 
Our findings can be summarized as follows:
\begin{itemize}

\item The typical redshift completeness of any BL Lac sample is $<$50\,\%.
The {\it Fermi} sample is no exception with only 103 BL Lacs (out of 211)
having a spectroscopic redshift measurement\footnote{A similar fraction
holds as well for the 2LAC sample of 423 BL Lacs \citep{2LAC}.}. Using four different techniques
(described in Sec.~\ref{sec:z}) we were able to provide quantitative constraints
on the redshift of an additional 104 objects making this the largest and most
complete sample of BL Lacs available in the literature. We find that most
of the objects without a spectroscopic redshift (and thus $\sim$half of
the BL Lac population) lie at z$>$0.5--0.7 which
is larger than the typical spectroscopic limit reached for BL Lacs.

\item Independently of the functional form used to represent the data, we find
that the BL Lac population displays (as found for other classes of AGN) a speed of evolution which depends on luminosity, with high-luminosity sources evolving faster than low-luminosity ones. The negative evolution (i.e. more BL Lacs at lower than
higher redshifts) of the low-luminosity BL Lacs is a major result of this work. We find that HSPs are certainly responsible for most, if not all, of the  detected negative evolution. This confirms previous claims of negative evolution based on samples of X-ray selected  BL Lacs, which contained a large fraction
of HSPs \citep{rector00,beckmann03}.

\item This work allows us to explore the link between the BL Lac and the FSRQ families of blazars. 
The local (z$\approx$0) luminosity function of BL Lac overlaps and 
connects smoothly
to that of FSRQs, highlighting the similarity between the two classes with
BL Lacs having on average lower luminosity (and thus very likely lower
Lorentz factors) than FSRQs. This last aspect is confirmed by the analysis
of the intrinsic luminosity function which allows us, using a simple beaming
model, to derive the distributions of Lorentz factors and of viewing angles. 
FSRQs and BL Lacs have a similar distribution of Lorentz factors
(i.e. a power-law  distribution with index $\approx-$2.5), but the
one of BL Lacs extends to slower jet speeds implying that the jets of 
BL Lacs are, on average, seen under larger angles than those of FSRQs 
($\sim$5\,$^{\circ}$ for BL Lacs versus $\sim$2\,$^{\circ}$ for FSRQs).

\item One of the most interesting finding of this work is the evidence
 supporting the  {\it genetic} link between FSRQs and BL Lacs as proposed
by \cite{cavaliere02} and \cite{bottcher02}. In this scenario BL Lacs
represent the final (gas starved, inefficiently accreting) 
and long-lasting phase of an earlier, short-lived, merger-driven gas-rich epoch (the FSRQ).
The sudden increase in the space density of BL Lacs (driven in particular by the HSPs) at the same epoch as the turn off of FSRQs corroborates the idea of a 
transition from the FSRQ to the BL Lac class. To investigate further the
details of this transition would require, for both classes,
 a robust beaming correction and knowledge of the black hole mass and host galaxy environment, which are at present not well constrained.

\item The study of the luminosity function shows that 
the spectra of BL Lacs at GeV energies soften with increasing luminosity
even after correcting for the substantial selection effects. The effect
is not as dramatic as reported in the literature \citep[e.g.][]{ghisellini09b},
but might still be caused by the fact that particles in luminous BL Lacs
cool more efficiently than in low-luminosity objects.

\item Unresolved BL Lacs contribute $\sim$10--15\,\% of the IGRB measured 
by {\it Fermi} \citep{lat_edb}. However, the large density of hard sources
at low redshift, as implied from the luminosity function derived in this work,
will certainly increase the contribution of BL Lacs to the IGRB at $>10$\,GeV. 
A confirmation of this is already available in the study of the $>10$\,GeV
sources detected by {\it Fermi} \citep{1FHL}.

\end{itemize}

}

%%%%%%%%%%%%%%%%%%%%%%%%%%%%%%%%%%%%%%%%%%%%%%%%%%%%%%%%%%%%%%%%%%%%%%%%%%
%%%%%%%%%%%%%%%%%%%%%%%%%%%%%%%%%%%%%%%%%%%%%%%%%%%%%%%%%%%%%%%%%%%%%%%%%%
\clearpage
\acknowledgments
The comments from an expert referee are gratefully acknowledged.
MA acknowledges generous support from  NASA grant NNH09ZDA001N for the study
of the origin of the extragalactic gamma-ray background
and hospitality at Goddard and NRL 
while writing part of this work. RWR acknowledges NASA grants NNX08AW30G and
NNX11A044G and extensive consultation with the OVRO Fermi group. 
D.G. acknowledges financial contribution from the agreement ASI-INAF I/009/10/0.
MA is in debt to J.~Wall for originally sharing the details of the ML method used in this and other works.

The \textit{Fermi} LAT Collaboration acknowledges generous ongoing support
from a number of agencies and institutes that have supported both the
development and the operation of the LAT as well as scientific data analysis.
These include the National Aeronautics and Space Administration and the
Department of Energy in the United States, the Commissariat \`a l'Energie Atomique
and the Centre National de la Recherche Scientifique / Institut National de Physique Nucl\'eaire et de Physique des Particules in France, the Agenzia 
Spaziale Italiana and the Istituto Nazionale di Fisica Nucleare in Italy, 
the Ministry of Education, Culture, Sports, Science and Technology (MEXT), 
High Energy Accelerator Research Organization (KEK) and Japan Aerospace 
Exploration Agency (JAXA) in Japan, and the K.~A.~Wallenberg Foundation, 
the Swedish Research Council and the Swedish National Space Board in Sweden.
Additional support for science analysis during the operations phase 
is gratefully acknowledged from the Istituto Nazionale di Astrofisica in 
Italy and the Centre National d'\'Etudes Spatiales in France.

{\it Facilities:} \facility{Fermi/LAT}

%%%%%%%%%%%%%%%%%%%%%%%%%%%%%%%%%%%%%%%%%%%%%%%%%%%%%%%%%%%%%%%%%%%%%%%%%%
%%%%%%%%%%%%%%%%%%%%%%%%%%%%%%%%%%%%%%%%%%%%%%%%%%%%%%%%%%%%%%%%%%%%%%%%%%
%%
%%
%% APPENDIX
%%
%%
%%%%%%%%%%%%%%%%%%%%%%%%%%%%%%%%%%%%%%%%%%%%%%%%%%%%%%%%%%%%%%%%%%%%%%%%%%%
%%%%%%%%%%%%%%%%%%%%%%%%%%%%%%%%%%%%%%%%%%%%%%%%%%%%%%%%%%%%%%%%%%%%%%%%%%%

\appendix
\section{Appendix}
\subsection{Table with Redshift Constraints}
\label{sec:table}
Table~\ref{tab:cat} reports the 211 BL Lacs used in this work
with all the  available redshift constraints (with the 
exception of the exclusion functions).
The sample and the nature of the redshift constraints are described
in $\S$~\ref{sec:sample} and $\S$~\ref{sec:z}.

%%%%%%%%%%%%%%%%%%%%%%%%%%%%%%%%%%%%%%% tables
\begin{deluxetable}{lcccccccc}
\tablewidth{0pt}
\tabletypesize{\scriptsize}
%\rotate
\tablecaption{The 211 BL Lac Objects detected by {\it Fermi} used for this analysis. The nature of the redshift constraints is described in $\S$~\ref{sec:z}.
\label{tab:cat}}
\tablehead{
%%%%%%%% column names
\colhead{NAME}       & \colhead{Flux$_{100}$\tablenotemark{a}}       &
\colhead{Photon Index}     & \colhead{z\tablenotemark{b}}                  &
\colhead{photo-z\tablenotemark{c}}          & \colhead{z$_{LL}$\tablenotemark{d}}            &
\colhead{z$_{MAX}$\tablenotemark{e}}        & \colhead{photo-z$_{UL}$\tablenotemark{f} }      &
\colhead{SED CLASS\tablenotemark{g}} \\
%
%%%%%%%%%  units
%
\colhead{}                     & \colhead{}  & 
%\colhead{}\scriptsize(10$^{-8}$ ph cm$^{-2}$ s$^{-1}$)}    & 
\colhead{}  &   
\colhead{}                        & \colhead{}                            &
\colhead{}                        & \colhead{} &
\colhead{}                        & \colhead{} 
}
\startdata

1FGL J0006.9+4652  & 4.35$\pm0.79$ & 2.50$\pm0.12$  & \nodata & \nodata  & \nodata & \nodata & \nodata & ISP \\ 
1FGL J0021.7-2556  & 1.22$\pm0.41$ & 2.13$\pm0.16$  & \nodata & \nodata  & 0.56 & 1.63 & 1.44 & ISP \\ 
1FGL J0022.2-1850  & 0.45$\pm0.16$ & 1.64$\pm0.13$  & \nodata & \nodata  & 0.77 & 1.64 & 1.38 & HSP \\ 
1FGL J0033.5-1921  & 2.03$\pm0.32$ & 1.89$\pm0.07$  & \nodata & \nodata  & 0.50\tablenotemark{h} & 1.77 & \nodata & HSP \\ 
1FGL J0035.1+1516  & 0.70$\pm0.23$ & 1.72$\pm0.12$  & \nodata & 1.28  & \nodata & 1.65 & \nodata & HSP \\ 
1FGL J0038.0+1236  & 1.41$\pm0.59$ & 2.23$\pm0.19$  & 0.089 & \nodata  & \nodata & 1.76 & \nodata & HSP \\ 
1FGL J0045.3+2127  & 1.39$\pm0.38$ & 1.86$\pm0.11$  & \nodata & \nodata  & \nodata & 1.78 & 1.06 & HSP \\ 
1FGL J0050.6-0928  & 7.39$\pm0.64$ & 2.20$\pm0.05$  & 0.635 & \nodata  & \nodata & 2.18 & \nodata & ISP \\ 
1FGL J0100.2+0747  & 1.90$\pm0.36$ & 1.90$\pm0.09$  & \nodata & \nodata  & \nodata & 4.01 & \nodata & ISP/HSP \\ 
1FGL J0105.7+3930  & 4.80$\pm0.90$ & 2.70$\pm0.14$  & 0.440 & \nodata  & \nodata & 2.68 & \nodata & \nodata \\ 
1FGL J0109.0+1816  & 0.76$\pm0.35$ & 2.00$\pm0.19$  & 0.443 & \nodata  & \nodata & 2.48 & \nodata & HSP \\ 
1FGL J0114.4+1327  & 3.81$\pm0.76$ & 2.66$\pm0.14$  & \nodata & \nodata  & \nodata & 1.63 & 1.22 & ISP/HSP \\ 
1FGL J0115.5+2519  & 1.49$\pm0.42$ & 2.02$\pm0.12$  & \nodata & \nodata  & 0.27 & 1.63 & 1.45 & HSP \\ 
1FGL J0115.7+0357  & 1.19$\pm0.40$ & 2.03$\pm0.15$  & 0.913 & \nodata  & 0.14 & 1.62 & 1.25 & \nodata \\ 
1FGL J0120.5-2700  & 3.90$\pm0.45$ & 2.03$\pm0.06$  & \nodata & \nodata  & 0.56 & 1.76 & \nodata & ISP \\ 
1FGL J0136.5+3905  & 2.86$\pm0.39$ & 1.80$\pm0.06$  & \nodata & \nodata  & \nodata & 1.65 & \nodata & HSP \\ 
1FGL J0141.7-0929  & 1.85$\pm0.45$ & 2.16$\pm0.12$  & 0.735 & \nodata  & 0.50 & 2.17 & \nodata & ISP \\ 
1FGL J0144.6+2703  & 4.77$\pm0.65$ & 2.22$\pm0.08$  & \nodata & \nodata  & 0.71 & 1.66 & \nodata & ISP \\ 
1FGL J0154.1+0823  & 1.68$\pm0.39$ & 1.97$\pm0.10$  & 0.681 & \nodata  & 0.34 & 1.64 & 1.37 & ISP \\ 
1FGL J0155.0+4433  & 0.95$\pm0.55$ & 2.10$\pm0.23$  & \nodata & \nodata  & 0.39 & 1.63 & \nodata & ISP/HSP \\ 
1FGL J0158.0-3931  & 2.19$\pm0.52$ & 2.34$\pm0.14$  & \nodata & \nodata  & \nodata & 2.15 & 1.35 & ISP \\ 
1FGL J0159.5+1047  & 1.28$\pm0.19$ & 1.95$\pm0.06$  & 0.195 & \nodata  & \nodata & 1.76 & \nodata & HSP \\ 
1FGL J0159.7-2741  & 0.98$\pm0.30$ & 2.06$\pm0.14$  & \nodata & \nodata  & 0.58 & 1.78 & 1.05 & ISP \\ 
1FGL J0203.5+3044  & 4.91$\pm0.85$ & 2.74$\pm0.13$  & 0.761 & \nodata  & \nodata & 2.72 & \nodata & \nodata \\ 
1FGL J0209.3-5229  & 1.80$\pm0.48$ & 1.94$\pm0.11$  & \nodata & \nodata  & \nodata & 2.18 & 1.18 & HSP \\ 
1FGL J0210.6-5101  & 14.63$\pm0.95$ & 2.37$\pm0.04$  & 0.999 & \nodata  & \nodata & \nodata & \nodata & ISP \\ 
1FGL J0211.2+1049  & 3.86$\pm0.69$ & 2.27$\pm0.09$  & \nodata & \nodata  & \nodata & 1.67 & \nodata & ISP \\ 
1FGL J0213.2+2244  & 1.29$\pm0.40$ & 1.95$\pm0.13$  & 0.459 & \nodata  & \nodata & 2.66 & \nodata & HSP \\ 
1FGL J0217.9-6630  & 1.19$\pm0.46$ & 2.07$\pm0.17$  & \nodata & \nodata  & 0.67 & 1.88 & 1.25 & HSP \\ 
1FGL J0222.6+4302  & 21.51$\pm1.03$ & 1.94$\pm0.02$  & \nodata & \nodata  & \nodata & 1.67 & \nodata & ISP \\ 
1FGL J0238.6+1637  & 43.54$\pm1.10$ & 2.15$\pm0.02$  & 0.940 & \nodata  & \nodata & \nodata & \nodata & ISP \\ 
1FGL J0238.6-3117  & 0.96$\pm0.35$ & 2.07$\pm0.17$  & 0.232 & \nodata  & \nodata & 1.63 & 1.02 & HSP \\ 
1FGL J0250.4+1715  & 1.28$\pm0.40$ & 2.13$\pm0.13$  & 0.612 & \nodata  & \nodata & \nodata & 3.10 & \nodata \\ 
1FGL J0303.5-2406  & 4.71$\pm0.41$ & 2.00$\pm0.05$  & 0.260 & \nodata  & \nodata & \nodata & \nodata & HSP \\ 
1FGL J0315.9-2609  & 0.32$\pm0.16$ & 1.62$\pm0.17$  & 0.443 & \nodata  & \nodata & \nodata & \nodata & HSP \\ 
1FGL J0316.1+0904  & 1.73$\pm0.42$ & 1.78$\pm0.09$  & \nodata & \nodata  & \nodata & 1.66 & \nodata & HSP \\ 
1FGL J0319.7+1847  & 0.51$\pm0.23$ & 1.65$\pm0.16$  & 0.190 & \nodata  & \nodata & \nodata & \nodata & HSP \\ 
1FGL J0322.1+2336  & 4.26$\pm0.92$ & 2.41$\pm0.12$  & \nodata & \nodata  & \nodata & \nodata & \nodata & HSP \\ 
1FGL J0323.7-0106  & 0.34$\pm0.16$ & 1.59$\pm0.17$  & 0.392 & \nodata  & \nodata & 2.17 & 1.54 & HSP \\ 
1FGL J0326.2+0222  & 1.94$\pm0.53$ & 2.21$\pm0.13$  & 0.147 & \nodata  & \nodata & \nodata & \nodata & HSP \\ 
1FGL J0334.2-4010  & 7.85$\pm0.10$ & 2.34$\pm0.01$  & 1.357 & \nodata  & 1.21 & 2.05 & \nodata & ISP \\ 
1FGL J0334.4-3727  & 2.59$\pm0.18$ & 2.09$\pm0.04$  & \nodata & \nodata  & \nodata & 1.92 & 1.34 & ISP \\ 
1FGL J0354.6+8009  & 7.76$\pm0.90$ & 2.58$\pm0.08$  & \nodata & \nodata  & \nodata & \nodata & \nodata & ISP \\ 
1FGL J0416.8+0107  & 0.80$\pm0.50$ & 1.96$\pm0.24$  & 0.287 & \nodata  & \nodata & \nodata & \nodata & HSP \\ 
1FGL J0428.6-3756  & 31.07$\pm0.89$ & 2.13$\pm0.02$  & 1.111 & \nodata  & \nodata & \nodata & \nodata & ISP \\ 
1FGL J0434.1-2018  & 1.67$\pm0.46$ & 2.31$\pm0.15$  & 0.928 & \nodata  & \nodata & 2.43 & \nodata & ISP \\ 
1FGL J0448.5-1633  & 1.05$\pm0.37$ & 1.97$\pm0.15$  & \nodata & \nodata  & \nodata & 1.63 & 1.25 & HSP \\ 
1FGL J0449.5-4350  & 10.40$\pm0.55$ & 1.99$\pm0.03$  & 0.205 & \nodata  & \nodata & \nodata & \nodata & HSP \\ 
1FGL J0507.9+6738  & 1.37$\pm0.20$ & 1.73$\pm0.06$  & 0.340 & \nodata  & \nodata & 2.51 & \nodata & HSP \\ 
1FGL J0509.3+0540  & 8.18$\pm0.87$ & 2.31$\pm0.06$  & 0.336 & \nodata  & \nodata & 1.66 & 1.24 & ISP \\ 
1FGL J0516.7-6207  & 5.56$\pm0.01$ & 2.28$\pm0.00$  & 1.300 & \nodata  & \nodata & 1.87 & \nodata & ISP \\ 
1FGL J0536.2-3348  & 5.54$\pm0.70$ & 2.37$\pm0.08$  & \nodata & \nodata  & \nodata & 2.17 & 1.16 & HSP \\ 
1FGL J0538.8-4404  & 38.22$\pm1.08$ & 2.28$\pm0.02$  & 0.892 & \nodata  & \nodata & \nodata & \nodata & ISP \\ 
1FGL J0543.8-5531  & 0.97$\pm0.31$ & 1.75$\pm0.12$  & 0.271\tablenotemark{h} & \nodata  & \nodata & 2.57 & 1.08 & HSP \\ 
1FGL J0616.9+5701  & 1.63$\pm0.48$ & 2.06$\pm0.13$  & \nodata & \nodata  & 0.80 & 3.94 & \nodata & ISP \\ 
1FGL J0617.7-1718  & 1.14$\pm0.45$ & 1.98$\pm0.15$  & 0.098 & \nodata  & \nodata & 1.75 & \nodata & ISP \\ 
1FGL J0700.4-6611  & 5.61$\pm0.01$ & 2.13$\pm0.00$  & \nodata & \nodata  & \nodata & 1.92 & 1.46 & ISP \\ 
1FGL J0706.5+3744  & 1.82$\pm0.55$ & 2.19$\pm0.14$  & \nodata & \nodata  & \nodata & 1.63 & \nodata & HSP \\ 
1FGL J0707.3+7742  & 2.48$\pm0.29$ & 2.28$\pm0.06$  & \nodata & \nodata  & \nodata & 1.76 & \nodata & ISP \\ 
1FGL J0710.6+5911  & 0.24$\pm0.12$ & 1.50$\pm0.18$  & 0.125 & \nodata  & \nodata & \nodata & \nodata & HSP \\ 
1FGL J0711.4+4731  & 2.91$\pm0.68$ & 2.52$\pm0.14$  & 1.292 & \nodata  & \nodata & \nodata & \nodata & ISP \\ 
1FGL J0712.7+5033  & 2.86$\pm0.47$ & 2.07$\pm0.08$  & 0.502 & \nodata  & \nodata & 1.67 & \nodata & ISP \\ 
1FGL J0721.9+7120  & 17.39$\pm0.80$ & 2.15$\pm0.03$  & \nodata & \nodata  & \nodata & 2.61 & \nodata & ISP \\ 
1FGL J0738.2+1741  & 5.08$\pm0.52$ & 2.06$\pm0.05$  & \nodata & \nodata  & 0.42 & 1.80 & 1.30\tablenotemark{i} & HSP \\ 
1FGL J0752.8+5353  & 0.88$\pm0.34$ & 1.95$\pm0.16$  & 0.730 & \nodata  & \nodata & 1.94 & \nodata & ISP \\ 
1FGL J0757.2+0956  & 5.30$\pm0.69$ & 2.44$\pm0.08$  & 0.266 & \nodata  & \nodata & \nodata & \nodata & ISP \\ 
1FGL J0804.7+7534  & 0.71$\pm0.30$ & 1.79$\pm0.15$  & 0.121 & \nodata  & \nodata & \nodata & \nodata & HSP \\ 
1FGL J0809.5+5219  & 2.07$\pm0.48$ & 1.99$\pm0.11$  & 0.137 & \nodata  & \nodata & 2.22 & \nodata & HSP \\ 
1FGL J0811.1-7527  & 1.39$\pm0.39$ & 1.86$\pm0.11$  & \nodata & \nodata  & 0.69 & 1.91 & 1.40 & ISP \\ 
1FGL J0811.2+0148  & 3.58$\pm0.70$ & 2.56$\pm0.13$  & 1.148 & \nodata  & \nodata & \nodata & \nodata & ISP \\ 
1FGL J0815.0+6434  & 3.10$\pm0.63$ & 2.31$\pm0.11$  & 0.239 & \nodata  & \nodata & 1.64 & \nodata & ISP \\ 
1FGL J0818.2+4222  & 12.19$\pm0.71$ & 2.17$\pm0.04$  & \nodata & \nodata  & \nodata & 2.47 & \nodata & ISP \\ 
1FGL J0825.9+0309  & 0.51$\pm0.28$ & 1.88$\pm0.21$  & 0.505 & \nodata  & \nodata & 3.21 & \nodata & ISP \\ 
1FGL J0831.6+0429  & 7.18$\pm0.76$ & 2.49$\pm0.07$  & 0.174 & \nodata  & \nodata & 2.19 & \nodata & ISP \\ 
1FGL J0844.0+5314  & 0.51$\pm0.23$ & 1.90$\pm0.18$  & \nodata & \nodata  & \nodata & 2.51 & \nodata & ISP \\ 
1FGL J0847.2+1134  & 0.23$\pm0.10$ & 1.49$\pm0.16$  & 0.198 & \nodata  & \nodata & 2.17 & \nodata & HSP \\ 
1FGL J0854.8+2006  & 5.37$\pm0.55$ & 2.20$\pm0.06$  & 0.306 & \nodata  & \nodata & \nodata & \nodata & ISP \\ 
1FGL J0856.6-1105  & 5.70$\pm0.71$ & 2.34$\pm0.07$  & \nodata & \nodata  & 1.40 & 2.18 & 1.54 & ISP \\ 
1FGL J0902.4+2050  & 1.65$\pm0.44$ & 2.11$\pm0.13$  & \nodata & \nodata  & \nodata & 2.18 & 1.21 & ISP \\ 
1FGL J0905.5+1356  & 0.90$\pm0.35$ & 1.94$\pm0.16$  & \nodata & \nodata  & \nodata & 1.64 & 1.35 & HSP \\ 
1FGL J0910.7+3332  & 1.69$\pm0.48$ & 2.26$\pm0.14$  & 0.354 & \nodata  & \nodata & 1.77 & \nodata & HSP \\ 
1FGL J0915.7+2931  & 1.67$\pm0.11$ & 1.95$\pm0.03$  & \nodata & \nodata  & \nodata & 1.69 & \nodata & HSP \\ 
1FGL J0945.6+5754  & 1.50$\pm0.46$ & 2.21$\pm0.15$  & 0.229 & \nodata  & \nodata & 2.17 & \nodata & ISP/HSP \\ 
1FGL J0953.0-0838  & 2.22$\pm0.40$ & 1.93$\pm0.08$  & \nodata & \nodata  & \nodata & 1.64 & 1.28 & HSP \\ 
1FGL J1000.9+2915  & 1.95$\pm0.43$ & 2.14$\pm0.11$  & 0.558 & \nodata  & \nodata & \nodata & \nodata & ISP \\ 
1FGL J1007.9+0619  & 3.02$\pm0.70$ & 2.38$\pm0.12$  & \nodata & \nodata  & \nodata & 2.17 & 1.44 & ISP \\ 
1FGL J1012.2+0634  & 1.61$\pm0.76$ & 2.31$\pm0.21$  & 0.727 & \nodata  & 0.52 & 2.16 & \nodata & ISP \\ 
1FGL J1015.1+4927  & 6.44$\pm0.48$ & 1.92$\pm0.04$  & 0.212 & \nodata  & \nodata & \nodata & \nodata & HSP \\ 
1FGL J1031.0+5051  & 0.57$\pm0.23$ & 1.78$\pm0.16$  & \nodata & \nodata  & \nodata & 2.17 & \nodata & HSP \\ 
1FGL J1032.7+3737  & 1.38$\pm0.42$ & 2.27$\pm0.16$  & \nodata & \nodata  & 0.53 & 2.17 & \nodata & ISP \\ 
1FGL J1037.7+5711  & 3.22$\pm0.47$ & 2.03$\pm0.07$  & \nodata & \nodata  & \nodata & 1.64 & \nodata & ISP \\ 
1FGL J1053.6+4927  & 0.41$\pm0.14$ & 1.56$\pm0.13$  & 0.140 & \nodata  & \nodata & 2.17 & \nodata & HSP \\ 
1FGL J1054.5+2212  & 3.67$\pm0.13$ & 2.32$\pm0.02$  & \nodata & \nodata  & \nodata & 1.64 & 1.36 & ISP \\ 
1FGL J1058.1-8006  & 7.50$\pm0.43$ & 2.56$\pm0.02$  & 0.581 & \nodata  & \nodata & \nodata & \nodata & ISP \\ 
1FGL J1058.4+0134  & 13.88$\pm0.07$ & 2.32$\pm0.00$  & 0.888 & \nodata  & \nodata & \nodata & \nodata & ISP \\ 
1FGL J1058.6+5628  & 5.62$\pm0.53$ & 2.01$\pm0.05$  & 0.143 & \nodata  & \nodata & 2.18 & \nodata & HSP \\ 
1FGL J1059.3-1132  & 4.37$\pm0.02$ & 2.23$\pm0.00$  & \nodata & \nodata  & \nodata & 1.65 & \nodata & ISP \\ 
1FGL J1104.4+0734  & 2.17$\pm0.56$ & 2.30$\pm0.13$  & \nodata & \nodata  & \nodata & 1.65 & \nodata & ISP/HSP \\ 
1FGL J1104.4+3812  & 17.09$\pm0.57$ & 1.81$\pm0.02$  & 0.031 & \nodata  & \nodata & \nodata & \nodata & HSP \\ 
1FGL J1107.8+1502  & 0.86$\pm0.04$ & 2.01$\pm0.02$  & \nodata & \nodata  & 0.60 & 2.16 & \nodata & HSP \\ 
1FGL J1117.1+2013  & 1.36$\pm0.27$ & 1.77$\pm0.08$  & 0.138 & \nodata  & \nodata & 2.17 & \nodata & HSP \\ 
1FGL J1121.0+4209  & 0.39$\pm0.16$ & 1.64$\pm0.15$  & 0.124 & \nodata  & \nodata & 2.17 & \nodata & HSP \\ 
1FGL J1133.1+0033  & 2.66$\pm0.52$ & 2.15$\pm0.10$  & 0.678 & \nodata  & \nodata & 1.86 & \nodata & ISP \\ 
1FGL J1136.6+7009  & 1.14$\pm0.27$ & 1.87$\pm0.10$  & 0.046 & \nodata  & \nodata & \nodata & \nodata & HSP \\ 
1FGL J1150.2+2419  & 2.10$\pm0.50$ & 2.28$\pm0.13$  & \nodata & \nodata  & \nodata & 2.21 & \nodata & ISP \\ 
1FGL J1150.5+4152  & 1.76$\pm0.37$ & 1.93$\pm0.10$  & \nodata & \nodata  & 0.85 & 1.66 & \nodata & HSP \\ 
1FGL J1151.6+5857  & 1.31$\pm0.56$ & 2.23$\pm0.19$  & \nodata & \nodata  & \nodata & 1.76 & \nodata & ISP \\ 
1FGL J1154.0-0008  & 0.38$\pm0.34$ & 1.72$\pm0.33$  & 0.254 & \nodata  & \nodata & 2.22 & \nodata & HSP \\ 
1FGL J1202.9+6032  & 2.71$\pm0.75$ & 2.44$\pm0.16$  & 0.065 & \nodata  & \nodata & 2.18 & \nodata & ISP \\ 
1FGL J1204.4+1139  & 1.33$\pm0.46$ & 2.23$\pm0.17$  & 0.296 & \nodata  & \nodata & 2.17 & \nodata & HSP \\ 
1FGL J1217.7+3007  & 5.86$\pm0.58$ & 1.98$\pm0.05$  & 0.130 & \nodata  & \nodata & \nodata & \nodata & HSP \\ 
1FGL J1218.4-0128  & 1.11$\pm0.31$ & 1.96$\pm0.12$  & \nodata & \nodata  & 0.64 & 1.64 & 1.23 & ISP \\ 
1FGL J1221.3+3008  & 2.02$\pm0.36$ & 1.76$\pm0.07$  & 0.184 & \nodata  & \nodata & 2.18 & \nodata & HSP \\ 
1FGL J1221.5+2814  & 8.12$\pm0.64$ & 2.09$\pm0.04$  & 0.103 & \nodata  & \nodata & 2.22 & \nodata & ISP \\ 
1FGL J1226.7-1332  & 0.60$\pm0.21$ & 1.74$\pm0.13$  & \nodata & \nodata  & \nodata & 1.76 & 1.30\tablenotemark{i} & ISP \\ 
1FGL J1230.4+2520  & 1.21$\pm0.35$ & 2.07$\pm0.13$  & 0.135 & \nodata  & \nodata & 1.78 & \nodata & ISP \\ 
1FGL J1231.6+2850  & 2.46$\pm0.36$ & 1.94$\pm0.07$  & 0.236 & \nodata  & \nodata & 2.18 & \nodata & HSP \\ 
1FGL J1243.1+3627  & 1.25$\pm0.28$ & 1.79$\pm0.09$  & \nodata & \nodata  & 0.48 & 1.77 & \nodata & HSP \\ 
1FGL J1248.2+5820  & 6.35$\pm0.61$ & 2.17$\pm0.06$  & \nodata & \nodata  & \nodata & 1.64 & \nodata & ISP \\ 
1FGL J1249.8+3706  & 0.54$\pm0.21$ & 1.80$\pm0.15$  & \nodata & \nodata  & \nodata & 2.19 & \nodata & HSP \\ 
1FGL J1253.0+5301  & 3.74$\pm0.51$ & 2.13$\pm0.08$  & \nodata & \nodata  & 0.66 & 1.64 & \nodata & ISP \\ 
1FGL J1303.0+2433  & 4.88$\pm0.52$ & 2.17$\pm0.06$  & \nodata & \nodata  & 0.77 & 1.69 & \nodata & ISP \\ 
1FGL J1304.3-4352  & 3.85$\pm0.64$ & 2.06$\pm0.07$  & \nodata & \nodata & \nodata & 2.12 & 1.30\tablenotemark{i} & HSP \\ 
1FGL J1309.5+4304  & 1.45$\pm0.32$ & 1.94$\pm0.10$  & 0.691 & \nodata  & 0.69 & 1.80 & \nodata & HSP \\ 
1FGL J1314.7+2346  & 1.76$\pm0.40$ & 2.10$\pm0.11$  & \nodata & \nodata  & \nodata & 4.68 & 1.30 & ISP \\ 
1FGL J1338.9+1153  & 1.17$\pm0.05$ & 2.08$\pm0.02$  & \nodata &  1.61$^{+0.04}_{-0.10}$\tablenotemark{i} & 1.59 & 1.94 & \nodata & ISP \\ 
1FGL J1351.5+1115  & 0.17$\pm0.02$ & 1.49$\pm0.04$  & \nodata & \nodata  & 0.62 & 1.64 & 1.12 & HSP \\ 
1FGL J1418.3-0235  & 1.31$\pm0.33$ & 1.88$\pm0.10$  & \nodata & \nodata  & \nodata & 1.64 & 1.37 & HSP \\ 
1FGL J1421.0+5421  & 3.69$\pm0.88$ & 2.76$\pm0.17$  & 0.153 & \nodata  & \nodata & \nodata & \nodata & ISP \\ 
1FGL J1425.0+3614  & 0.78$\pm0.39$ & 2.05$\pm0.20$  & \nodata & \nodata  & \nodata & 2.17 & \nodata & ISP \\ 
1FGL J1426.9+2347  & 7.47$\pm0.49$ & 1.85$\pm0.03$  & \nodata & \nodata  & \nodata & 1.66 & 1.11 & HSP \\ 
1FGL J1428.7+4239  & 0.38$\pm0.17$ & 1.60$\pm0.16$  & 0.129 & \nodata  & \nodata & 2.18 & \nodata & HSP \\ 
1FGL J1437.0+5640  & 0.20$\pm0.12$ & 1.46$\pm0.21$  & \nodata & \nodata  & \nodata & 2.08 & \nodata & HSP \\ 
1FGL J1440.9+0613  & 5.66$\pm0.85$ & 2.63$\pm0.11$  & \nodata & \nodata  & 0.32 & 1.63 & 1.31 & ISP \\ 
1FGL J1442.8+1158  & 0.44$\pm0.26$ & 1.73$\pm0.23$  & 0.163 & \nodata  & \nodata & 2.17 & \nodata & HSP \\ 
1FGL J1444.0-3906  & 2.72$\pm0.45$ & 1.90$\pm0.07$  & \nodata & \nodata  & \nodata & 2.20 & \nodata & HSP \\ 
1FGL J1447.9+3608  & 1.60$\pm0.39$ & 1.99$\pm0.11$  & \nodata & \nodata  & 0.74 & 1.76 & \nodata & HSP \\ 
1FGL J1454.6+5125  & 2.58$\pm0.53$ & 2.30$\pm0.10$  & \nodata & \nodata  & \nodata & 1.63 & \nodata & ISP \\ 
1FGL J1501.1+2237  & 1.16$\pm0.26$ & 1.77$\pm0.09$  & 0.235 & \nodata  & \nodata & 2.18 & \nodata & HSP \\ 
1FGL J1503.5-1544  & 0.89$\pm0.45$ & 1.79$\pm0.19$  & \nodata & \nodata  & 0.21 & 1.76 & \nodata & HSP \\ 
1FGL J1505.1-3435  & 1.85$\pm0.73$ & 2.19$\pm0.17$  & \nodata & \nodata  & 1.55 & 3.13 & \nodata & ISP \\ 
1FGL J1517.8-2423  & 7.47$\pm0.83$ & 2.13$\pm0.06$  & 0.048 & \nodata  & \nodata & \nodata & \nodata & ISP \\ 
1FGL J1521.0-0350  & 1.67$\pm0.50$ & 2.04$\pm0.13$  & \nodata & \nodata  & 0.87 & 1.80 & \nodata & HSP \\ 
1FGL J1522.6-2732  & 5.94$\pm0.84$ & 2.30$\pm0.08$  & 1.294 & \nodata  & \nodata & \nodata & \nodata & ISP \\ 
1FGL J1542.9+6129  & 7.08$\pm0.62$ & 2.16$\pm0.05$  & \nodata & \nodata  & \nodata & 1.76 & \nodata & ISP \\ 
1FGL J1548.7-2250  & 2.36$\pm0.85$ & 2.19$\pm0.16$  & 0.192 & \nodata  & \nodata & 1.65 & \nodata & HSP \\ 
1FGL J1553.5-3116  & 0.50$\pm0.21$ & 1.71$\pm0.14$  & \nodata & \nodata  & \nodata & 1.97 & \nodata & HSP \\ 
1FGL J1555.7+1111  & 6.77$\pm0.45$ & 1.68$\pm0.03$  & \nodata & \nodata  & \nodata & 1.77 & 1.35 & HSP \\ 
1FGL J1558.9+5627  & 2.60$\pm0.75$ & 2.19$\pm0.14$  & 0.300 & \nodata  & 1.05 & 2.47 & \nodata & ISP \\ 
1FGL J1607.1+1552  & 4.62$\pm0.66$ & 2.32$\pm0.08$  & 0.496 & \nodata  & \nodata & \nodata & \nodata & ISP \\ 
1FGL J1643.5-0646  & 4.11$\pm0.86$ & 2.27$\pm0.10$  & 0.082 & \nodata  & \nodata & 2.07 & \nodata & HSP \\ 
1FGL J1649.6+5241  & 1.61$\pm0.48$ & 2.16$\pm0.14$  & \nodata & \nodata  & \nodata & 2.47 & \nodata & \nodata \\ 
1FGL J1653.9+3945  & 5.67$\pm0.45$ & 1.81$\pm0.04$  & 0.034 & \nodata  & \nodata & \nodata & \nodata & HSP \\ 
1FGL J1719.2+1745  & 4.33$\pm0.52$ & 2.02$\pm0.06$  & \nodata & \nodata  & \nodata & 1.64 & \nodata & ISP \\ 
1FGL J1725.0+1151  & 2.48$\pm0.50$ & 1.89$\pm0.09$  & \nodata & \nodata  & \nodata & 1.65 & \nodata & HSP \\ 
1FGL J1725.5+5854  & 1.31$\pm0.35$ & 2.03$\pm0.12$  & \nodata & \nodata  & \nodata & 1.66 & \nodata & ISP \\ 
1FGL J1727.9+5010  & 0.79$\pm0.33$ & 1.94$\pm0.17$  & 0.055 & \nodata  & \nodata & \nodata & \nodata & HSP \\ 
1FGL J1744.2+1934  & 0.74$\pm0.32$ & 1.83$\pm0.16$  & 0.083 & \nodata  & \nodata & \nodata & \nodata & HSP \\ 
1FGL J1748.5+7004  & 2.29$\pm0.20$ & 2.05$\pm0.04$  & 0.770 & \nodata  & \nodata & \nodata & \nodata & ISP \\ 
1FGL J1749.0+4323  & 2.39$\pm0.10$ & 2.09$\pm0.02$  & \nodata & \nodata  & 0.57 & 1.65 & \nodata & ISP \\ 
1FGL J1751.5+0937  & 11.15$\pm1.37$ & 2.32$\pm0.06$  & 0.322 & \nodata  & \nodata & \nodata & \nodata & ISP \\ 
1FGL J1754.3+3212  & 3.06$\pm0.53$ & 2.10$\pm0.09$  & \nodata & \nodata  & \nodata & 1.63 & \nodata & HSP \\ 
1FGL J1800.4+7827  & 6.11$\pm0.04$ & 2.35$\pm0.00$  & 0.684 & \nodata  & \nodata & \nodata & \nodata & ISP \\ 
1FGL J1807.0+6945  & 6.33$\pm0.89$ & 2.53$\pm0.09$  & 0.051 & \nodata  & \nodata & \nodata & \nodata & ISP \\ 
1FGL J1809.6+2908  & 1.20$\pm0.47$ & 2.07$\pm0.16$  & \nodata & \nodata  & \nodata & 1.63 & \nodata & ISP \\ 
1FGL J1811.0+1607  & 3.35$\pm0.69$ & 2.22$\pm0.10$  & \nodata & \nodata  & \nodata & 1.74 & \nodata & ISP \\ 
1FGL J1813.4+3141  & 2.77$\pm0.49$ & 2.11$\pm0.09$  & 0.117 & \nodata  & \nodata & \nodata & \nodata & ISP \\ 
1FGL J1824.0+5651  & 6.55$\pm0.73$ & 2.36$\pm0.07$  & 0.664 & \nodata  & \nodata & 2.48 & \nodata & ISP \\ 
1FGL J1829.8+5404  & 1.96$\pm0.71$ & 2.39$\pm0.19$  & \nodata & \nodata  & \nodata & 2.46 & \nodata & HSP \\ 
1FGL J1832.6-5700  & 2.40$\pm0.74$ & 2.22$\pm0.15$  & \nodata & \nodata  & 1.23 & 1.96 & \nodata & HSP \\ 
1FGL J1838.6+4756  & 1.09$\pm0.36$ & 1.92$\pm0.13$  & \nodata & \nodata  & \nodata & \nodata & \nodata & HSP \\ 
1FGL J1849.6-4314  & 2.04$\pm0.56$ & 2.17$\pm0.13$  & \nodata & \nodata  & \nodata & 1.94 & \nodata & ISP/HSP \\ 
1FGL J1903.0+5539  & 2.93$\pm0.46$ & 1.97$\pm0.07$  & \nodata & \nodata  & 0.73 & 1.63 & \nodata & ISP \\ 
1FGL J1918.4-4108  & 2.06$\pm0.42$ & 1.91$\pm0.09$  & \nodata & \nodata  & 1.59 & 2.11 & \nodata & ISP \\ 
1FGL J1926.8+6153  & 2.76$\pm0.55$ & 2.13$\pm0.10$  & \nodata & \nodata  & \nodata & 1.65 & \nodata & HSP \\ 
1FGL J1936.9-4720  & 0.73$\pm0.36$ & 1.82$\pm0.18$  & 0.265 & \nodata  & \nodata & \nodata & \nodata & HSP \\ 
1FGL J1958.4-3013  & 2.07$\pm0.82$ & 2.23$\pm0.17$  & 0.119 & \nodata  & \nodata & \nodata & \nodata & HSP \\ 
1FGL J2000.0+6508  & 7.22$\pm0.67$ & 2.05$\pm0.05$  & 0.049 & \nodata  & \nodata & \nodata & \nodata & HSP \\ 
1FGL J2006.0+7751  & 3.14$\pm0.91$ & 2.44$\pm0.16$  & 0.342 & \nodata  & \nodata & \nodata & \nodata & ISP \\ 
1FGL J2009.1+7228  & 4.32$\pm1.15$ & 2.58$\pm0.15$  & \nodata & \nodata  & 1.74 & 2.03 & \nodata & ISP \\ 
1FGL J2009.5-4849  & 3.87$\pm0.49$ & 1.88$\pm0.06$  & 0.071 & \nodata  & \nodata & \nodata & \nodata & HSP \\ 
1FGL J2015.3-0129  & 2.26$\pm0.62$ & 2.19$\pm0.13$  & \nodata & \nodata  & \nodata & 1.78 & 1.22 & ISP \\ 
1FGL J2016.2-0903  & 2.21$\pm0.01$ & 2.18$\pm0.00$  & \nodata & \nodata  & 0.60 & 1.63 & \nodata & ISP \\ 
1FGL J2031.5+1219  & 4.11$\pm0.04$ & 2.42$\pm0.01$  & 1.213 & \nodata  & 0.85 & \nodata & \nodata & ISP \\ 
1FGL J2039.0-1047  & 2.80$\pm0.12$ & 2.18$\pm0.02$  & \nodata & \nodata  & \nodata & 1.63 & \nodata & ISP \\ 
1FGL J2131.7-0914  & 0.88$\pm0.39$ & 1.97$\pm0.18$  & 0.449 & \nodata  & \nodata & \nodata & \nodata & HSP \\ 
1FGL J2139.3-4235  & 9.71$\pm0.69$ & 2.12$\pm0.04$  & \nodata & \nodata  & \nodata & 1.91 & \nodata & ISP \\ 
1FGL J2143.1-3927  & 1.34$\pm0.48$ & 2.07$\pm0.16$  & 0.429 & \nodata  & \nodata & 2.00 & \nodata & ISP/HSP \\ 
1FGL J2146.6-1345  & 1.09$\pm0.35$ & 1.85$\pm0.13$  & \nodata & \nodata  & \nodata & 1.64 & \nodata & HSP \\ 
1FGL J2149.7+0327  & 3.19$\pm0.82$ & 2.60$\pm0.16$  & \nodata & \nodata  & 0.72 & 1.62 & 1.42 & ISP \\ 
1FGL J2158.8-3013  & 21.73$\pm0.71$ & 1.91$\pm0.02$  & 0.116 & \nodata  & \nodata & \nodata & \nodata & HSP \\ 
1FGL J2223.3+0103  & 0.46$\pm0.26$ & 1.85$\pm0.21$  & \nodata & \nodata  & \nodata & 1.63 & \nodata & \nodata \\ 
1FGL J2236.2+2828  & 10.57$\pm0.79$ & 2.38$\pm0.05$  & 0.790 & \nodata  & \nodata & 1.64 & \nodata & ISP \\ 
1FGL J2236.4-1432  & 6.93$\pm0.71$ & 2.37$\pm0.07$  & \nodata & \nodata  & 0.61 & 2.53 & 1.55 & ISP \\ 
1FGL J2243.1-2541  & 2.90$\pm0.52$ & 2.27$\pm0.10$  & 0.774 & \nodata  & \nodata & \nodata & \nodata & ISP \\ 
1FGL J2244.0+2021  & 3.06$\pm0.43$ & 1.90$\pm0.07$  & \nodata & \nodata  & 0.40 & 1.64 & \nodata & HSP \\ 
1FGL J2247.3+0000  & 1.19$\pm0.37$ & 2.08$\pm0.14$  & 0.949 & \nodata  & \nodata & 1.85 & \nodata & ISP \\ 
1FGL J2250.1+3825  & 0.98$\pm0.27$ & 1.80$\pm0.10$  & 0.119 & \nodata  & \nodata & \nodata & \nodata & HSP \\ 
1FGL J2251.7+4030  & 3.98$\pm0.81$ & 2.45$\pm0.11$  & 0.229 & \nodata  & \nodata & 1.67 & \nodata & ISP \\ 
1FGL J2256.3-2009  & 0.73$\pm0.28$ & 1.95$\pm0.16$  & \nodata & \nodata  & \nodata & 1.93 & \nodata & ISP \\ 
1FGL J2307.3+1452  & 2.00$\pm0.55$ & 2.16$\pm0.13$  & \nodata & \nodata  & \nodata & 1.66 & \nodata & ISP \\ 
1FGL J2323.5+4211  & 2.00$\pm0.50$ & 1.97$\pm0.11$  & \nodata & \nodata  & 0.27 & 1.70 & \nodata & HSP \\ 
1FGL J2325.2+3957  & 3.32$\pm0.49$ & 2.03$\pm0.07$  & \nodata & \nodata  & 1.05 & 1.85 & \nodata & ISP \\ 
1FGL J2325.8-4043  & 2.44$\pm0.87$ & 2.22$\pm0.15$  & \nodata & \nodata  & \nodata & \nodata & \nodata & HSP \\ 
1FGL J2329.2+3755  & 0.50$\pm0.22$ & 1.66$\pm0.15$  & \nodata & \nodata  & \nodata & 1.76 & \nodata & HSP \\ 
1FGL J2334.7+1429  & 0.80$\pm0.05$ & 2.04$\pm0.02$  & \nodata & \nodata  & \nodata & 2.66 & 1.30\tablenotemark{i} & ISP \\ 
1FGL J2339.0+2123  & 0.23$\pm0.15$ & 1.57$\pm0.23$  & 0.291 & \nodata  & \nodata & \nodata & \nodata & HSP \\ 
1FGL J2341.6+8015  & 4.51$\pm0.72$ & 2.23$\pm0.08$  & 0.274 & \nodata  & \nodata & \nodata & \nodata & HSP \\ 
1FGL J2343.6+3437  & 0.31$\pm0.19$ & 1.68$\pm0.22$  & 0.366 & \nodata  & \nodata & \nodata & \nodata & HSP \\ 
1FGL J2352.1+1752  & 0.74$\pm0.29$ & 1.96$\pm0.16$  & \nodata & 1.45  & 0.65 & 1.63 & \nodata & HSP \\ 
1FGL J2359.0-3035  & 0.70$\pm0.27$ & 1.95$\pm0.16$  & 0.165 & \nodata  & \nodata & \nodata & \nodata & HSP \\

\enddata
\tablenotetext{a}{Flux in the 0.1--100\,GeV band in units of 10$^{-8}$ ph cm$^{-2}$ s$^{-1}$.}
\tablenotetext{b}{Spectroscopic redshift as reported in \cite{agn_cat}, \cite{2LAC},
\cite{shaw12} and \cite{shaw13}.}
\tablenotetext{c}{Photometric redshift estimates from \cite{rau12}.}
\tablenotetext{d}{Spectroscopic redshift lower limits from \cite{shaw13} and \cite{shaw13b}.}
\tablenotetext{e}{Spectroscopic redshift upper limits from \cite{shaw13}.}
\tablenotetext{f}{Photometric redshift upper limits from \cite{rau12}.}
\tablenotetext{g}{Blazar classification based on the frequency
of the peak of the synchrotron component as reported in \cite{2LAC}
and \cite{shaw13}.}
\tablenotetext{h}{From \cite{pita12}.}
\tablenotetext{i}{Photometric redshift or upper limits from the work of Bolmer et al. (2013, in prep.).}
\end{deluxetable}

\subsection{Best-fit Parameters to sub-classes of BL Lacs}
\label{sec:app}

Tables~\ref{tab:ple_classes} and \ref{tab:ldde_classes} report
the best-fit parameters to the HSP, ISP and LSP sub-classes
as described in $\S$~\ref{sec:hsp} and $\S$~\ref{sec:lsp}.

\begin{deluxetable}{lccccccccccc}
\tablewidth{0pt}
\tabletypesize{\scriptsize}
\rotate
\tablecaption{Best-fit parameters of the Pure Luminosity and Pure Density
Evolution LFs to sub-classes of BL Lacs. Parameters without an error estimate were kept fixed during the fit. Parameter values were computed as the median of all the best-fit parameters
to the Monte Carlo sample, while the uncertainties represent the 68\,\%
containment regions around the median value.
\label{tab:ple_classes}}
\tablehead{\colhead{Model}   & 
\colhead{A\tablenotemark{a}} & \colhead{$\gamma_1$} & 
\colhead{L$_*$\tablenotemark{b}}              & \colhead{$\gamma_2$} &
\colhead{k}                  & \colhead{$\tau$}     &
\colhead{$\xi$}              &
\colhead{$\mu^*$}              & \colhead{$\beta$}    & 
\colhead{$\sigma$}           & -2$\ln$L\tablenotemark{c}
}
\startdata 

PLE$_{HSP}$   & $7.40^{+9.46}_{-3.37}\times10^{2}$ & $1.47^{+0.88}_{-0.19}$ & $6.45^{+5.39}_{-2.94}\times10^{-2}$ & $7.62^{+2.38}_{-5.94}$ & $3.82^{+1.29}_{-1.17}$ & $1.35^{+0.17}_{-0.33}$ & $-0.41^{+0.08}_{-0.14}$ & $1.97^{+0.09}_{-0.04}$ & $4.47^{+5.25}_{-3.79}\times10^{-2}$ & $0.25^{+0.08}_{-0.03}$ & -$607.3$\\ 

PLE$_{ISP+LSP}$   & $2.72^{+6.93}_{-2.34}\times10^{2}$ & $1.60^{+1.40}_{-0.31}$ & $4.24^{+7.23}_{-2.10}\times10^{-2}$ & $ .08^{+5.92}_{-2.24}$ & $7.86^{+1.41}_{-1.86}$ & $0.98^{+0.29}_{-0.32}$ & $-0.25^{+0.05}_{-0.09}$ & $2.27^{+0.04}_{-0.03}$ & $-3.32^{+2.46}_{-3.02}\times10^{-2}$ & $0.20^{+0.03}_{-0.02}$ & -$272.0$\\

PLE$_{LSP}$   & $86.57^{+232.56}_{-58.31}$ & $1.51^{+0.77}_{-0.36}$ & $8.05^{+9.41}_{-4.34}\times10^{-2}$ & $8.14^{+1.86}_{-5.26}$ & $7.59^{+1.78}_{-2.09}$ & $1.30^{+0.26}_{-0.39}$ & $-0.23^{+0.05}_{-0.08}$ & $2.32^{+0.28}_{-0.08}$ & $-3.23^{+6.71}_{-7.25}\times10^{-2}$ & $0.23^{+0.21}_{-0.04}$ & -$81.3$\\

PLE$_{HSP+ISP}$   & $1.22^{+0.75}_{-0.55}\times10^{3}$ & $1.48^{+0.15}_{-0.13}$ & $3.68^{+2.37}_{-1.14}\times10^{-2}$ & $5.39^{+1.44}_{-1.32}$ & $5.11^{+1.03}_{-1.08}$ & $1.26^{+0.18}_{-0.21}$ & $-0.34^{+0.05}_{-0.09}$ & $2.06^{+0.03}_{-0.02}$ & $4.86^{+2.50}_{-1.90}\times10^{-2}$ & $0.25^{+0.02}_{-0.02}$ & -$715.8$\\

\enddata
\tablenotetext{a}{In units of $10^{-13}$\,Mpc$^{-3}$  erg$^{-1}$ s.}
\tablenotetext{b}{In units of $10^{48}$\,erg s$^{-1}$.}
\tablenotetext{c}{Value of the -2$\times$log-likelihood when the function
is minimized.}
\end{deluxetable}

\begin{deluxetable}{lccccccccccccc}
\tablewidth{0pt}
\tabletypesize{\tiny}
\rotate
\tablecaption{Best-fit parameters of the LDDE LFs to sub-classes of BL Lacs. Parameters without an error
estimate were kept fixed during the fit. Parameter values were computed as the median of all the best-fit parameters
to the Monte Carlo sample, while the uncertainty represent the 68\,\%
containment region around the median value.
\label{tab:ldde_classes}}
\tablehead{\colhead{Model}   & 
\colhead{A\tablenotemark{a}}          & \colhead{$\gamma_1$} & 
\colhead{L$_*$\tablenotemark{b}}      & \colhead{$\gamma_2$} &
\colhead{z$_c^*$}                     & 
\colhead{p1$^*$}                  & \colhead{$\tau$}     &
\colhead{p2}              & \colhead{$\alpha$} &
\colhead{$\mu^*$}              & \colhead{$\beta$}    & 
\colhead{$\sigma$}           & -2$\ln$L\tablenotemark{c}
}
\startdata 
LDDE$_{HSP}$   & $9.59^{+11.77}_{-5.36}$ & $0.28^{+0.25}_{-0.29}$ & $0.42^{+0.26}_{-0.20}$ & $3.47^{+16.5 }_{-1.20}$ & $1.60^{+0.20}_{-0.40}$ & $0.48^{+1.63}_{-0.48}$ & $6.76^{+2.33}_{-1.82}$ & $-11.12^{+6.10}_{-3.88}$ & $0.11^{+0.05}_{-0.08}$ & $1.97^{+0.09}_{-0.04}$ & $4.40^{+4.18}_{-3.55}\times10^{-2}$ & $0.24^{+0.08}_{-0.04}$ & -$619.4$\\ 

LDDE$_{ISP+LSP}$   & $17.1^{+212.3}_{-14.5}$ & $0.48^{+0.36}_{-1.26}$ & $0.45^{+1.65}_{-0.42}$ & $1.98^{+10.49}_{-0.71}$ & $1.15^{+0.22}_{-0.20}$ & $4.54^{+2.64}_{-2.58}$ & $3.82^{+1.66}_{-1.61}$ & $-5.89^{+2.59}_{-3.81}$ & $4.69^{+68.47}_{-106.12}\times10^{-3}$ & $2.26^{+0.04}_{-0.03}$ & $-2.81^{+2.21}_{-2.58}\times10^{-2}$ & $0.20^{+0.03}_{-0.02}$ & -$275.8$\\ 

LDDE$_{LSP}$   & $3.34^{+36.99}_{-2.05}$ & $0.48^{+0.31}_{-0.67}$ & $1.48^{+0.70}_{-1.11}$ & $6.33^{+13.67}_{-4.91}$ & $0.96^{+0.30}_{-0.12}$ & $4.10^{+5.90}_{-3.35}$ & $5.34^{+4.66}_{-2.70}$ & $-5.53^{+2.12}_{-4.97}$ & $-1.73^{+93.76}_{-206.12}\times10^{-3}$ & $2.32^{+0.20}_{-0.09}$ & $-3.24^{+7.53}_{-9.38}\times10^{-2}$ & $0.23^{+0.21}_{-0.04}$ & -$87.7$\\

LDDE$_{HSP+ISP}$   & $29.1^{+28.6}_{-16.0}$ & $0.22^{+0.24}_{-0.29}$ & $0.26^{+0.25}_{-0.13}$ & $2.10^{+1.09}_{-0.49}$ & $1.46^{+0.17}_{-0.18}$ & $1.98^{+1.46}_{-1.20}$ & $6.38^{+1.58}_{-1.66}$ & $-8.29^{+3.05}_{-5.28}$ & $9.41^{+3.81}_{-4.09}\times10^{-2}$ & $2.05^{+0.03}_{-0.02}$ & $5.55^{+2.34}_{-2.17}\times10^{-2}$ & $0.24^{+0.03}_{-0.02}$ & -$733.9$\\

\enddata
\tablenotetext{a}{In unit of $10^{-10}$\,Mpc$^{-3}$  erg$^{-1}$ s.}
\tablenotetext{b}{In unit of $10^{48}$\,erg s$^{-1}$.}
\tablenotetext{c}{Value of the -2$\times$log-likelihood when the function
is minimized.}
\end{deluxetable}

%%%%%%%%%%%%%%%%%%%%%%%%%%%%%%%%%%%%%%%%%%%%%%%%%% biblio
\bibliographystyle{apj}
\bibliography{/Users/majello/Work/Papers/BiblioLib/biblio}

\end{document}